\documentclass[epj,twocolumn]{svjour}
\usepackage{geometry}                % See geometry.pdf to learn the layout options. There are lots.
\usepackage{graphicx}
\usepackage{amssymb}
\usepackage{amsmath}
\usepackage{epstopdf}
\usepackage{epsfig}
\usepackage{epsf}
\usepackage{dcolumn}
\usepackage[FIGTOPCAP,nooneline]{subfigure} 
\usepackage{threeparttable}

\title{Active transport and cluster formation on 2D networks}
\author{Philip Greulich \and Ludger Santen}
\institute{Fachrichtung Theoretische Physik, Universit\"at des Saarlandes, Saarbr\"ucken, Germany}
\date{\today}                                           % Activate to display a given date or no date

\begin{document}
\bibliographystyle{epj}

\abstract{
We introduce a model for active transport on inhomogeneous networks embedded in a diffusive environment which is motivated by vesicular transport on actin filaments. In the presence of a hard-core interaction, particle clusters are observed that exhibit an algebraically decaying distribution in a large parameter regime, indicating the existence of clusters on all scales. The scale free behavior can be understood by a mechanism promoting preferential attachment of particles to large clusters. The results are compared with a diffusion limited aggregation model and active transport on a regular network. For both models we observe aggregation of particles to clusters which are characterized by a finite size-scale if the relevant time-scales and particle densities are considered. 
 % end of PACS codes
} %end of abstract

\maketitle

%%%%%%%%%%%%%%%%%%%%%%%%%%%%%%%%%%%%%%%%%%%%%%%%%%%%%%%%%%%%%%%%%%%%%%%%%%%%%%%%%%%%
\section{Introduction}
%%%%%%%%%%%%%%%%%%%%%%%%%%%%%%%%%%%%%%%%%%%%%%%%%%%%%%%%%%%%%%%%%%%%%%%%%%%%%%%%%%%

Active transport processes play an important role in biological and engineered systems. Examples are road traffic or active intracellular transport of vesicles and organelles by motor proteins that perform directed movement along the cytoskeleton. In recent years stochastic systems of self-driven particles have been applied to model intracellular transport of motor proteins in a number of works \cite{pff1,ludger_pff,kif1a1,kif1a2,motors_klumpp2,motors_klumpp4,2spec_klumpp}. Most of these models are variants of the \emph{totally asymmetric simple exclusion process (TASEP)} (with and without Langmuir kinetics) which serves as a paradigmatic system for driven non-equi\-librium systems.  While the microtubules usually arrange in an ordered pattern  (e.g. a radial structure in  most mammalian cells, though longitudinal in neuronal axons), actin filaments often form randomly structured undirected networks. Therefore the investigation of transport on networks arises to be an interesting object of research. 

Active transport on an undirected but regular network has been investigated by Klumpp et al. \cite{lipowski_network} who studied the dynamic properties of diffusive non-interacting particles on a lattice with an embedded regular square network, which consists of active stripes where particles perform biased motion. They showed that though movement of particles remains globally diffusive on long time-scales, the diffusion constant is enhanced by the presence of the network. 

In transport systems considering steric exclusion interaction between particles, aggregation, manifesting in the formation of jams, is a common phenomenon. Jams can form in one dimensional systems with single tracks due to boundary conditions (\emph{boundary induced phase transitions} \cite{krug1}), induced by defects \cite{lebowitz_1def_1,barma3,asep_def,pff_dis,PASEP_dis_santen,chou1} or they emerge spontaneously due to stochastic slow down of vehicles in highway traffic \cite{NaSch,traffic1}. In two dimensional regular street networks, mutual interference of vehicles at intersections lead to jamming \cite{street_network}.

Studies of transport on inhomogeneous topological networks (graphs with nodes and edges, no distances) revealed an interesting phenomenology. E.g. non-interacting particles performing a random walk exhibit an inhomogeneous density distribution on the nodes \cite{noh_rw_network}, while inclusion of an attractive zero range interaction even allows the particles to aggregate and form a condensate, corresponding to nodes containing a finite fraction of particles \cite{noh_netw_intpart}. These results show that the structure of a transport network strongly influences  transport properties. In order to model active transport on actin filament networks, it is therefore necessary to consider realistic network structures. 

For many biological processes, concentration gradients are crucial. 
One example is the aggregation of proteins inside the cell or in the cell membrane. Clusters of aggregated proteins can be observed and characterized experimentally for example by high resolution fluorescence microscopy \cite{sieber_memb_prot}. In some cases these clusters are essential for cell functionality but they can also lead to dysfunctions or even apoptosis. In yeast cell membranes for example one observes the aggregation of Erd2p-receptors which can promote the internalization of toxins \cite{schmitt}. Most recent works on membrane protein aggregation considered an attractive interaction between proteins as source for (reversible) aggregation \cite{memb_prot_clusters,sieber_memb_prot,gil_memb_prot}. For this kind of particle dynamics, the resulting clusters are governed by a well defined size scale. 

Jamming in vesicular transport may yield an alternative aggregation mechanism of proteins. Vesicles, transported on different filaments can block each other at filament crossing points, inducing queuing of vesicles. The existence of a quasi ~two-dimensional irregular actin filament network beneath the membrane \cite{alberts,actin_patches} suggests jamming of vesicles prior exocytosis, resulting in receptor clusters on the membrane surface. In this case large scale features of cluster distributions can vary from diffusion limited aggregation. The limits of resolution in optical microscopy \cite{methods} do not allow to distinguish clustered single receptors. By contrast the size of larger particle aggregates can in principle be given with relatively high precision \cite{sieber_memb_prot}. Therefore it is useful to relate the cluster size distribution with microscopic transport mechanisms by means of theoretical modelling. 

In this work we propose a model for active transport of extended hard-core particles (corresponding to vesicles) on a two-dimensional randomly disordered network embedded in a diffusive environment. The model is motivated by intracellular transport on submembranal networks, we therefore adapt the model parameters to this reference system. We check particle configurations in order to identify the formation of clusters and investigate cluster size distributions. The results are compared with a regular network in diffusive environment and a diffusive system without network where attractive particle-particle interactions promote cluster formation. The main focus will be on robust properties of clusters that serve as criteria to discriminate between different microscopic aggregation dynamics.

%%%%%%%%%%%%%%%%%%%%%%%%%%%%%%%%%%%%%%%%%%%%%%%%%%%%%%%%%%%%%%%%%%%%%%%%%%%%%%%%%%%%%%%%%%%%%%%%%%%%%%%%%%%%%%
\section{Network models}
%%%%%%%%%%%%%%%%%%%%%%%%%%%%%%%%%%%%%%%%%%%%%%%%%%%%%%%%%%%%%%%%%%%%%%%%%%%%%%%%%%%%%%%%%%%%%%%%%%%%%%%%%%%%%%%%

In the following, we introduce stochastic models in order to study the influence of the network structure on dynamical properties.
% While the presence of a network in a diffusive environment without interaction between particles merely leads to the enhancement of the diffusion process \cite{lipowski_network}, we are rather interested in collective patterns emerging due to interactions.  
Our simulations use stochastic dynamics in order to integrate the many-particle Master equation. At each time step, $N$ particles within the system of size $L$ are randomly chosen and updated (random sequential update) according to the rules given in Table \ref{reg_netw_dynamics} and \ref{particle_dyn_tab}, applying periodic boundary conditions. Time steps are normalized such that on average each free particle performs one diffusive step per time step $\Delta t$. Results are discussed for different \emph{particle densities}  $\rho_p^0:=N/L^2$ which, for biological reasons, is chosen as $\rho_p^0=0.04$ if not stated differently (see appendix).

\subsection{Regular networks}
\label{reg_netw_subsec}

%%%%%%%%%%%%%%%%%%%%%%%%%%%%%%%%%%%%%%%%%%%%%%%%%%%%%%%%%%%%%%%%%%%%%%%%%%%%%%%%%%%%%%%%%%
\begin{table*}
\begin{center}
\begin{tabular}{llll}
Process & Particle state(s) & Description & Probability  \\ \hline \\ \vspace{2mm} 
\emph{Diffusion} & D & \begin{minipage}[t]{8cm} Detached particles move to sites randomly chosen from the four neighbors \end{minipage} & $\omega_D=1$ \\ \vspace{2mm}
\emph{Forward Step} & A & \begin{minipage}[t]{8cm} Attached particles move to the next site in forward direction of filament \end{minipage} & $p=0.5$ \\ \vspace{2mm}
\emph{Attachment} & D$\to$A & \begin{minipage}[t]{8cm} Detached particles on active sites becomes bound \end{minipage} & $\omega_a=0.25$ \\ \vspace{2mm}
\emph{Detachment} & A$\to$D & \begin{minipage}[t]{8cm} Attached particles become detached \end{minipage} & $\omega_d=0.02$ \\ \vspace{2mm}
\emph{Blocking} & D & \begin{minipage}[t]{8cm} Forward movement of particles adjacent to intersection sites is inhibited if other particles occupy sites adjacent to intersection \end{minipage} & $b=1$
\end{tabular}
\caption{\label{reg_netw_dynamics} Brief prescription of the dynamic processes in the regular network model. Column 2 displays the particle states ``attached''(A) and ``detached''(D). The right column displays the probability that the respective process occurs within one time step. Numerical values given in the right column are default values which are used if not stated else and are chosen to fit the ones in \cite{lipowski_network}. System size $N=200$, mesh size $a=10$ sites.}
\end{center}

\end{table*}
%%%%%%%%%%%%%%%%%%%%%%%%%%%%%%%%%%%%%%%%%%%%%%%%%%%%%%%%%%%%%%%%%%%%%%%%%%%%%%%%%%%%%%%%%%%%%%%%%%%%

%%%%%%%%%%%%%%%%%%%%%%%%%%%%%%%%%%%%%%%%%%%%%%%%%%%%%%%%%%%%%%%%
\begin{center}
\begin{figure}
\resizebox{0.9\columnwidth}{!}{\includegraphics[width=8cm]{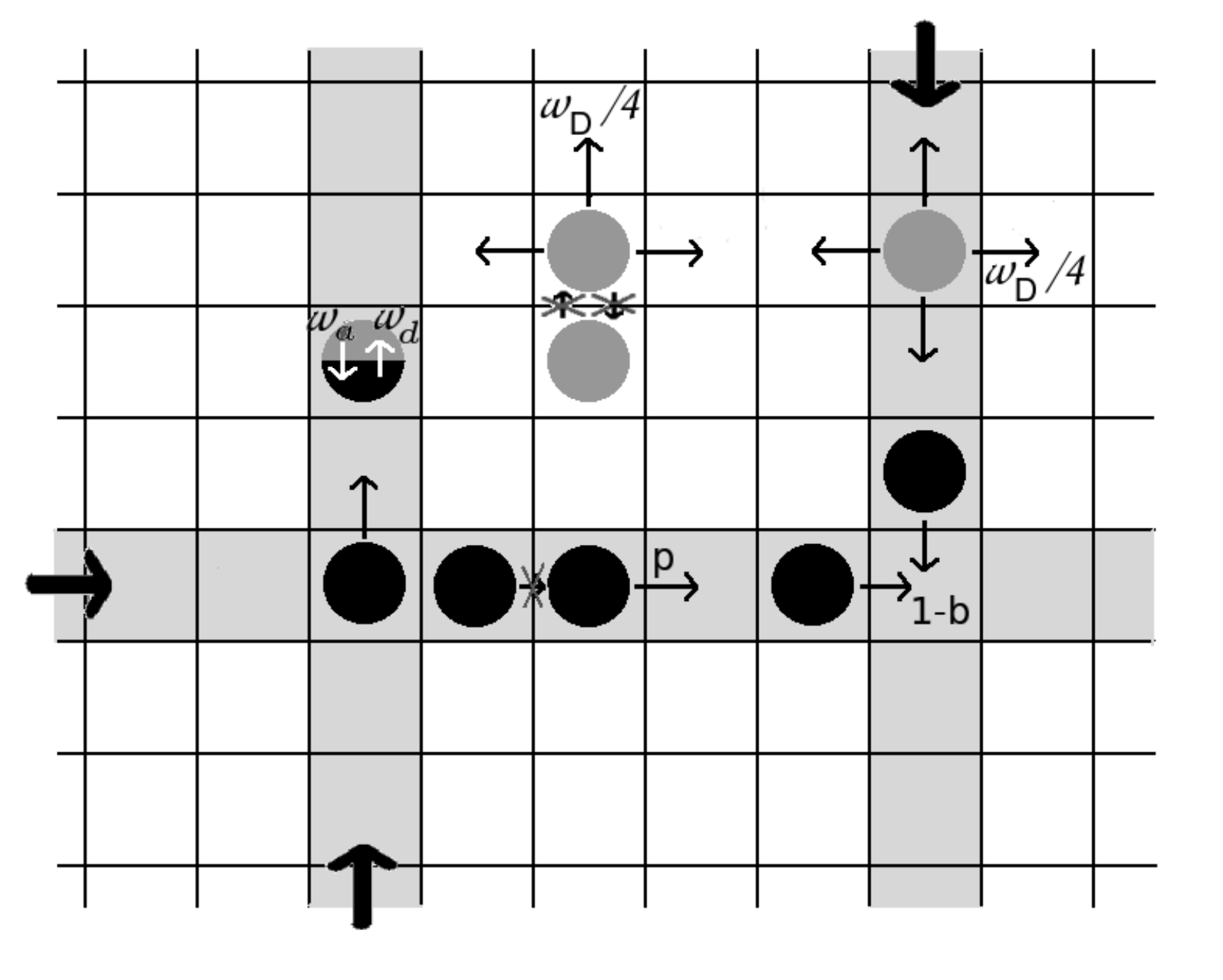}}
\caption{Illustration of the dynamics in a regular network. Dark gray discs are particles diffusing freely with rate $\omega_D$. Black discs represent attached particles that can only step in the preferred direction of the active stripe they occupy (bold arrows) with rate $p$. On filament sites (light gray), particles can interchange between attached and detached state with rates $\omega_a$ and $\omega_d$, respectively. Crossed arrows denote steps that are inhibited due to the exclusion principle.}
\label{reg_netw_illust}
\end{figure}
\end{center}

%%%%%%%%%%%%%%%%%%%%%%%%%%%%%%%%%%%%%%%%%%%%%%%%%%%%%%%%%%%%%%%%%%%%%%%%%%%%%

As a the first example for active transport on networks a discrete lattice gas model with a square network of active stripes, similar to the model investigated in \cite{lipowski_network}, is considered. $N \times N$ sites are arranged in a square lattice of edge length $L=N \Delta x$ where $\Delta x$ is the lattice spacing. Each site can either be empty or occupied by at most one particle. We distinguish the particle states \emph{attached}(A) and \emph{detached}(D). Detached particles always move diffusively. The system contains stripes of \emph{active} sites that constitute a regular square transport network. If particles are located at an active stripe, they can attach (if not yet attached) or detach (if attached). Attached particles perform a directed motion along stripes.  The orientation of stripes was chosen randomly with equal probability. Steps that would result in double occupation of a site are prohibited. 

%We distinguish between \emph{non-filament sites}, where particles perform a random walk and are always unbound and \emph{filament sites}, where particle perform a biased movement in a given direction if they are bound, or random walk if they are unbound. Particles on filament sites that are unbound can attach with a \emph{attachment rate} $\omega_a$ and bound particles can detach with probability $\omega_d$. A particle cannot step on an adjacent site if the target site is already occupied, corresponding to exclusion interaction. The explicite dynamics of the system are displayed in table \ref{reg_netw_dynamics} and illustrated in figure \ref{reg_netw_illust}. 

%One dimensional exclusion models with directed movement have been thoroughly investigated, mentioning the TASEP as the most prominent example, while there is a lack of knowledge on the influence of exclusion interaction in two dimenional networks \emph{[EVTL IN INTRO]}. 
Compared to the dynamics of non-interacting self-driven particles qualitatively new ~features arise due to the steric particle-particle interactions at intersections of the network.  Here we introduce an additional parameter, the \emph{blocking probability}: If at least two particles are at sites adjacent to an intersection site, each  particle may only access the intersection site with the probability $1-b$ (cf. figure \ref{reg_netw_illust}). Particles on intersection sites retain their moving direction. 

%%%%%%%%%%%%%%%%%%%%%%%%%%%%%%%%%%%%%%%%%%%%%%%%%%%%%%%%%%%%%%%%%%%%%%%%%%%%%%%%%%%%
\begin{table*}
\begin{center}
\begin{tabular}{cccc}
Process & Particle state(s) & Description & Probability \\ \hline \\
\emph{Diffusion} & D &
\begin{minipage}[t]{8cm} Detached particles move in a random direction. Step widths are uniformly distributed between $0$ and $2l_D$  \end{minipage} & $\omega_D=1$ \\
\emph{Step} & A & 
\begin{minipage}[t]{8cm} Attached particles move to adjacent subunit in (+)-direction.  \end{minipage} & $p=0.5$ \\
\emph{Attachment} & D$\to$A & 
\begin{minipage}[t]{8cm} Particles bind to subunits if their distance is less than $d_{b}$, becoming 'attached' \end{minipage} & $\omega_a=0.25$ \\
\emph{Detachment} & A$\to$D& 
\begin{minipage}[t]{8cm} Particles detach \end{minipage} & $\omega_d=0.02$ \\
\end{tabular}
\caption{\label{particle_dyn_tab} Particle dynamics in the inhomogeneous network. A='attached'; D='detached'. The given values in the right column mark default parameters chosen similar to \cite{lipowski_network}. The default particle density is $\rho_p^0$ (see Table \ref{default_parameters}).} 
\end{center} 
\end{table*}
%%%%%%%%%%%%%%%%%%%%%%%%%%%%%%%%%%%%%%%%%%%%%%%%%%%%%%%%%%%%%%%%%%%%%%%%%%%%%%%%%%%%

The explicit rules for the particle dynamics are displayed in Table \ref{reg_netw_dynamics} and illustrated in figure \ref{reg_netw_illust}. We have chosen 
the default parameter values analogous to \cite{lipowski_network} $\omega_{D}=1,\, p=0.5, \, \omega_d=0.02$. The particle density was chosen to be $\rho_p^0=0.04$, analogous to the value of the inhomogeneous network, and the system size $N=200$. In \cite{lipowski_network} the attachment rate is equal to one, which corresponds to an effective attachment rate $\omega_a=0.25$ if a particle is on an adjacent non-active site\footnote{In contrast to \cite{lipowski_network}, we allow for crossing of active stripes by diffusion.}. To be consistent with the subsequent continuous space model, we choose $b=1$.

\subsection{Inhomogeneous networks}
\label{disNetw_sec}

Generalizing to continuous space and allowing for arbitrary randomly distributed directions and lengths of active stripes we present a continuous model with randomly generated linear \emph{filaments} where hard-core particles can perform directed paths along these filaments. The model is motivated by vesicular transport on actin filament networks \cite{alberts}.

\subsubsection{General properties of the model}

The main components of our model are filaments and particles interacting via a spherical hard-core potential represented by a disc of radius $r_{p}$. This hard core potential is implemented by cancelling any steps that would result in an overlap of discs. Filaments are represented by linear sequences of subunits with a distance of $d_s$ in between. They are directed with a \emph{(-)-end} and a  \emph{(+)-end} at which new subunits can be generated to elongate the filament. Particles can attach to subunits that are within a distance less than $d_b$ and perform steps to adjacent subunits in the (+)-direction of the filament. The filaments are generated by a stochastic process that yields an isotropic random distribution of filament orientations and -lengths, which is mainly characterized by the number of subunits per area element $\rho_s$ and their length $d_s$. The properties of the stochastic process determine the structure of the filament network. Our simulations were performed on a network generated by dynamics that mimic the growth dynamics of real actin networks. The dynamics is discussed in the appendix (dynamics in Table \ref{dis_netw_dyn_tab}, default parameters in Table \ref{default_parameters}). This allows to adapt the model to real biological systems.

% However, some phenomena only emerge in a finite region of parameter space and a comparison with the biological parameters appears to be relevant. If not stated differently, we will use default parameters displayed in Table \ref{default_parameters} for our simulations. These are motivated by comparison with experimental data and model fits. Particle dynamics are mainly orientated on the dynamics in \cite{lipowski_network}. 

%There are two types of particles differing in their moving direction on filaments. Particles of type 1 move towards the (+)-end of filaments, while those of type 2 move to the (-)-end. In most considerations the hard core exclusion interaction of particles that inhibits overlap of particles is included. 

\subsubsection{Particle dynamics}

%system dynamics depend on parameters, displayed in \ref{table parameters} which correspond to biological features: \newline

%%%%%%%%%%%%%%%%%%%%%%%%%%%%%%%%%%%%%%%%%%%%%%%%%%%%%%%%%%%%%%%%
\begin{center}
\begin{figure}
\resizebox{0.9\columnwidth}{!}{\includegraphics[width=9cm]{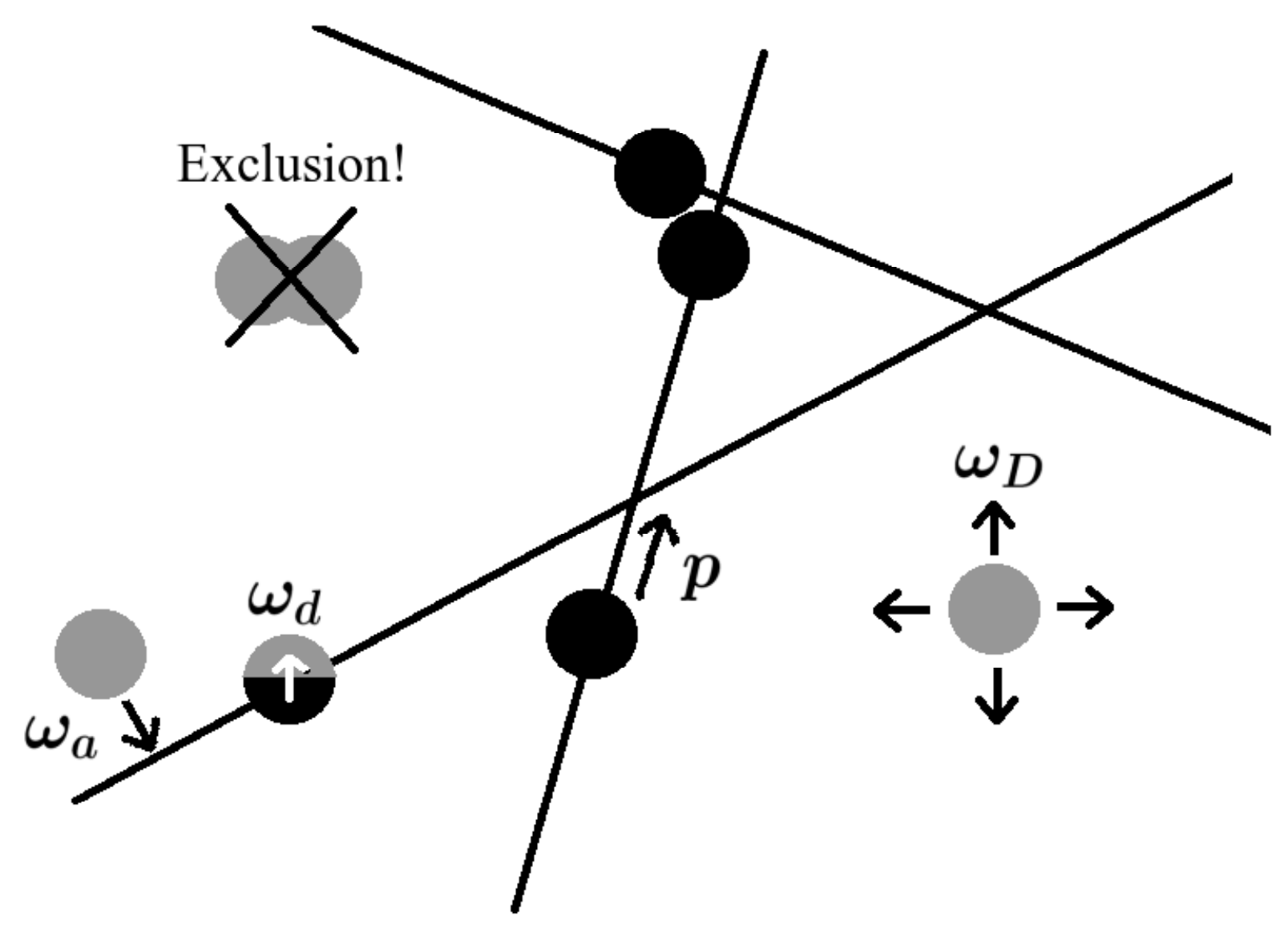}}
\caption{Illustration of the particle dynamics in an inhomogeneous network. Dark gray discs are free particles, black discs represent particles attached to filaments stepping to adjacent subunits (distance $d_s$) with rate $p$. Particles can attach to filaments with rate $\omega_a$ if they are within the binding distance $d_b$ and detach with rate $\omega_d$. Overlapping is inhibited due to exclusion.}
\label{dis_netw_part_illust}
\end{figure}
\end{center}

%%%%%%%%%%%%%%%%%%%%%%%%%%%%%%%%%%%%%%%%%%%%%%%%%%%%%%%%%%%%%%

After construction of the network particles obeying the exclusion principle are fed into the system at random positions ($\rho_p^0=0.04$ if not stated else, see Table \ref{default_parameters}). As mentioned above, the particle positions are updated following a random sequential update scheme, whereby the particle-particle as well as the  interactions between particles and the generated static network are considered . Like in the regular network model particles can freely diffuse in the 'detached' state and perform directed movement in the 'attached' state. The rules of the particle dynamics are prescribed in Table \ref{particle_dyn_tab} and illustrated in fig. \ref{dis_netw_part_illust}. We chose default parameters to fit the model in \cite{lipowski_network} (adjusting to the altered length and time scale) as displayed in Table \ref{default_parameters}.

%Of course also for random distribution of particles, clusters can form, though we want to distinguish dynamic driven clustering from random clustering. The probability, that two particle neighbourhoods %overlap is $P_{ol}=a_0/A_1\approx a_0/L^2$, where $a_0=\pi((2\lambda)^2-(2r_{p})^2)$ and $A_N=L^2-n\pi(2r_{p})^2$ is the effective configuration space of a particle added to an existing $n$-particle %distribution. The event that $n$ given particles are in the same clusters correponds to the occurence of at least $n-1$ overlapping events, which occurs with probability $P(n)={\mathcal %O}\left(\left(\frac{a_0}{L^2}\right)^{n-1}\right)$. Since the number of possibilities to combine $n$ out of $N$ particles is ${\mathcal O}(N^n)$, the probability for the existence of a $n$-particle-cluster %in a system with $N$ particles is $\tilde P(n)\sim N{\mathcal O}(\rho^{n-1})$ with the density of particle neighbourhoods $\rho=\frac{N a_0}{L^2}$. If we consider dilute systems with %$\rho^{-k}<N<\rho^{-k-1}$ particles and claim that the probability of a random $n$-cluster should be smaller than ${\mathcal O}(\rho)$, we can only consider clusters with more than $k$ particles as ``real'' %clusters of dynamic origin. 

%Stable dynamic clusters can only emerge if the flux of particles approaching the crossing on a filament is larger than the outflow of particles due to detachment and bypassing. In this case the lifetime of clusters will be large, while globally particles tend to be in clusters rather than moving free. Therefore we introduce two quantities to characterize clustering:

\section{Numerical results}

\subsection{Characterization of Clusters}
\label{cluster_char_sec}

Our aim is to relate the microscopic particle dynamics to the size distribution of their aggregates. In this section we discuss the definition of clusters for the different model systems. 

%%%%%%%%%%%%%%%%%%%%%%%%%%%%%%%%%%%%%%%%%%%%%%%%%%%%%%%%%%%%%%%%%%%%%%%%%%%%%%%%%%%%
\begin{center}
\begin{figure}
\resizebox{0.9\columnwidth}{!}{\includegraphics{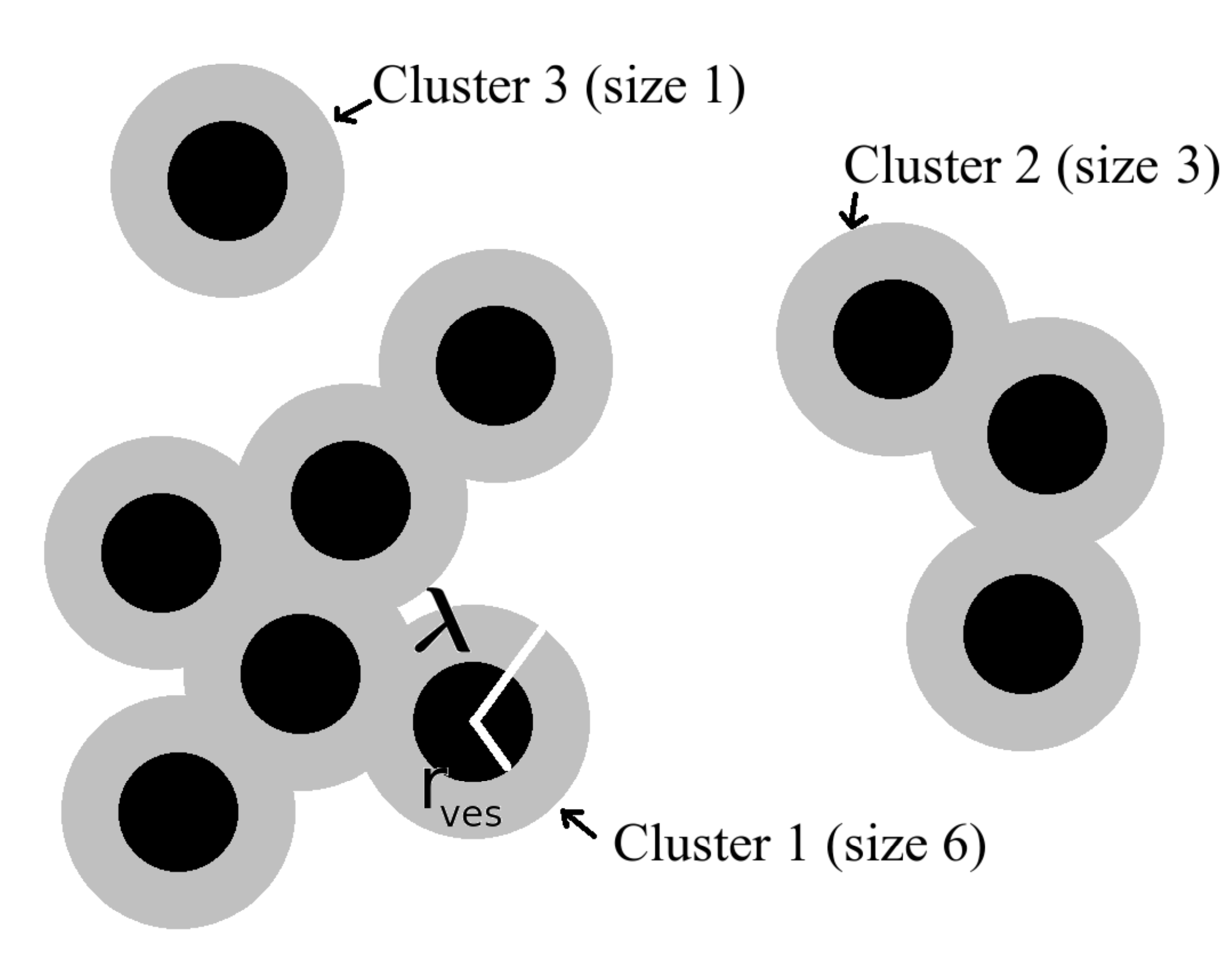}}
\caption{\label{illust_clusterdef} Illustration of particle clusters. Black discs represent particles, while gray discs are the $\lambda$-neighborhoods of each particle. Connected gray areas are clusters; the size of a given cluster is the number of particles on it.}
\end{figure}
\end{center}
%%%%%%%%%%%%%%%%%%%%%%%%%%%%%%%%%%%%%%%%%%%%%%%%%%%%%%%%%%%%%%%%%%%%%%%%%%%%%%%%%%%%

%%%%%%%%%%%%%%%%%%%%%%%%%%%%%%%%%%%%%%%%%%%%%%%%%%%%%%%%%%%%%%%%%%%%%%%%%%%%%%%%%%%%
\begin{figure}
\begin{center}
\subfigure[continuous space]{\label{random_clusters_cont}\resizebox{\columnwidth}{!}{\includegraphics{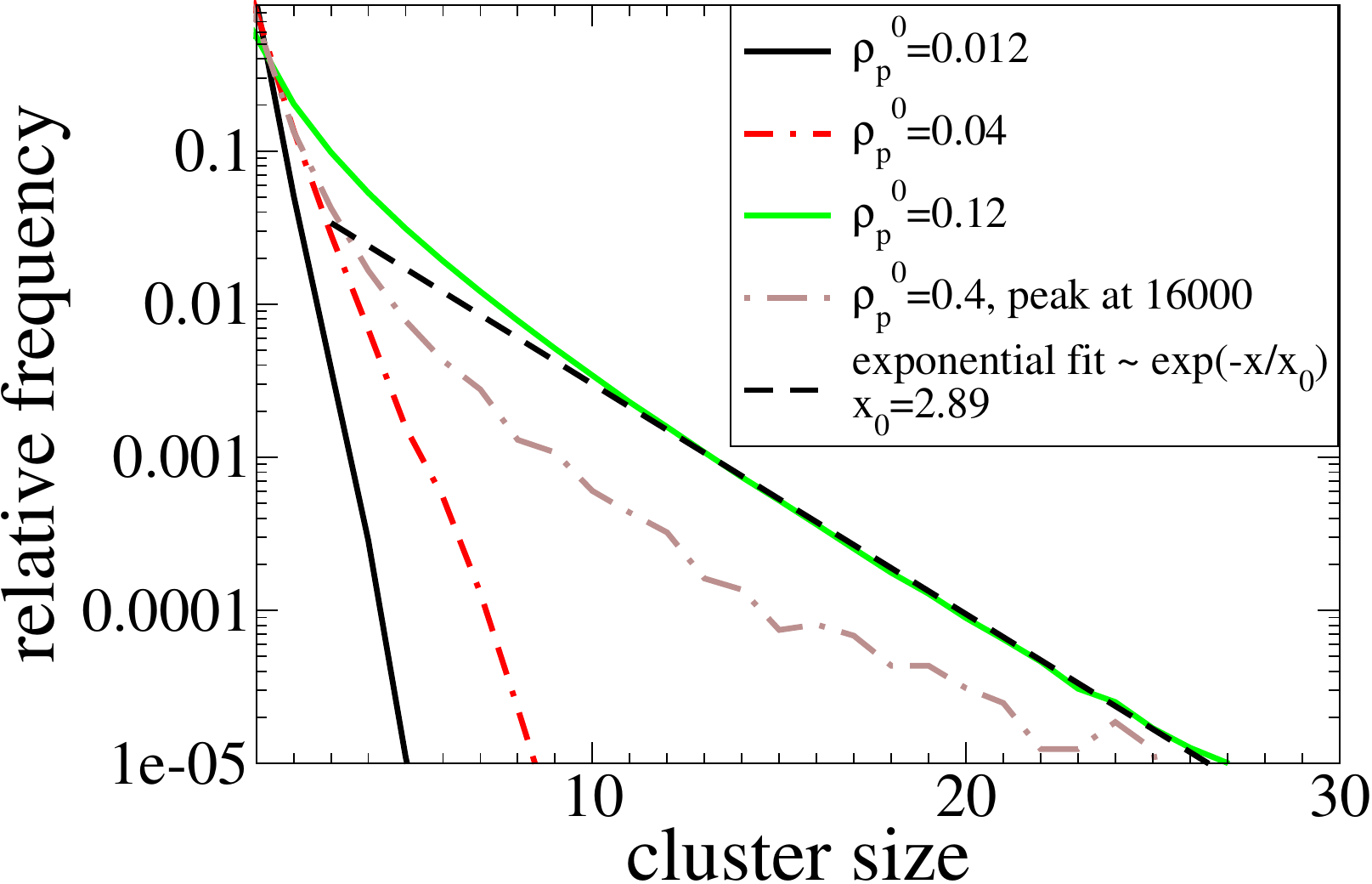}}}
\hspace{5mm}
\subfigure[discrete space]{\label{random_clusters_discr}\resizebox{\columnwidth}{!}{\includegraphics{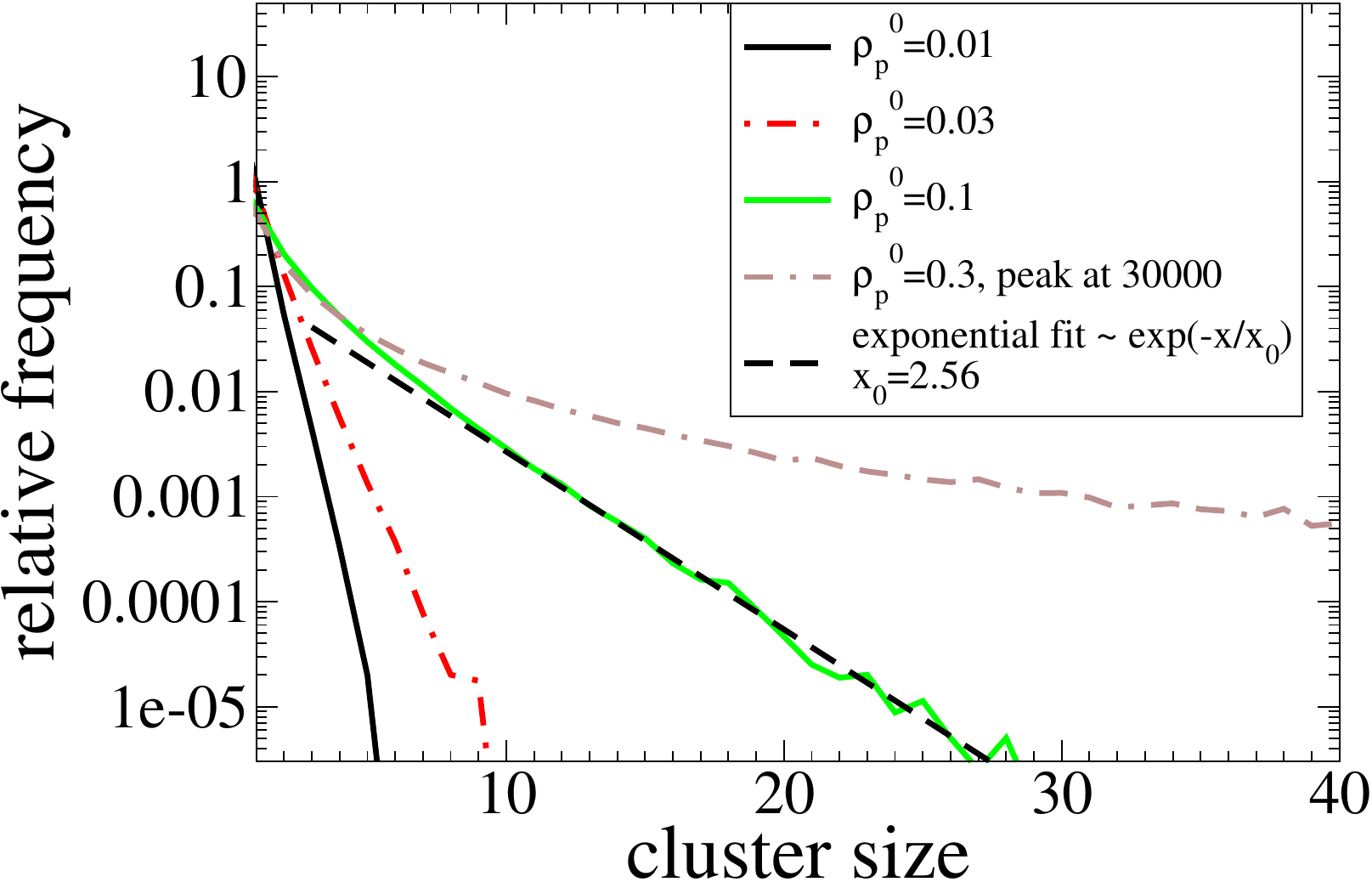}}}
\caption{\label{random_clusters} Cluster size distributions of random particle configurations in dependence on the particle density. For low densities the CD decays fast with a short size-scale. For large densities clusters on large size-scales and even such that span the whole system emerge (not visible in figure since on too large size-scale).}
\end{center}
\end{figure}
%%%%%%%%%%%%%%%%%%%%%%%%%%%%%%%%%%%%%%%%%%%%%%%%%%%%%%%%%%%%%%%%%%%%%%%%%%%%%%%%%%%%

%%%%%%%%%%%%%%%%%%%%%%%%%%%%%%%%%%%%%%%%%%%%%%%%%%%%%%%%%%%%%%%%%%%%%%%%%%%%%%%%%%%%
\begin{figure}
\begin{center}
\subfigure[continuous space]{\label{random_clusters_cont_lambdavar}\resizebox{\columnwidth}{!}{\includegraphics{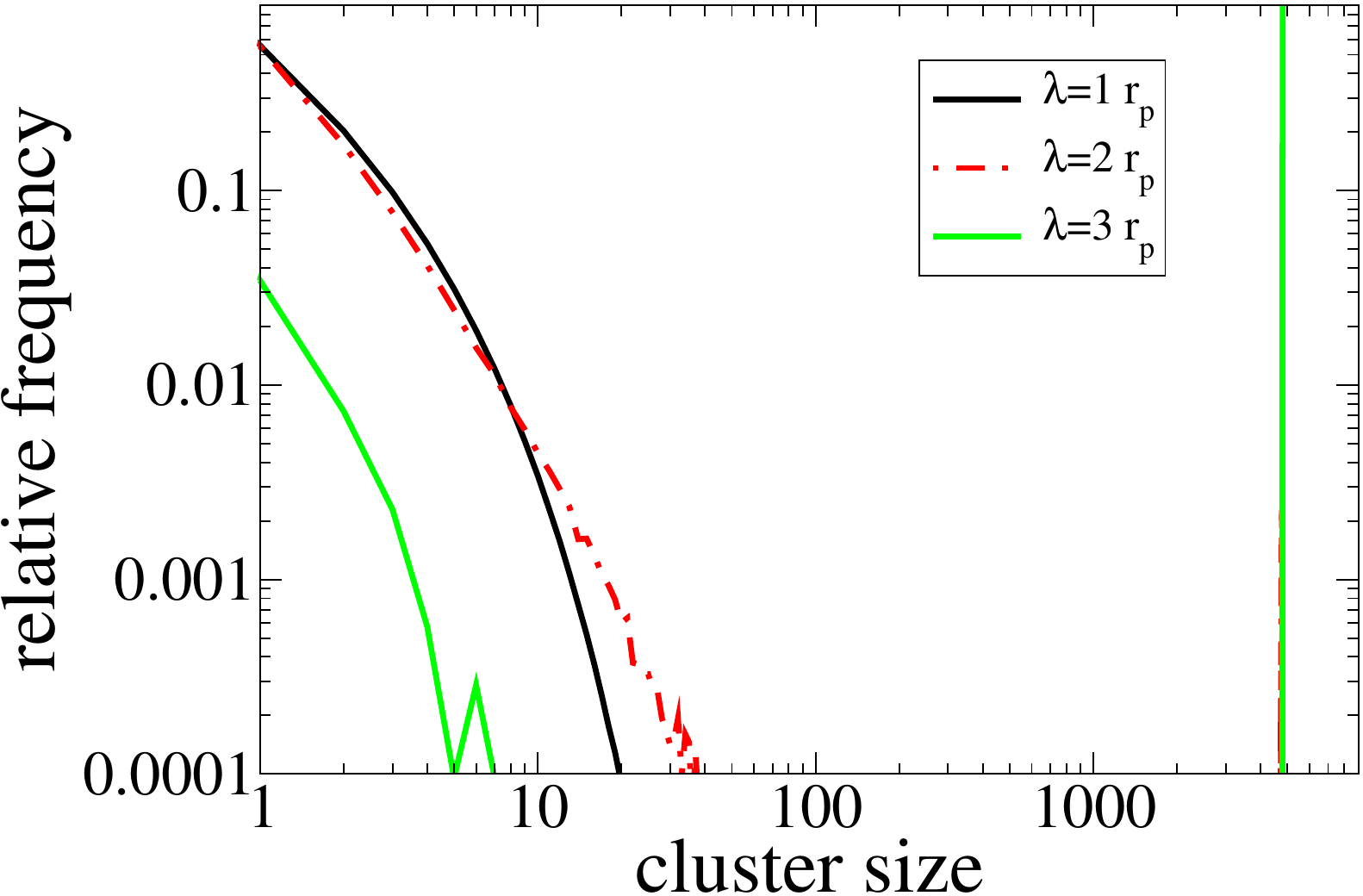}}}
\hspace{5mm}
\subfigure[discrete space]{\label{random_clusters_discr_lambdavar}\resizebox{\columnwidth}{!}{\includegraphics{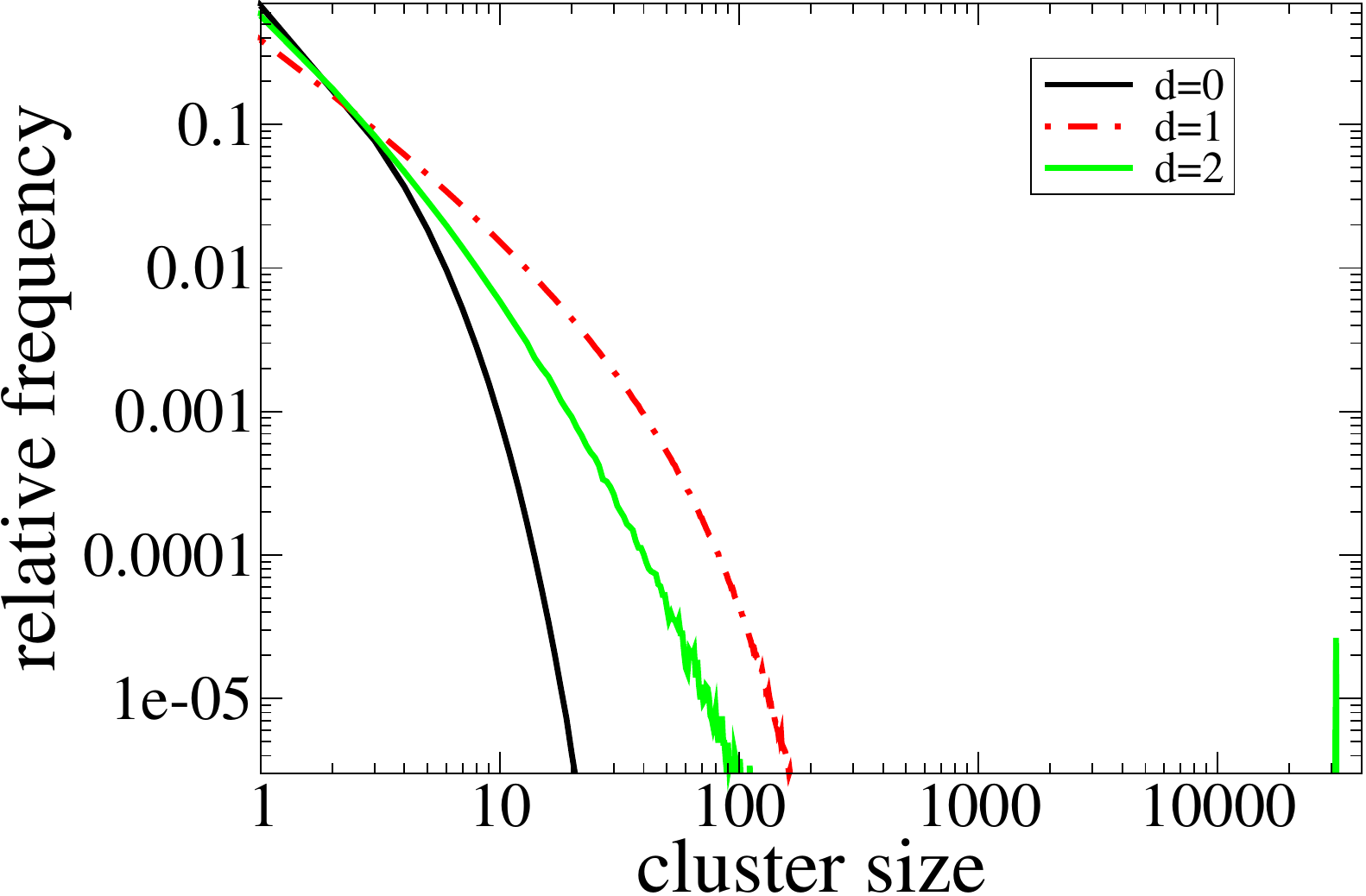}}}
\caption{\label{random_clusters_lambdavar} Cluster size distributions of random particle configurations for different coarse graining scales. $\lambda$ is the radius of the environment as defined in section \ref{cluster_char_sec}, while $d$ represents the maximum distance (number of vacancies) allowed between two particles connecting a cluster. One observes that for large coarse graining scales clusters spanning the whole system emerge.}
\end{center}
\end{figure}
%%%%%%%%%%%%%%%%%%%%%%%%%%%%%%%%%%%%%%%%%%%%%%%%%%%%%%%%%%%%%%%%%%%%%%%%%%%%%%%%%%%%

Clusters are groups of particles that are connected by overlapping neighborhoods. We therefore introduce the \emph{$\lambda-$neighborhood} of a particle representing a disc of radius $\lambda$ around the center of the particle. A cluster is defined as a set of particles included in a connected area of $\lambda-$neighborhoods (cf. fig. \ref{illust_clusterdef}). If continuous space variables are used, there exists no natural scale which identifies two particles as neighbors. We therefore have to specify the value of $\lambda$. In order to extract relevant results, we choose $\lambda$ such that qualitative results are robust on variation of $\lambda$. If not stated differently we choose $\lambda=2r_p$, which turns out to meet this condition (cf. fig \ref{CD_dis_netw_lambdavar}). 

In lattice models, static particle clusters are usually considered as connected sets of adjacent particles. However, this definition is not appropriate in this context since clusters move by propagation of vacancies. Therefore we consider particles separated by a single vacancy as belonging to the same cluster.  
Our main interest is in ensemble and time averages of \emph{cluster size distributions (CD)} and their asymptotic behavior. CDs display the relative frequency of cluster sizes emerging in the system. If not stated differently we averaged over 50000 time steps within individual runs, evaluating cluster distributions in distances of 500 time steps, taking an ensemble of 100 samples.

Clustering also occurs for random particle configurations. 
%in random systems in the absence of interactions or inhomogeneities
In fig. \ref{random_clusters} and \ref{random_clusters_lambdavar} we displayed cluster size distributions of random configurations in discrete and continuous space for different particle densities. Here the density $\rho_p^0$ is the particle number per area unit which corresponds to $(2r_p)^2$ in continuous space and one site in the lattice model. If 
%particles are randomly distributed and 
densities are not too large, the formation of large clusters is impeded resulting in an exponentially decaying cluster size distribution. For high densities one observes a small peak at the right end. At these densities clusters spanning the whole system emerge. In order to rule out these kinds of random clustering we will only consider densities below the regime of spanning clusters at relevant scales $\lambda$. In this work we are interested in cluster formation mechanisms beyond random clustering.

In the following, we will focus on particle configurations and cluster-size distributions in several transport models.

\subsection{Aggregation without network}
\label{attr_int_sec}

%%%%%%%%%%%%%%%%%%%%%%%%%%%%%%%%%%%%%%%%%%%%%%%%%%%%%%%%%%%%%%%%%%%%%%%%%%%%%%%%%%%%
\begin{figure*}
\begin{center}
\resizebox{0.65\columnwidth}{!}{\includegraphics{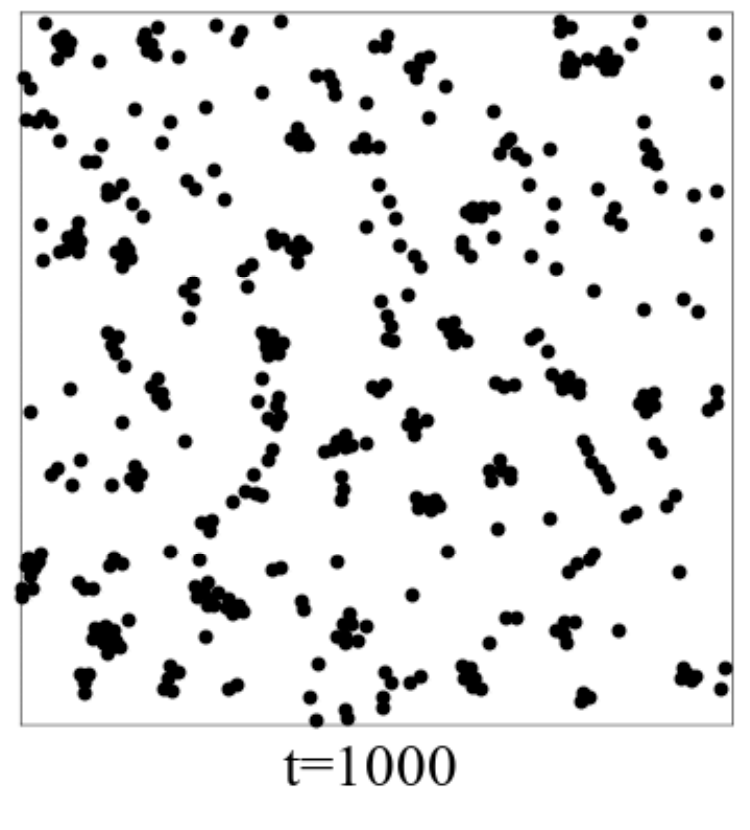}}
\hspace{5mm}
\resizebox{0.65\columnwidth}{!}{\includegraphics{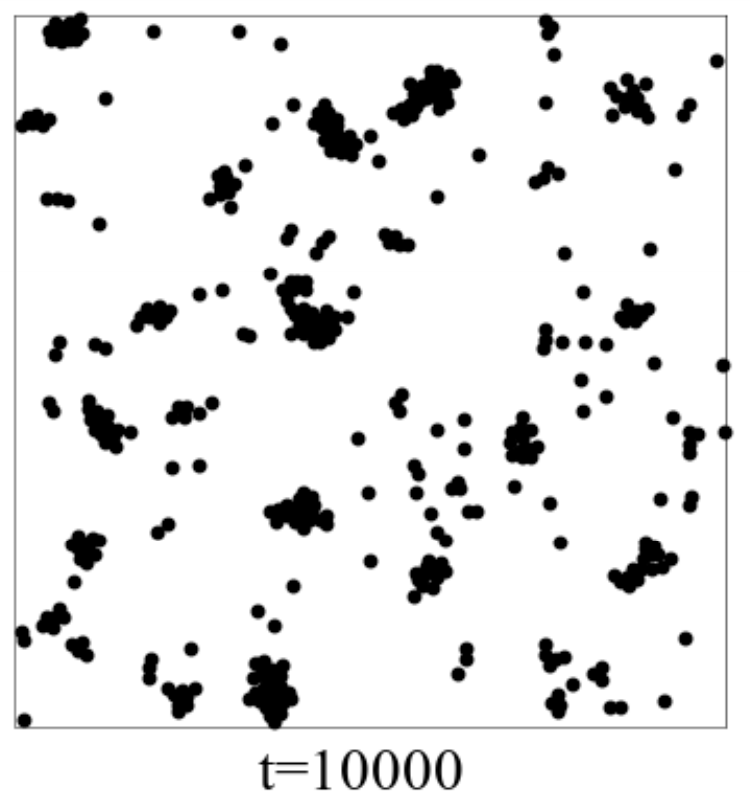}}
\vspace{5mm}
\resizebox{0.65\columnwidth}{!}{\includegraphics{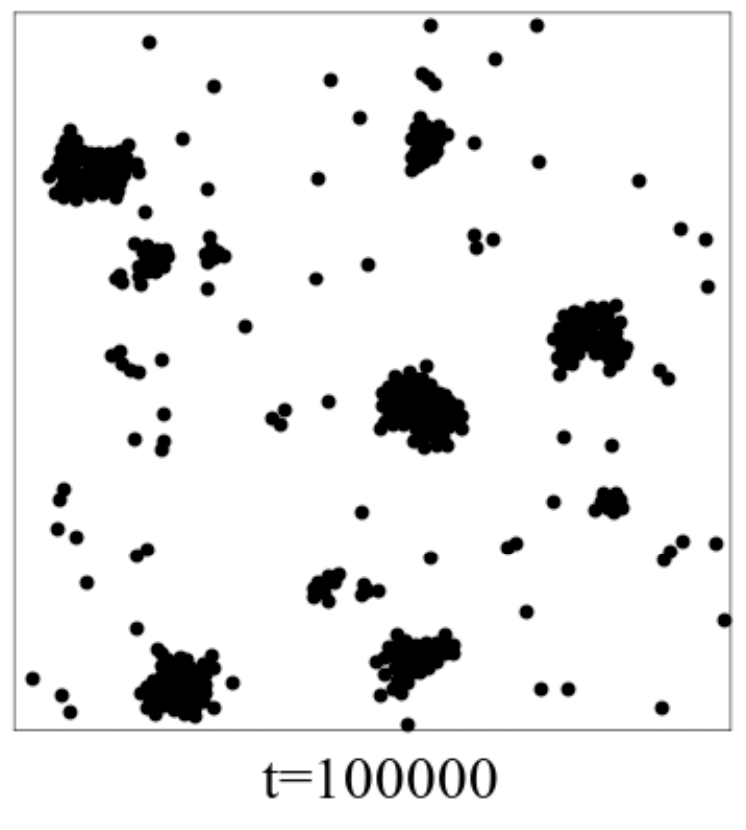}}
\hspace{5mm}
\resizebox{0.65\columnwidth}{!}{\includegraphics{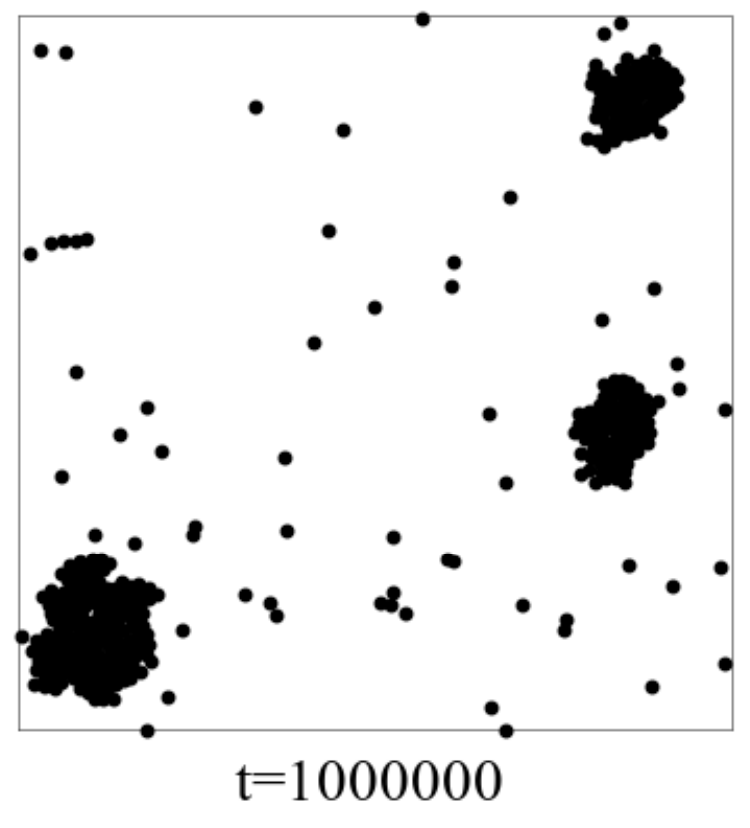}}
\caption{\label{config_no_netw} Configurations of particles (black discs = particle neighborhoods with radius $\lambda=2\,r_p$), exhibiting a mutual attractive interaction. Snapshots at different times for $V_0=2, d_{V}=3.5,\, \rho_p^0=0.04, \, L=200$, 1 timestep$\hat{=} 0.025sec$. One observes that already at small times clusters form and for long run times they the number of clusters decreases, while the average size of cluster increases.}
\end{center}
\end{figure*}
%%%%%%%%%%%%%%%%%%%%%%%%%%%%%%%%%%%%%%%%%%%%%%%%%%%%%%%%%%%%%%%%%%%%%%%%%%%%%%%%%%%%

%%%%%%%%%%%%%%%%%%%%%%%%%%%%%%%%%%%%%%%%%%%%%%%%%%%%%%%%%%%%%%%%%%%%%%%%%%%%%%%%%%%%
\begin{figure}
\begin{center}
\resizebox{\columnwidth}{!}{\includegraphics{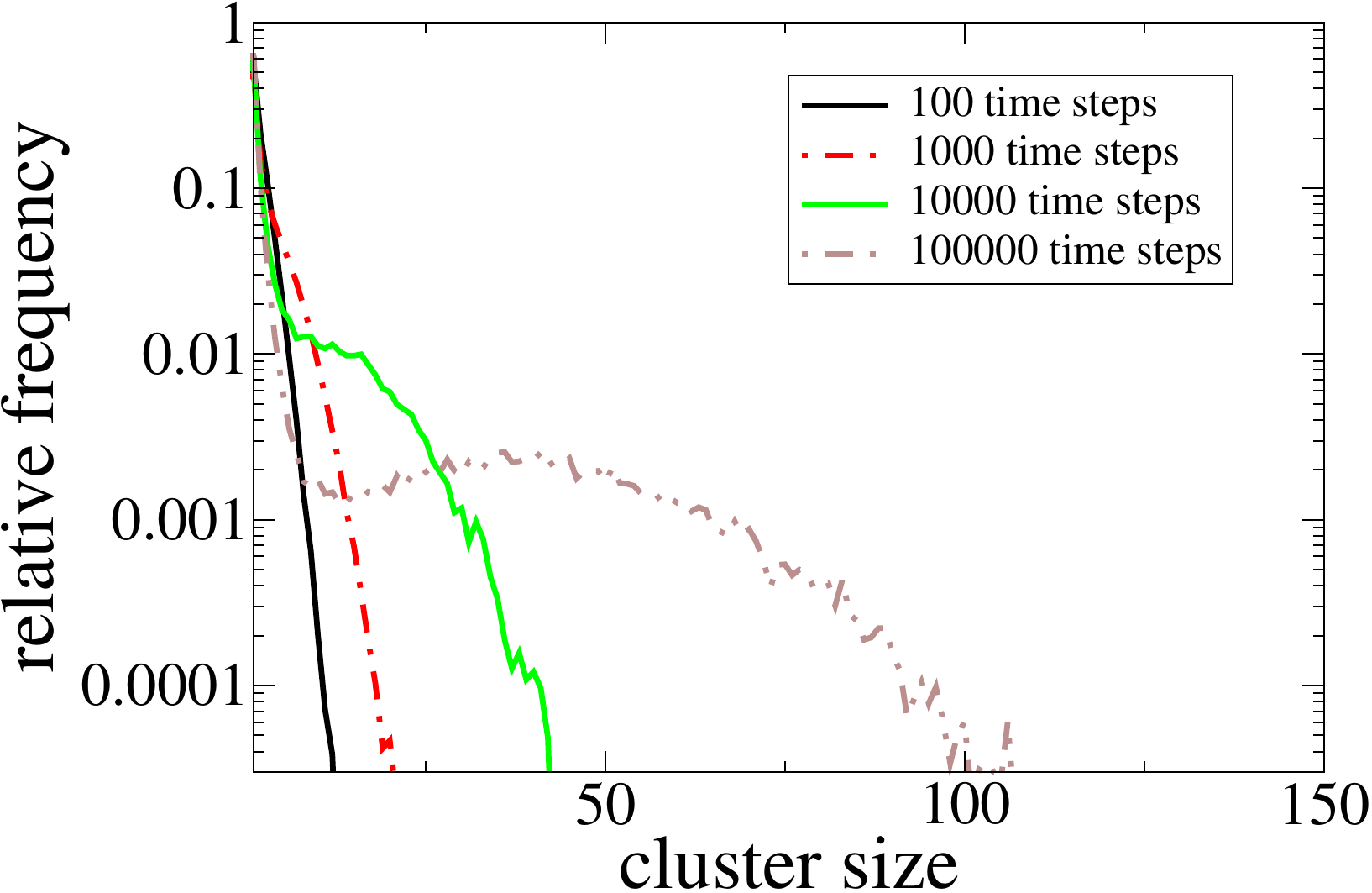}}
\caption{\label{CD_no_netw_tvar} Cluster distributions in dependence on the runtime in a system without network but attractive square well interaction potential. Parameters are $V_0=3, d_{V}=3.5,\, \rho_p^0=0.04, \, L=200\,r_p $, average over 200 runs. A maximum establishes, that moves slowly towards larger scales.}
\end{center}
\end{figure}
%%%%%%%%%%%%%%%%%%%%%%%%%%%%%%%%%%%%%%%%%%%%%%%%%%%%%%%%%%%%%%%%%%%%%%%%%%%%%%%%%%%%

%%%%%%%%%%%%%%%%%%%%%%%%%%%%%%%%%%%%%%%%%%%%%%%%%%%%%%%%%%%%%%%%%%%%%%%%%%%%%%%%%%%%%%%%%%%%%%%%5
\begin{figure}
\begin{center}
\subfigure[]{\label{CD_no_netw_potvar}\resizebox{\columnwidth}{!}{\includegraphics{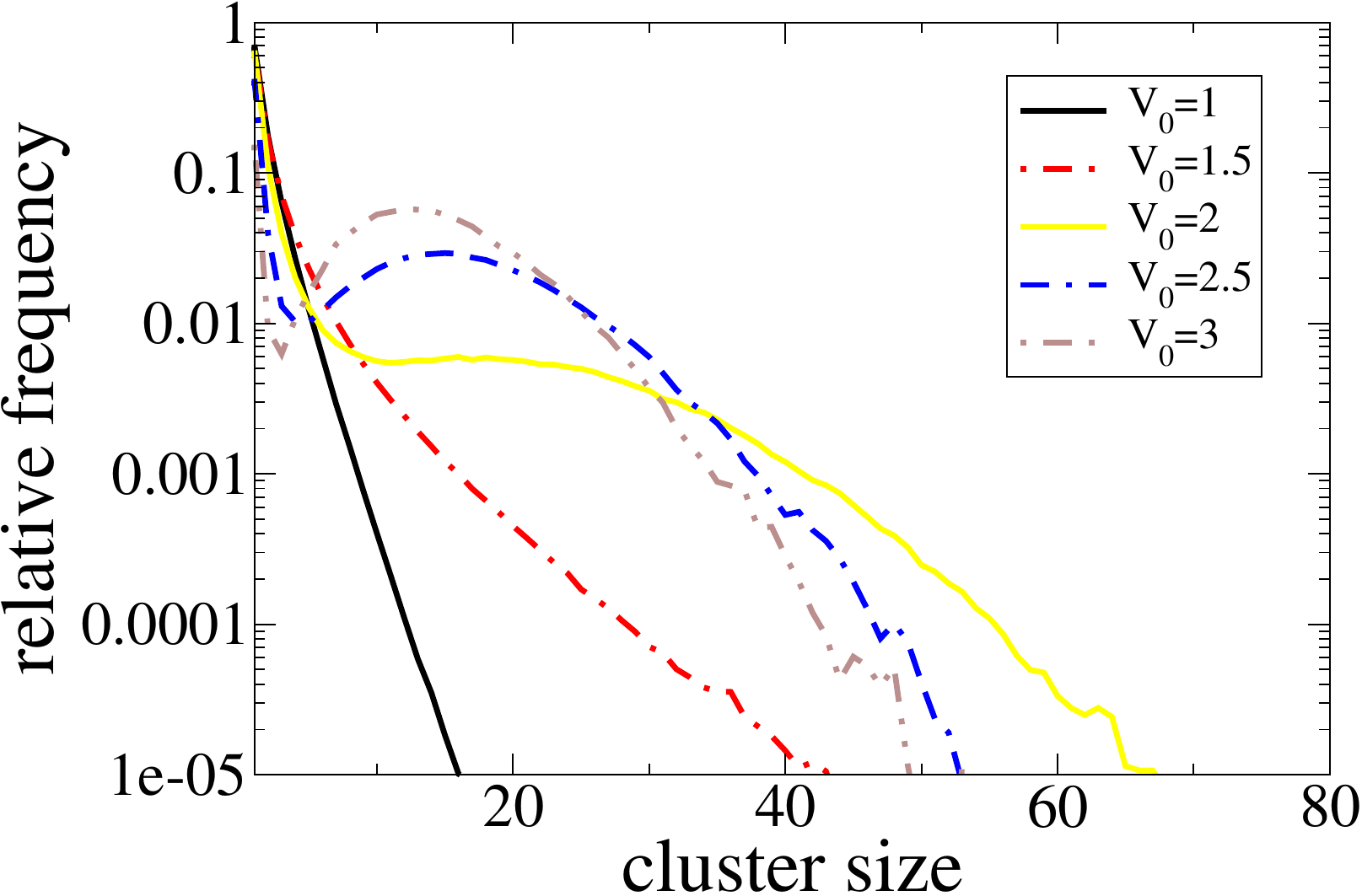}}}
\subfigure[]{\label{CD_no_netw_dintvar}\resizebox{\columnwidth}{!}{\includegraphics{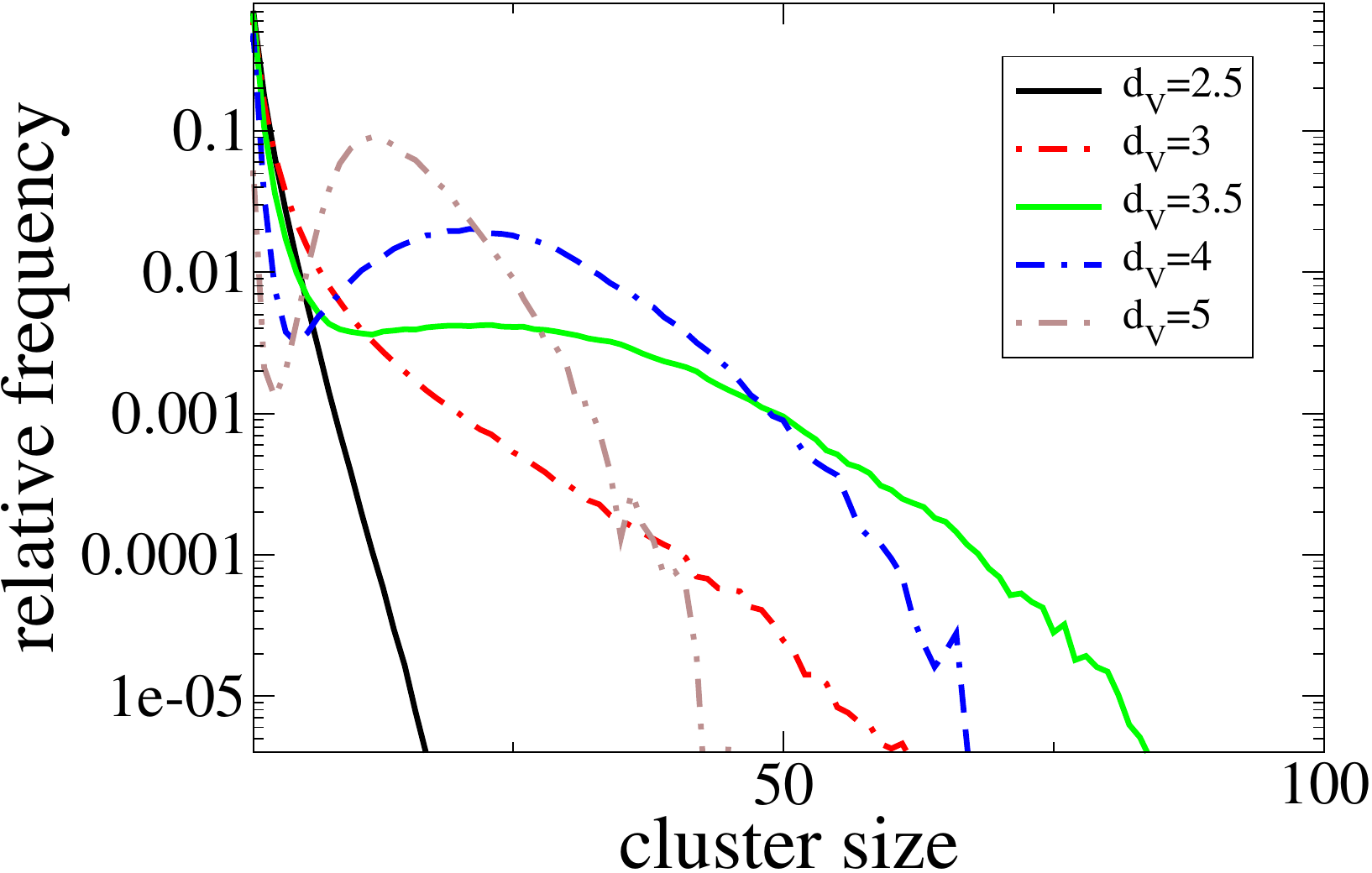}}}
\subfigure[]{\label{CD_no_netw_rhovar}\resizebox{\columnwidth}{!}{\includegraphics{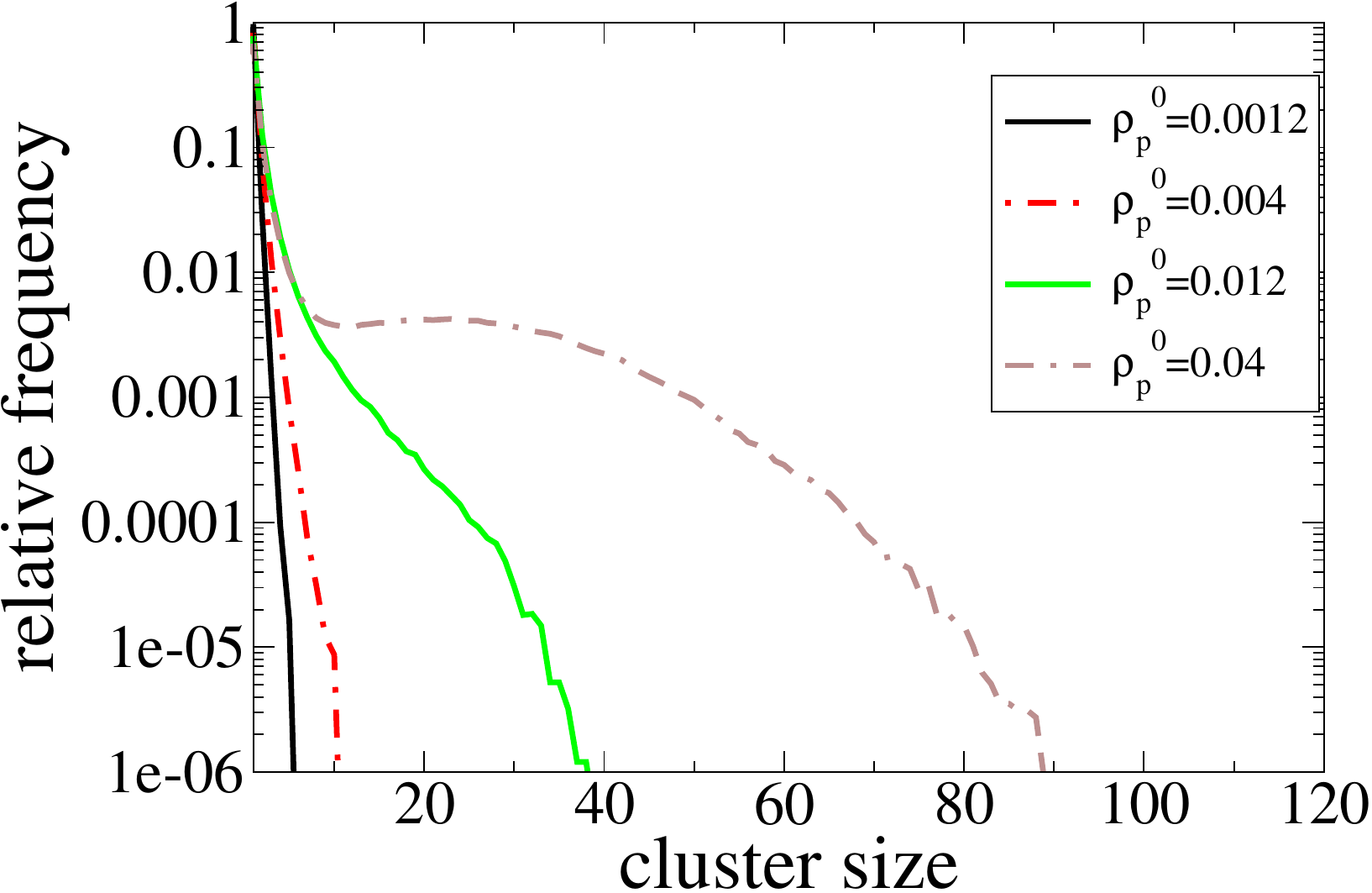}}}
%\resizebox{0.75\columnwidth}{!}{\includegraphics{cldistr_WWdist_var.pdf}}
\caption{\label{CD_no_netw} Plots of cluster distributions in the attractive particles model at intermediate times ($t=20000-30000$ time steps) in dependence on the potential depth $V_0$ (a), the potential width $d_V$ (b) and particle density $\rho_p^0$ (c). One observes the transition from an exponential decay (non clustering phase) to the formation of a maximum, corresponding to clusters at this size-scale (condensation). Default parameters are given in the text.}
\end{center}
\end{figure}
%%%%%%%%%%%%%%%%%%%%%%%%%%%%%%%%%%%%%%%%%%%%%%%%%%%%%%%%%%%%%%%%%%%%%%%%%%%%%%%%%%%%

As a first reference we investigate a diffusion limited aggregation model similar to the one introduced in \cite{memb_prot_clusters}\footnote{Here however, no additional long range repulsive force is assumed.}. Omitting filaments, we can use a variant of our model to mimic freely diffusing particles with an attractive interaction.
%A possible scenario for clustering of receptors on the cell surface is aggregation due to an attractive interaction between proteins diffusing in the cell membrane.
The corresponding process can be formulated as an equilibrium model consisting of diffusing hard-core particles (radius $r_p\sim 10nm$ \cite{sieber_memb_prot}); within the size-scale of membrane proteins) interacting via an attractive potential. We apply the particle dynamics discussed in sec. \ref{disNetw_sec} but do not consider filaments. In addition we introduce a particle-particle interaction realized by a square well potential of the form
\begin{equation}
V({\mathbf{x-x'}})=\left\lbrace 
\begin{array}{ll} 
-V_0 & \mbox{ for $|{\mathbf{x-x'}}|\leq d_V$} \\
0 & \mbox{ for $|{\mathbf{x-x'}}|>d_V$}
\end{array}
\right. \,
\end{equation}
where $\mathbf{x},\mathbf{x'}$ are particle positions. This potential can be implemented 
using a Metropolis acceptance probability $p=\min(e^{-\beta(V(x_{n+1})-V(x_n))},1)$ for a step from $x_n$ to $x_{n+1}$ ($n$ denotes the time index). In the following we use dimensionless quantities and put $\beta=1$. The default parameters are $V_0=2, d_V=3.5r_p$ and particle density $\rho_p^0=0.04$. Assuming a diffusion constant for membrane proteins $D\approx$0.0025$\mu$m$^2$/s \cite{membrane_diff} we choose a time step corresponding to $\Delta t=0.02$ seconds so that one diffusive step of length $r_p=10nm$ is performed per time step $\Delta t$.

In fig. \ref{config_no_netw} typical particle configurations at several runtimes are displayed, while in fig. \ref{CD_no_netw_tvar} ensemble averages of cluster size distributions are shown. Initial clustering already occurs on a rather small time-scale. Regarding the particle configurations we see that the number of clusters decreases with increasing runtime while the average size of remaining clusters increases. This is due to diffusion and merging of existing clusters after long times. Movement of large clusters is strongly suppressed, so that merging occurs quite slowly. The coarsening process can also be observed in the cluster size distribution. We observe a characteristic scale for larger clusters, manifesting in the emergence of a maximum, indicating a characteristic scale for cluster sizes. The dominant clusters always are within the same size-scale which increases with time. 

Since the cell membrane changes its structure steadily, patterns arising at time-scales corresponding to a finite fraction of a cell cycle cannot be assumed to be in a stationary state. Computing time averages we therefore focus on intermediate times and fix the averaging interval starting at 20000 time steps (corresponding to $\sim$7 minutes in real time) after random initialization of particles and ending at 30000 time steps. The time interval lies in the transient regime for default parameters. 
Within this interval we computed cluster size distributions (time and ensemble averages, 200 samples) for different parameter regimes and displayed them in figure \ref{CD_no_netw}.
%by averaging over 100 samples at given runtimes. 
One observes that for weak interaction no significant clustering occurs manifesting in an exponentially decaying CD, while for strong interaction $V_0$, including the default parameters, a maximum emerges hallmarking the formation of clusters.

One has to emphasize that for this kind of dynamics, clustering is reversible, i.e. in general particles can detach from a cluster due to thermal fluctuations and move to another one, such that non-vanishing particle currents between clusters may be present. This is in contrast to the irreversible clustering process discussed by Meakin and Family \cite{meakin_family} where a power law distribution of clusters was found at transient times\footnote{The stationary state of irreversible clustering is a single cluster if phase space is not separated.}. The interaction mechanism proposed in \cite{gil_memb_prot} however indicates a finite strength of protein-protein attraction so that thermal fluctuations allow detachment of particles. Destainville introduced an aggregation model claiming an additional long range repulsive force that stabilizes clusters such that a stationary state with a characteristic cluster size scale is reached \cite{memb_prot_clusters}. Here we see that at transient times, that might be more relevant for cell membrane dynamics, this intrinsic size scale is present even without a long range repulsive force
%However, after very large times we expect that all clusters tend to merge, forming one single large cluster containing a finite fraction of particles. While this state indeed represent the stationary state of the system, the time needed to arrive there is far to large to make it accessible to numerical computations of ensemble averages.

\subsection{Directed transport on regular networks}

%%%%%%%%%%%%%%%%%%%%%%%%%%%%%%%%%%%%%%%%%%%%%%%%%%%%%%%%%%%%%%%%%%%%%%%%%%%%%%%%%%%%
\begin{figure}
\begin{center}
\subfigure[$\rho_p^0=0.04$]{\label{config_SqNetw_rho=0,04}\resizebox{0.7\columnwidth}{!}{\includegraphics{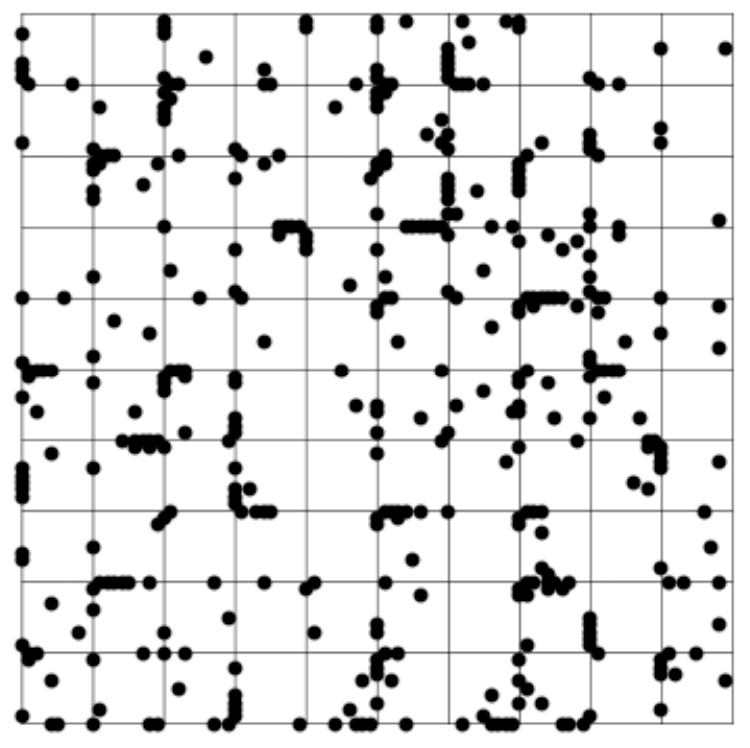}}}
\subfigure[$\rho_p^0=0.15$]{\label{config_SqNetw_rho=0,15}\resizebox{0.7\columnwidth}{!}{\includegraphics{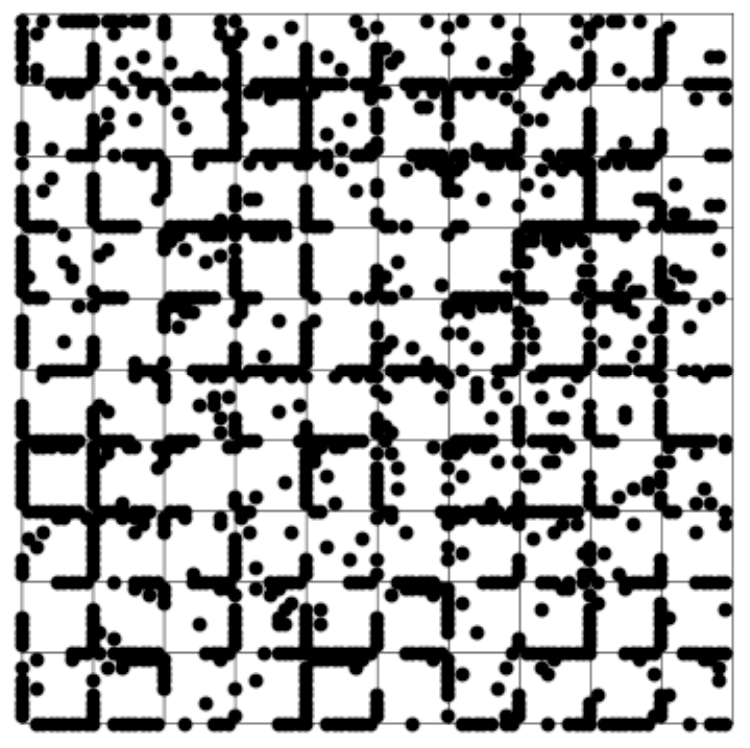}}}
\caption{\label{config_reg_netw} Particle configurations in a regular square network with system length $N=100$ sites. The black discs represent particle neighborhoods with radius 1.1 sites so that discs of particles with one vacancy or less in between overlap. One observes the formation of small L-shaped clusters at intersection points for moderate densities (a). For large densities clusters merge, forming cluster meshes on all size-scales (b).}
\end{center}
\end{figure}
%%%%%%%%%%%%%%%%%%%%%%%%%%%%%%%%%%%%%%%%%%%%%%%%%%%%%%%%%%%%%%%%%%%%%%%%%%%%%%%%%%%%

%%%%%%%%%%%%%%%%%%%%%%%%%%%%%%%%%%%%%%%%%%%%%%%%%%%%%%%%%%%%%%%%%%%%%%%%%%%%%%%%%%%%
\begin{figure}
\begin{center}
\resizebox{\columnwidth}{!}{\includegraphics{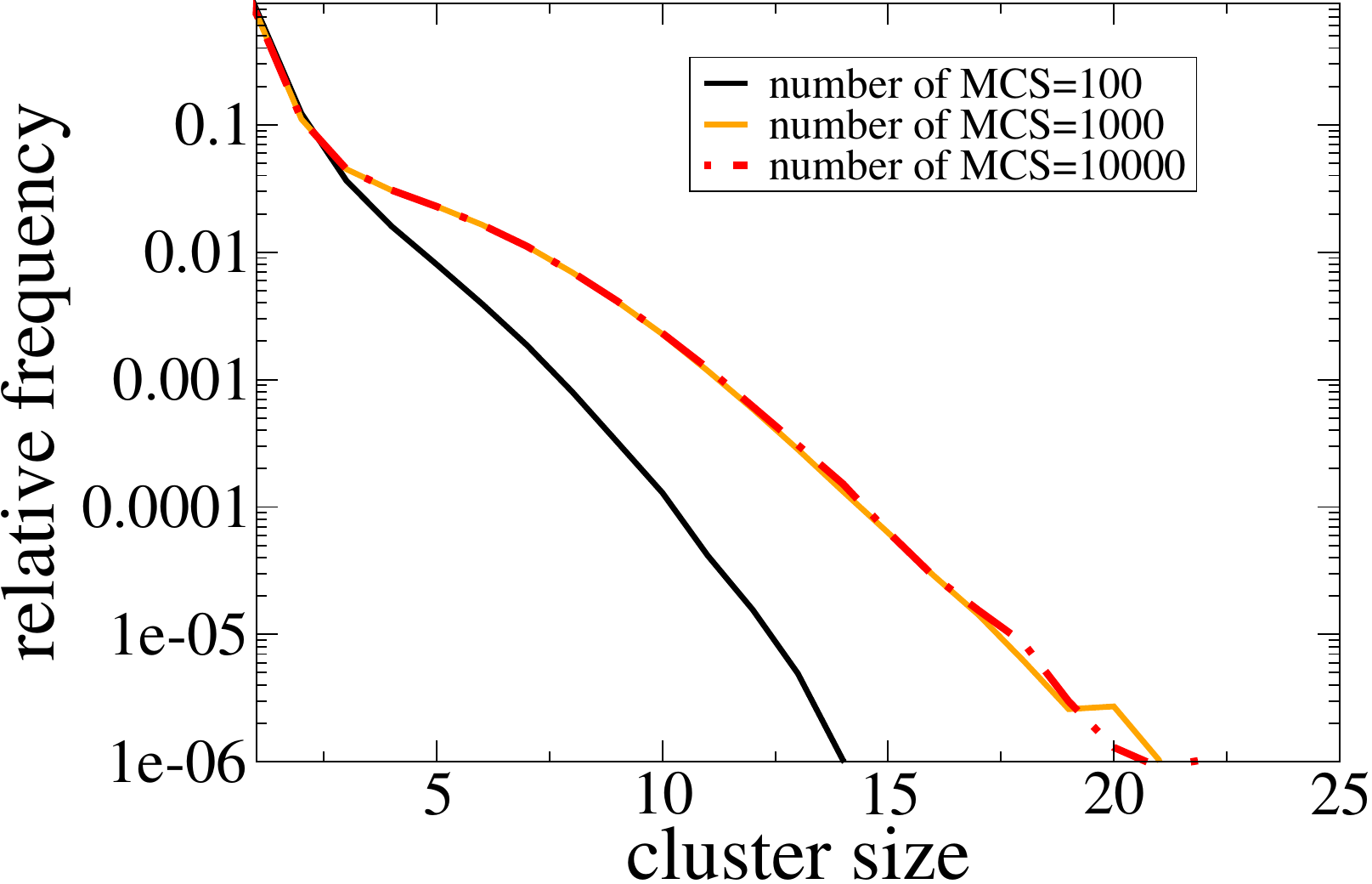}}
\caption{\label{CD_reg_netw_tvar} Cluster size distributions in a regular network for different runtimes (given in time steps). For a given runtime, we chose the last 100 steps to perform the measurement, taking 100 samples. The CD does not change after 1000 time steps, indicating that the system is in a stationary state.}
\end{center}
\end{figure}
%%%%%%%%%%%%%%%%%%%%%%%%%%%%%%%%%%%%%%%%%%%%%%%%%%%%%%%%%%%%%%%%%%%%%%%%%%%%%%%%%%%%

%%%%%%%%%%%%%%%%%%%%%%%%%%%%%%%%%%%%%%%%%%%%%%%%%%%%%%%%%%%%%%%%%%%%%%%%%%%%%%%%%%%%
\begin{figure*}
\begin{center}
\subfigure[]{\label{CD_reg_netw_rhovar_log}\resizebox{\columnwidth}{!}{\includegraphics{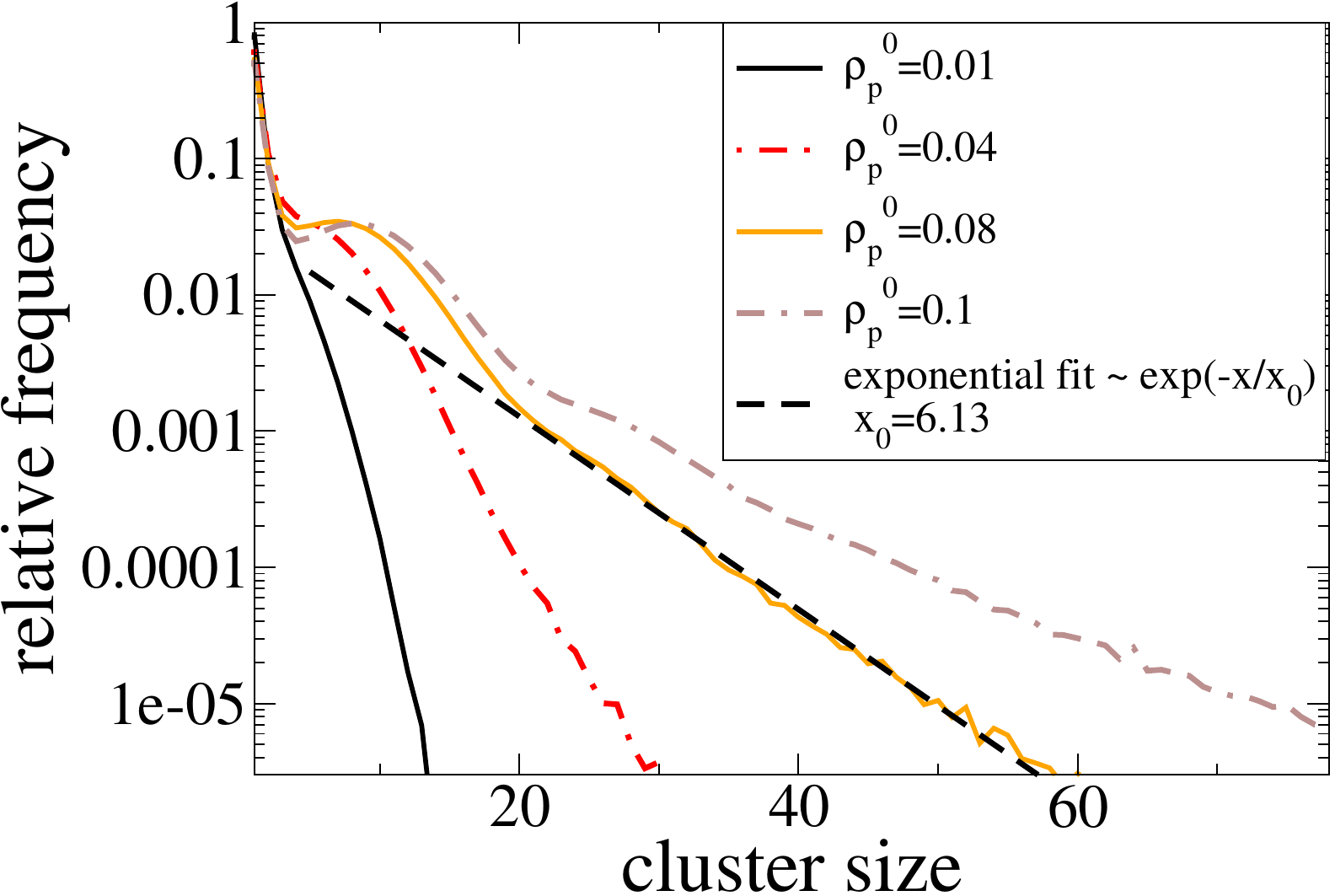}}}
\subfigure[]{\label{CD_reg_netw_rhovar_loglog}\resizebox{\columnwidth}{!}{\includegraphics{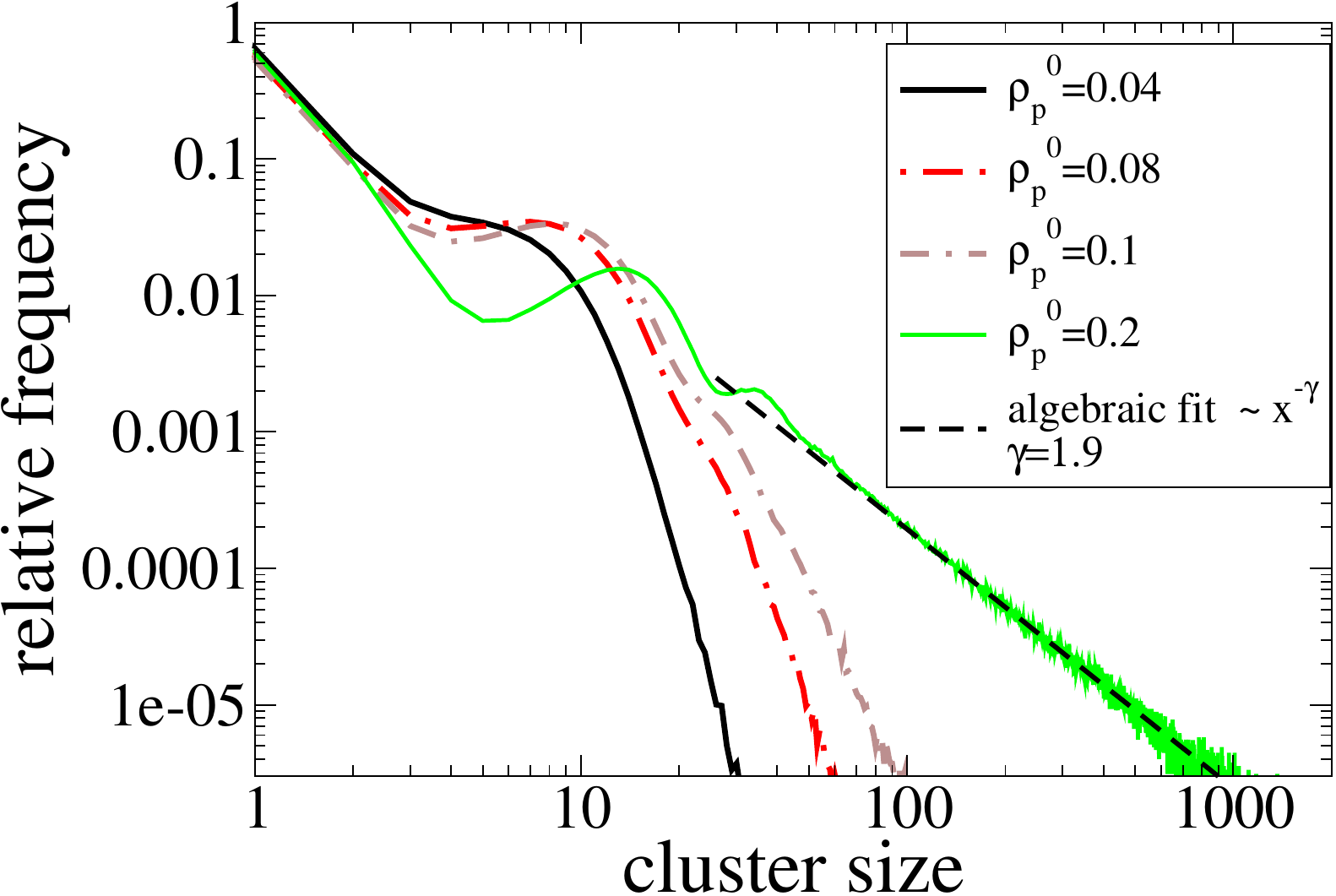}}}
\subfigure[]{\label{CD_reg_netw_blockrate_var}\resizebox{\columnwidth}{!}{\includegraphics{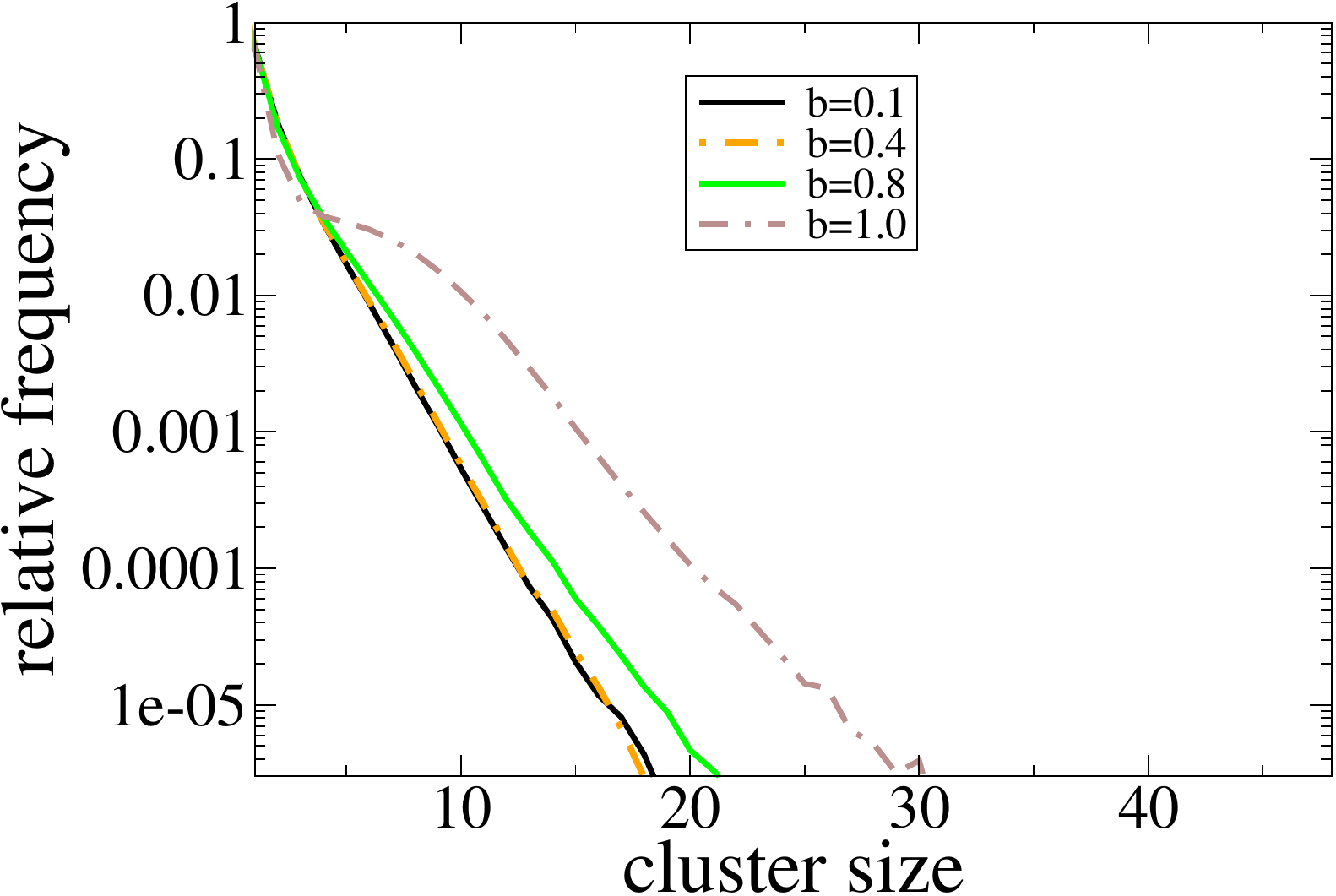}}}
%\subfigure[]{\label{CD_reg_netw_L_var}\resizebox{0.45\columnwidth}{!}{\includegraphics{CD_SqNetw_L_var.pdf}}}
\subfigure[]{\label{CD_reg_netw_lat_var}\resizebox{\columnwidth}{!}{\includegraphics{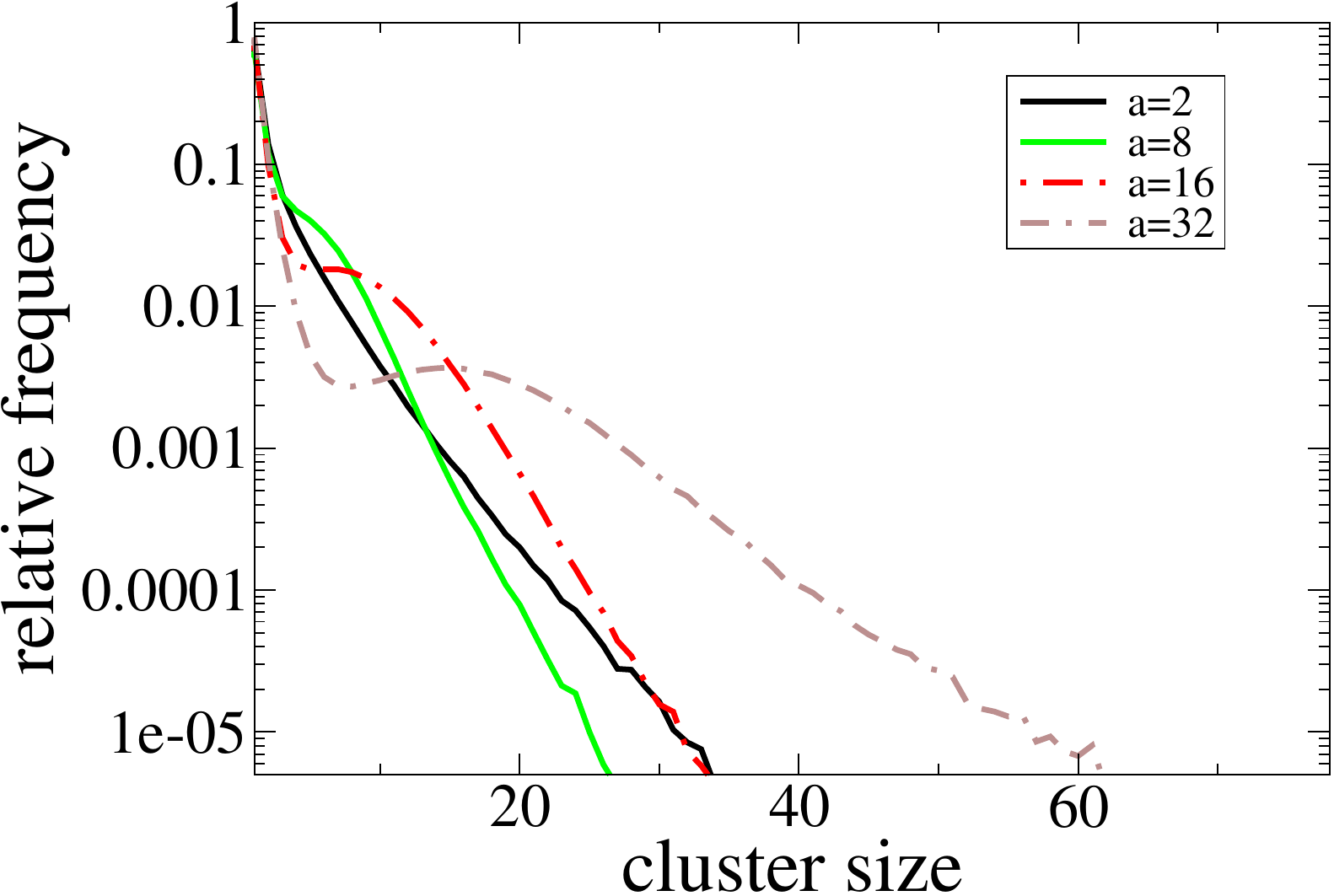}}}
\caption{\label{CD_reg_netw} Cluster distributions in a lattice gas model with exclusion interaction and a regular square network of active stripes in dependence on particle density $\rho_p^0$, logarithmic plot (a), double logarithmic plot (b), blocking rate $b$ (c)  % system size $L$ (\ref{CD_reg_netw_L_var}), 
and mesh size $a$ (d). One observes that cluster size distributions decay exponentially for moderate densities resulting in a finite size-scale of clusters, while at small densities, bulges emerge. For very large densities decay is algebraic indicating the emergence of clusters on all size-scales (see also fig. \ref{config_reg_netw}). Default parameters: see Table \ref{dis_netw_dyn_tab}.}
\end{center}
\end{figure*}
%%%%%%%%%%%%%%%%%%%%%%%%%%%%%%%%%%%%%%%%%%%%%%%%%%%%%%%%%%%%%%%%%%%%%%%%%%%%%%%%%%%%

%%%%%%%%%%%%%%%%%%%%%%%%%%%%%%%%%%%%%%%%%%%%%%%%%%%%%%%%%%%%%%%%%%%%%%%%%%%%%%%%%%%%
\begin{figure}
\begin{center}
\resizebox{\columnwidth}{!}{\includegraphics{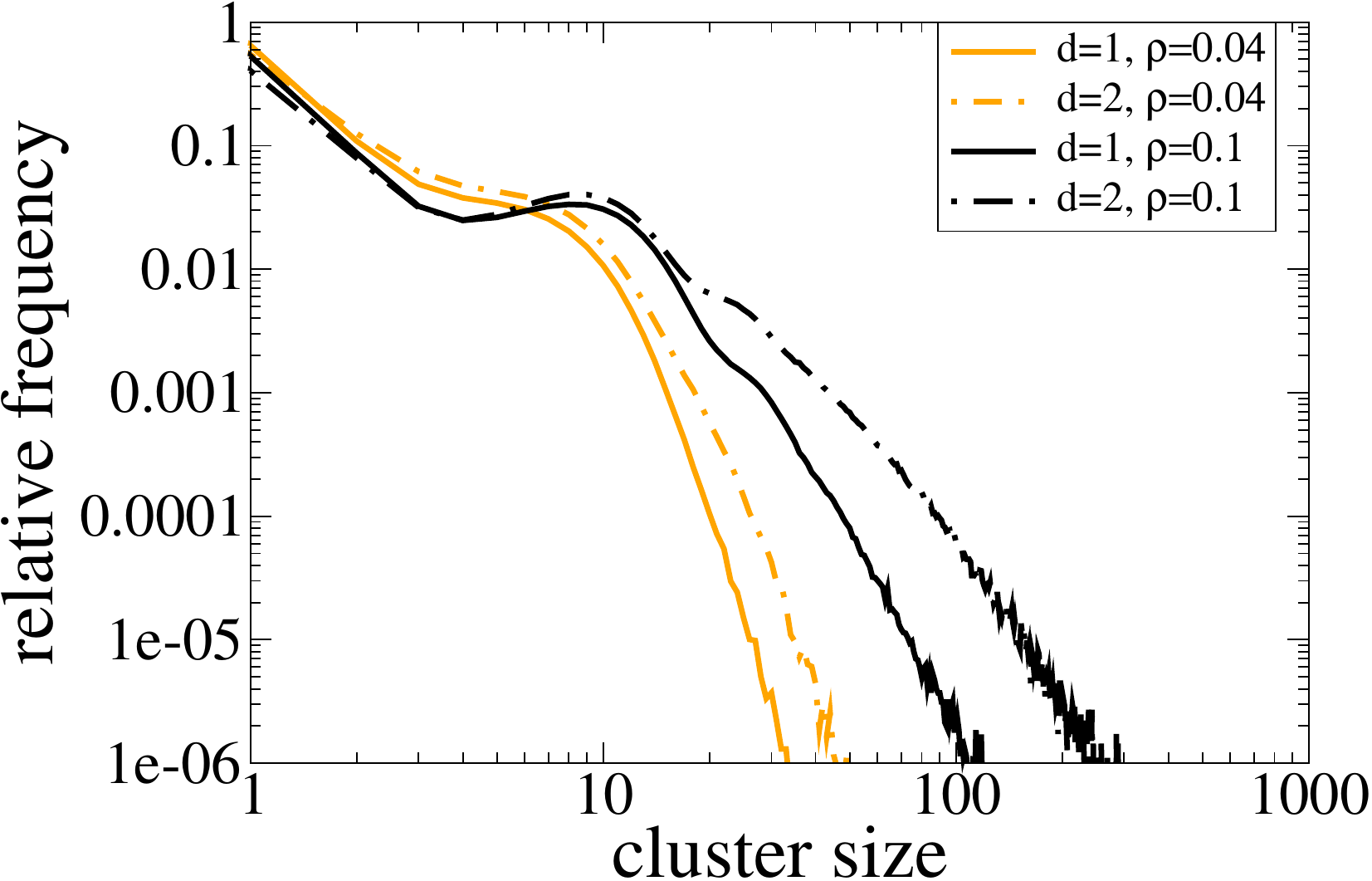}}
\caption{\label{CD_reg_netw_lambdavar} Cluster size distributions in the regular network obtained by using different definitions of the coarse graining scales. $d$ is the distance allowed between two particles to connect a cluster. One observes that while variations are small for moderate density, there is a significant influence on the scale for larger densities, indicating that clusters are not well separated.}
\end{center}
\end{figure}
%%%%%%%%%%%%%%%%%%%%%%%%%%%%%%%%%%%%%%%%%%%%%%%%%%%%%%%%%%%%%%%%%%%%%%%%%%%%%%%%%%%%

%In this section we briefly review the transport on regular networks. 
In this section we examine features of particle configurations and cluster distributions in the model introduced in section \ref{reg_netw_subsec}, i.e. a regular network of active stripes. As in the last section we start time averaging after $t_s=20000$ time steps. We carefully checked that a stationary state has been reached at this point. (cf. fig. \ref{CD_reg_netw_tvar}). As time averaging interval we choose $50000$ time steps.  
In fig. \ref{config_reg_netw} particle configurations for moderate and high densities are displayed. For particle density $\rho_p^0=0.04$ one observes small L-shaped clusters centering at intersections. For higher densities it appears that clusters are becoming larger and merge with each other to form large mesh-shaped clusters (cf. fig \ref{config_reg_netw}(b)). However, in this case clusters are hardly distinguishable and not well separated which results in sensitive dependence on the coarse graining scale (cf. fig. \ref{CD_reg_netw_lambdavar}).
In fig. \ref{CD_reg_netw} we plotted the cluster size distributions averaged over time and 100 individual runs. Examining the cluster size distributions in fig. \ref{CD_reg_netw}, one observes similar to random clustering an exponential decay for densities which are biologically relevant (see also the configuration in figure \ref{config_reg_netw}(a)). However, here they are overlapped by one or more bulges which appear to be in the size scale of the L-shaped clusters at intersections. A more detailed discussion of these profiles will be explicated in sec. \ref{theory_sec}. 
%For $b=1$ and moderate densities like in fig. \ref{config_SqNetw_meddens} one observes a a bulge and a fast exponential decrease indicating that clusters only exist on a small finite scale for moderate density regimes. 

For large densities ($\gtrsim 0.1$) the decay of the cluster size distribution becomes algebraic, indicating that clusters on all size-scales exist. These large clusters correspond to the ones generated by merged small clusters as displayed in fig. \ref{config_reg_netw}(b).

\subsection{Inhomogeneous networks}
\label{actin_netw_sec}

\begin{figure}
\begin{center}
\resizebox{0.8\columnwidth}{!}{\includegraphics{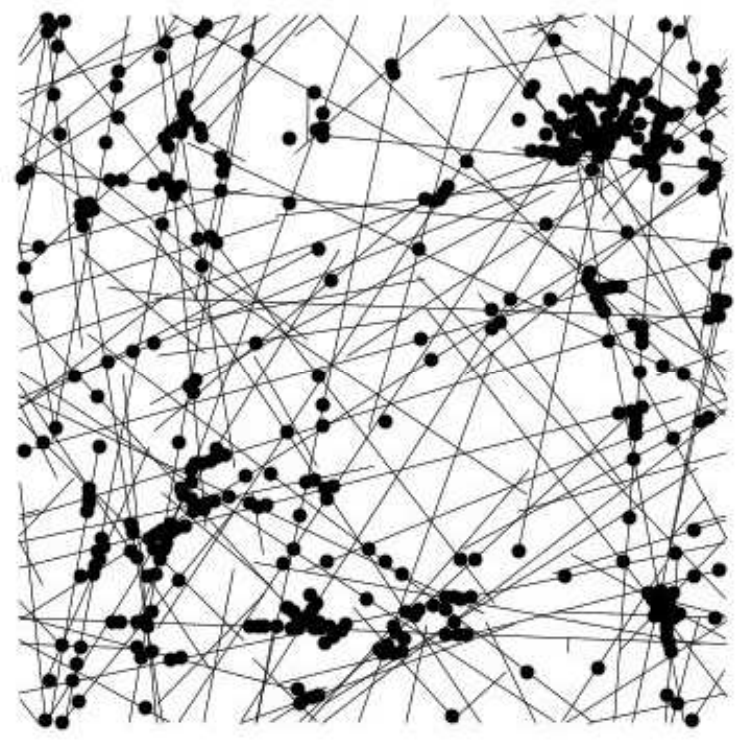}}
\caption{\label{config_disNetw_cl} Particle configurations (black discs = neighborhoods with $\lambda=2\,r_p$) for default parameters and $\rho_p^0=0.04$ and system size $L=200\,r_p$. One observes big and small clusters. This general picture is predominant for a large parameter regime and moderate densities.}
\end{center}
\end{figure}

\begin{figure}
\begin{center}
%\resizebox{0.75\columnwidth}{!}{\includegraphics{CD_rho_var_dil.pdf}}
\resizebox{\columnwidth}{!}{\includegraphics{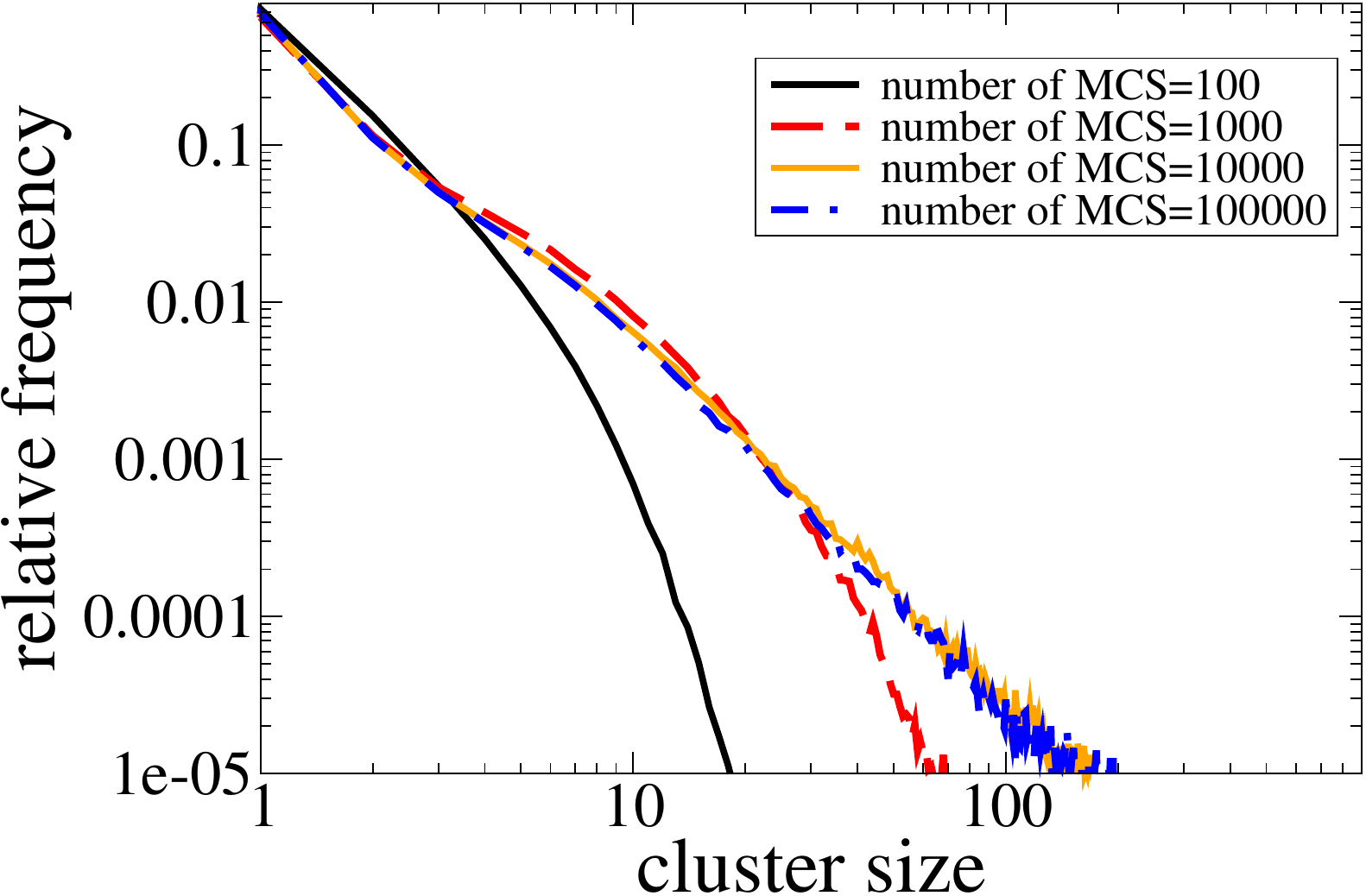}}
\caption{\label{CD_disNetw_t} Cluster size distributions for different runtimes. For 10000 time steps a stationary state is reached.}
%Comparison of two neighbourhood radii $\lambda=2$ and $\lambda=4$ shows no big difference, i.e. the selection of the coarse graining scale with $\lambda=2.0=2\,r_{p}$ appears to be appropriate.
\end{center}
\end{figure}

\begin{figure}
\begin{center}
%\resizebox{0.75\columnwidth}{!}{\includegraphics{CD_rho_var_dil.pdf}}
\resizebox{\columnwidth}{!}{\includegraphics{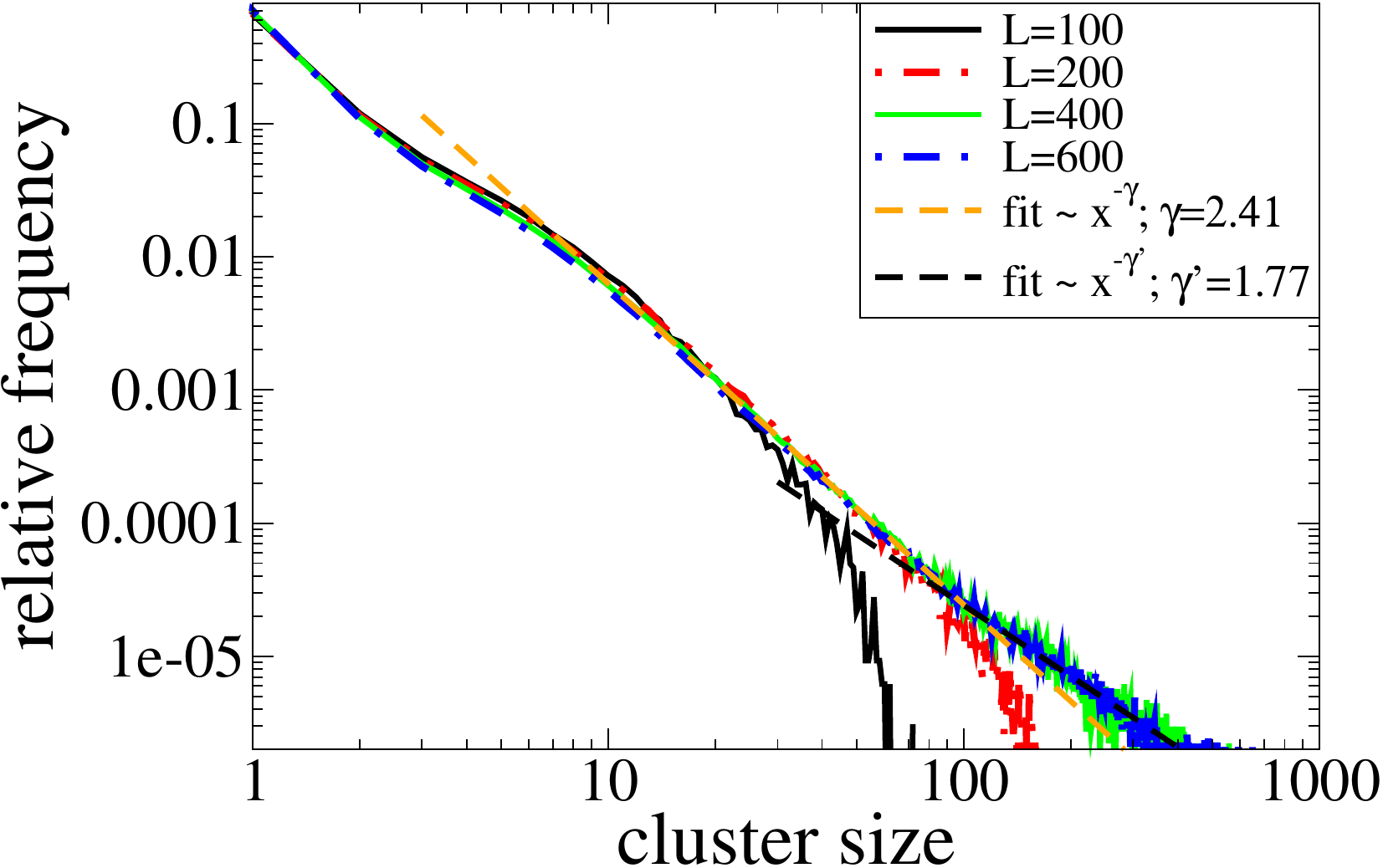}}

\caption{\label{CD_disNetw_Lvar} Cluster size distributions in dependence on the systems size. The cluster distribution decays algebraically in a large regime indicating the absence of a intrinsic size-scale. Fitting the curves, one observes a crossover from exponent $\gamma$ to a larger exponent $\gamma'$ for large cluster sizes. This crossover is well approximated by the formula $\gamma'=\gamma/2+0.5$ (\ref{gamma_crossover}) .  Default parameters: see Table \ref{default_parameters}}
%Comparison of two neighbourhood radii $\lambda=2$ and $\lambda=4$ shows no big difference, i.e. the selection of the coarse graining scale with $\lambda=2.0=2\,r_{p}$ appears to be appropriate.
\end{center}
\end{figure}

\begin{figure*}
\begin{center}
%\resizebox{0.75\columnwidth}{!}{\includegraphics{CD_rho_var_dil.pdf}}
\subfigure[]{\resizebox{\columnwidth}{!}{\includegraphics{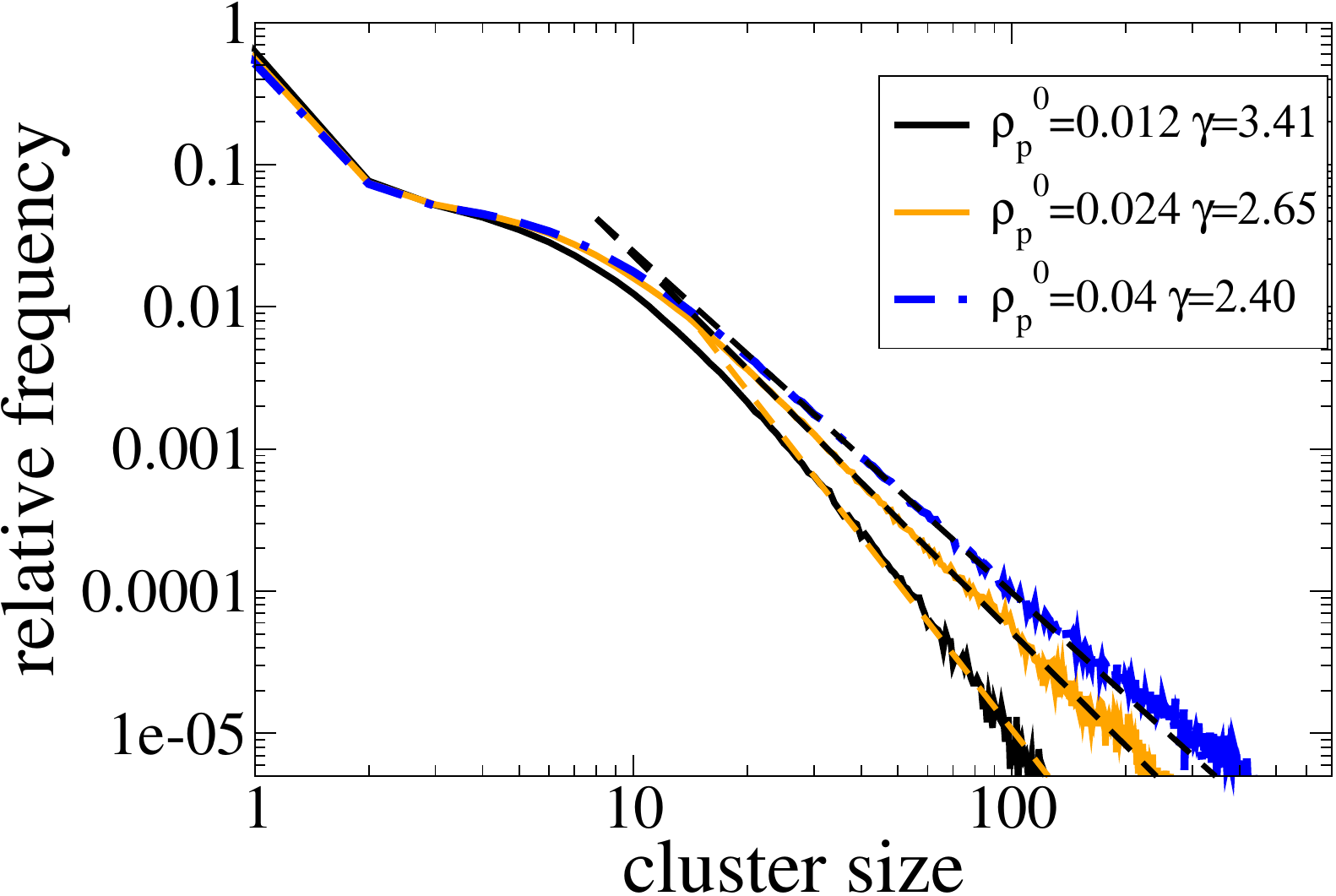}}}
\subfigure[]{\resizebox{\columnwidth}{!}{\includegraphics{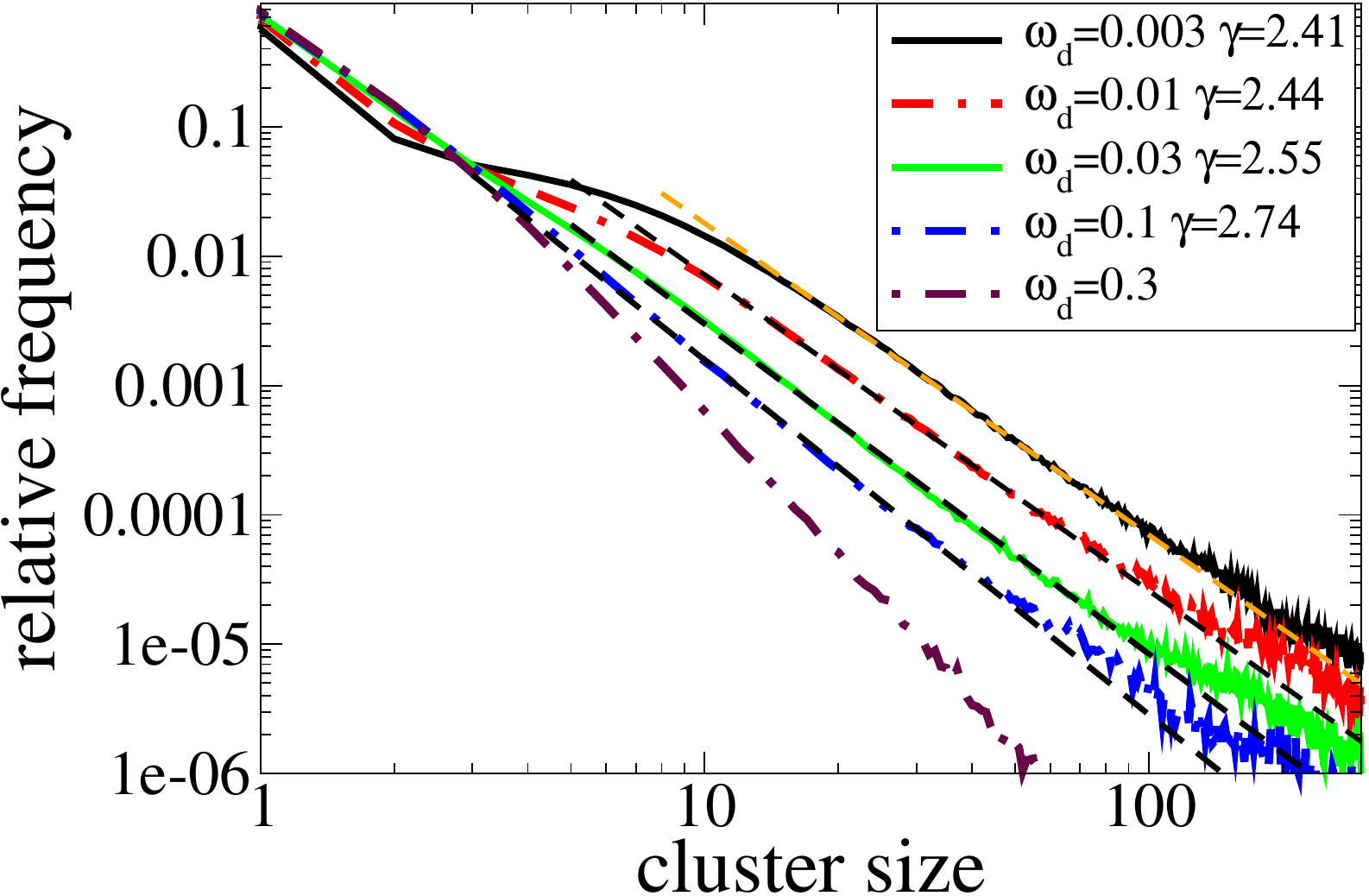}}}
\subfigure[]{\resizebox{\columnwidth}{!}{\includegraphics{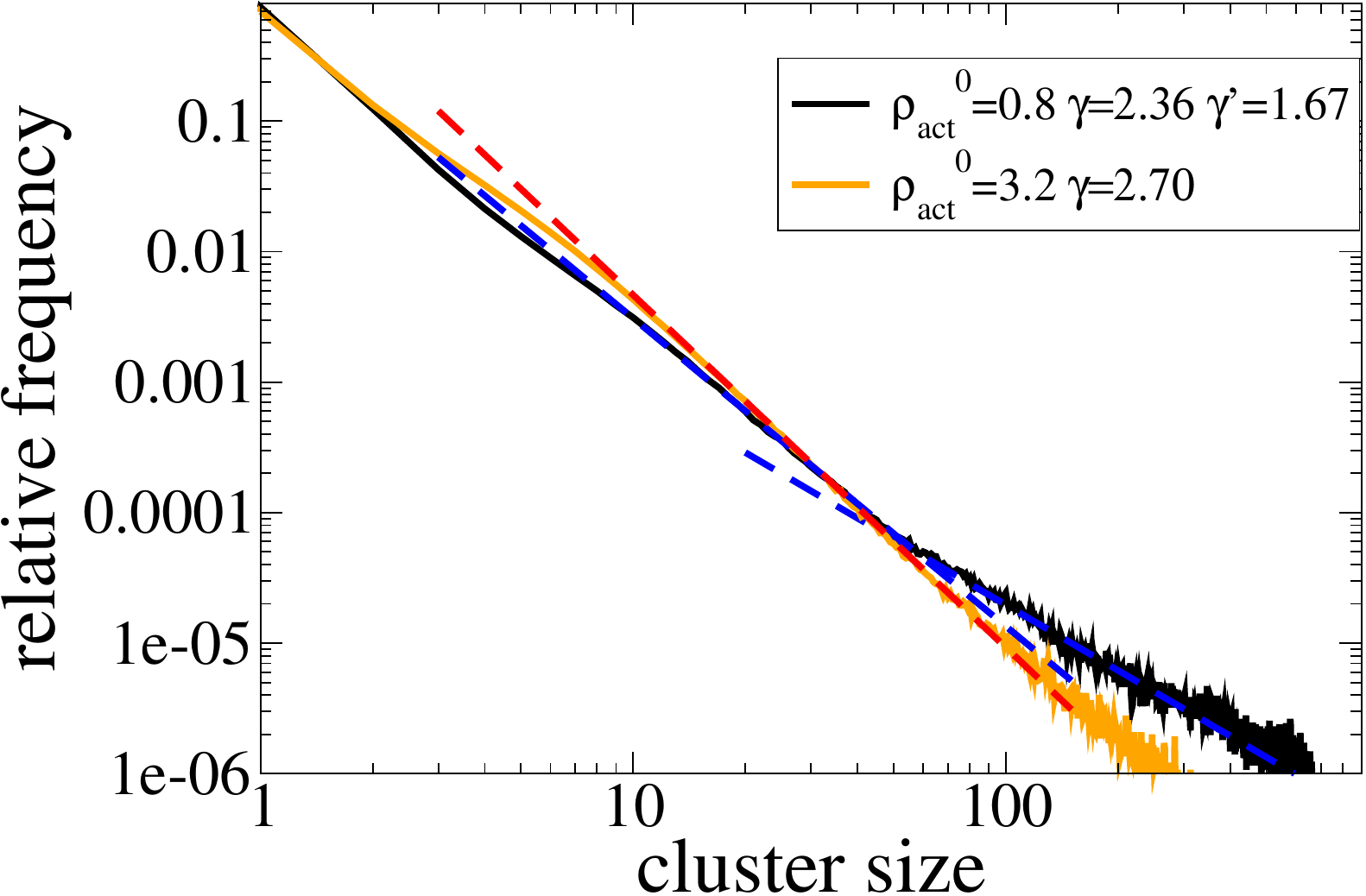}}}
\subfigure[]{\resizebox{\columnwidth}{!}{\includegraphics{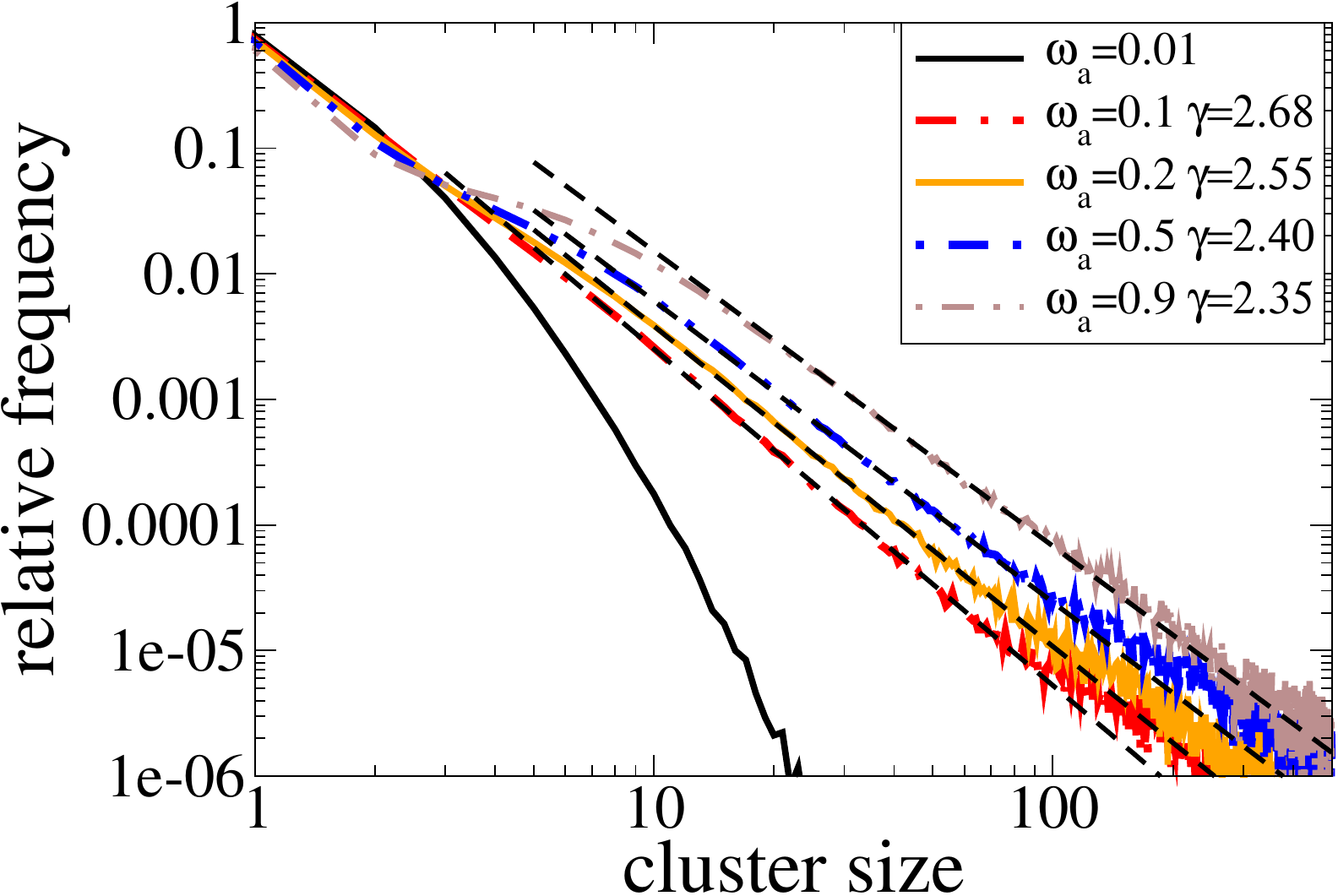}}}
\caption{\label{CD_disNetw} Cluster size distributions in a diffusive system with a inhomogeneous active transport network, single parameters varied: (a) particle density $\rho_p^0$ for small detachment rate $\omega_d=0.002$, (b) detachment rate $\omega_d$, (c) actin density ($\sim$ network density $\rho_s$) $\rho_{act}^0$, (d) attachment rate $\omega_a$. The exponent of the algebraic fit mainly depends only on $\rho_p^0$ and $\rho_{act}^0$. The dependence on $\omega_d$ and $\omega_a$ is weak, as long as $\omega_d \ll \omega_a$. Default parameters: see Table \ref{default_parameters}}
%Comparison of two neighbourhood radii $\lambda=2$ and $\lambda=4$ shows no big difference, i.e. the selection of the coarse graining scale with $\lambda=2.0=2\,r_{p}$ appears to be appropriate.
\end{center}
\end{figure*}

\begin{figure}
\begin{center}
\resizebox{\columnwidth}{!}{\includegraphics{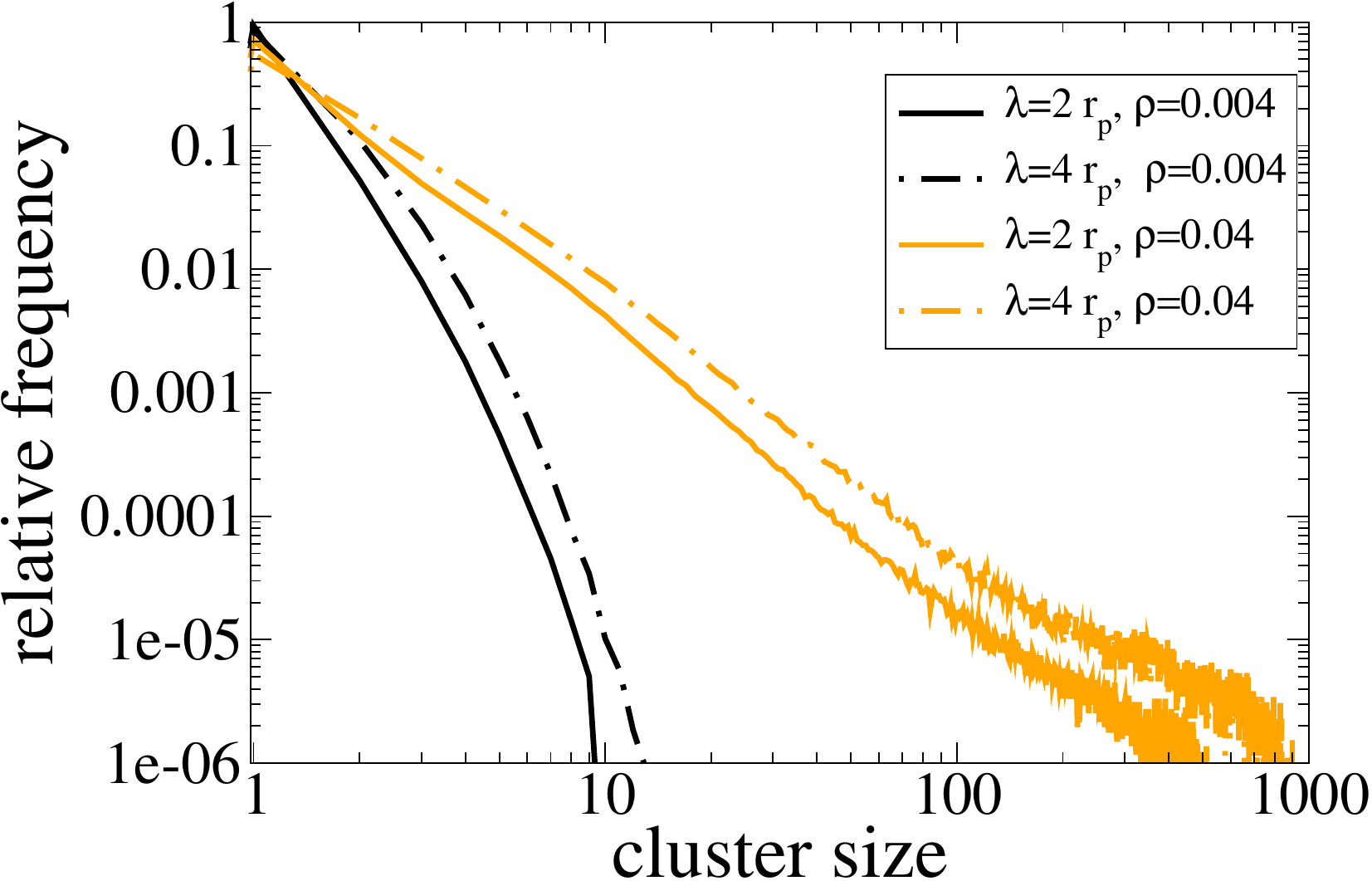}}
\caption{\label{CD_dis_netw_lambdavar} Cluster size distributions in the inhomogeneous network for different coarse graining scales. $\lambda$ is the radius of the neighborhood as defined in section \ref{cluster_char_sec}. As well in the regime where large clusters emerge and in the non-clustering regime, the dependence on the coarse graining scale is weak indicating well separated clusters.}
\end{center}
\end{figure}

The filament growth dynamics described in the appendix generate a network where single filaments have random length and direction.
%\subsubsection{Networks without branching filaments}
Particle configurations and cluster size distributions were obtained, applying steric interactions, which are shown in figures \ref{config_disNetw_cl}-\ref{CD_dis_netw_lambdavar}. The time evolution of the cluster size distribution (Fig. \ref{CD_disNetw_t}) shows that a stationary state is reached after 10000 time steps. Starting time averaging after 20000 time steps (averaging interval=50000 time steps) therefore captures the steady state dynamics. For low $\omega_d$ however, the transient time might be significantly prolonged. Therefore we use a much larger time of 400000$\Delta t$ for starting cluster evaluation. After that time we did not observe time dependence of CDs even for the smallest considered value of $\omega_d$. 

%%%%%%%%%%%%%%%%%%%%%%%%%%%%%%%%%%%%%%%%%%%%%%%%%%%%%%%%%%%%%%%%%%%%%%%%%%%%%%%%%%%%
%%%%%%%%%%%%%%%%%%%%%%%%%%%%%%%%%%%%%%%%%%%%%%%%%%%%%%%%%%%%%%%%%%%%%%%%%%%%%%%%%%%%

% %%%%%%%%%%%%%%%%%%%%%%%%%%%%%%%%%%%%%%%%%%%%%%%%%%%%%%%%%%%%%%%%%%%%%%%%%%%%%%%%%%%%
% \begin{figure}
% \begin{center}
% \subfigure[low $\rho_{act}$]{\resizebox{0.6\columnwidth}{!}{\includegraphics{config_netw_lowactdens.pdf}}}
% \subfigure[high $\omega_d$]{\resizebox{0.6\columnwidth}{!}{\includegraphics{config_netw_highwd.pdf}}}
% \subfigure[$\rho_p^0=0.008$]{\resizebox{0.6\columnwidth}{!}{\includegraphics{config_disNetw_lowdens.pdf}}}
% \caption{\label{config_disNetw_nocl} Particle configurations (neighborhoods with $\lambda=2\,r_p$ displayed) for system size $L=200\,r_p$. Parameter examples where clustering is suppressed. (a):  Very low actin density $\rho_{act}=0.04$; no intersections appear, thus no clusters can form. (b): High detachment rate $\omega_d=0.8$: Particles are only rarely attached to filaments so that influence of the network is small and brownian dynamics prevail. (c): Low particles density; particles meet only rarely so that cluster cannot nucleate. }
% \end{center}
% \end{figure}
% %%%%%%%%%%%%%%%%%%%%%%%%%%%%%%%%%%%%%%%%%%%%%%%%%%%%%%%%%%%%%%%%%%%%%%%%%%%%%%%%%%%%

%%%%%%%%%%%%%%%%%%%%%%%%%%%%%%%%%%%%%%%%%%%%%%%%%%%%%%%%%%%%%%%%%%%%%%%%%%%%%%%%%%%%
%%%%%%%%%%%%%%%%%%%%%%%%%%%%%%%%%%%%%%%%%%%%%%%%%%%%%%%%%%%%%%%%%%%%%%%%%%%%%%%%%%%%

%%%%%%%%%%%%%%%%%%%%%%%%%%%%%%%%%%%%%%%%%%%%%%%%%%%%%%%%%%%%%%%%%%%%%%%%%%%%%%%%%%%%
%%%%%%%%%%%%%%%%%%%%%%%%%%%%%%%%%%%%%%%%%%%%%%%%%%%%%%%%%%%%%%%%%%%%%%%%%%%%%%%%%%%%

%%%%%%%%%%%%%%%%%%%%%%%%%%%%%%%%%%%%%%%%%%%%%%%%%%%%%%%%%%%%%%%%%%%%%%%%%%%%%%%%%%%%
%%%%%%%%%%%%%%%%%%%%%%%%%%%%%%%%%%%%%%%%%%%%%%%%%%%%%%%%%%%%%%%%%%%%%%%%%%%%%%%%%%%%

%%%%%%%%%%%%%%%%%%%%%%%%%%%%%%%%%%%%%%%%%%%%%%%%%%%%%%%%%%%%%%%%%%%%%%%%%%%%%%%%%%%%
%%%%%%%%%%%%%%%%%%%%%%%%%%%%%%%%%%%%%%%%%%%%%%%%%%%%%%%%%%%%%%%%%%%%%%%%%%%%%%%%%%%%

\begin{figure}
\begin{center}
\resizebox{\columnwidth}{!}{\includegraphics{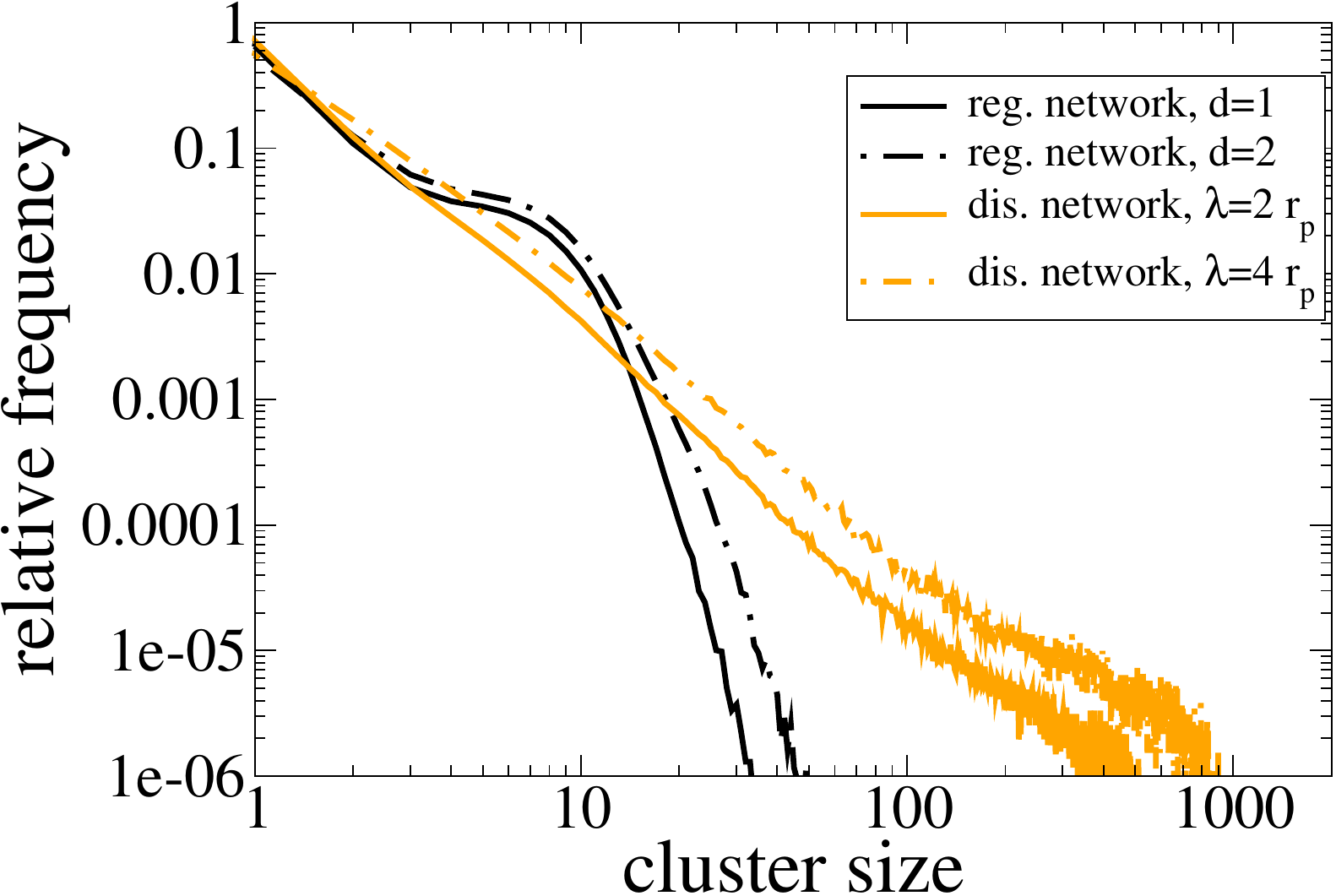}}
\caption{\label{CD_Sq_vs_dis} Comparison of regular and inhomogeneous network displaying cluster size distributions for particle density $\rho_p^0=0.04$. While the CD decays exponentially in the regular network, its slope is algebraic in the inhomogeneous one, demonstrating significant enhancement of clustering by the inhomogeneous network structure (see fig. \ref{CD_reg_netw_lambdavar} for definition of $d$).}
\end{center}
\end{figure}

\begin{figure}
\begin{center}
\resizebox{\columnwidth}{!}{\includegraphics{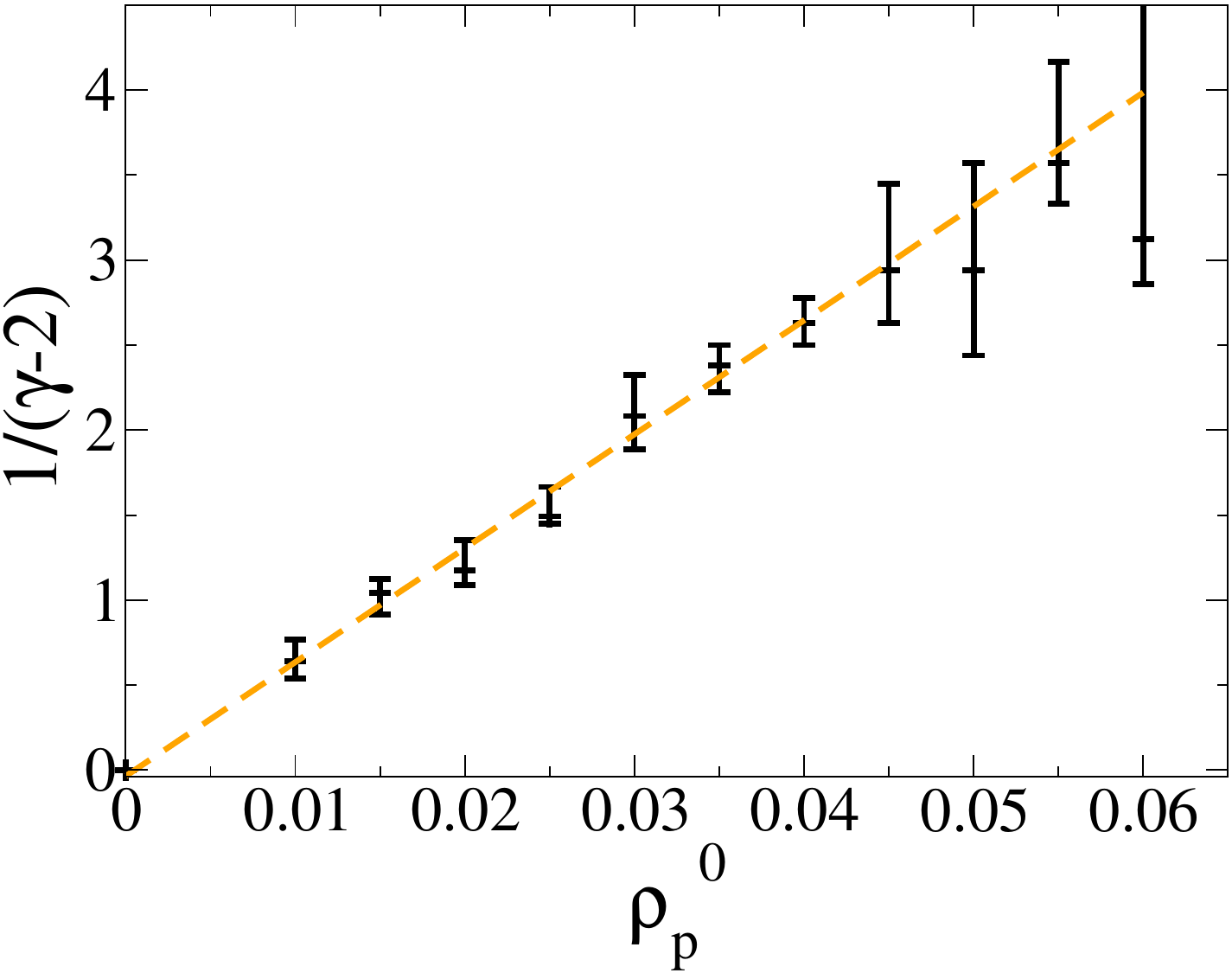}}

\caption{\label{gamma_rho_wd=0.002} Dependence of the exponent $\gamma$ of the cluster size distribution $p(m)$ on the particle density $\rho_p^0$ for $\omega_d=0.002$. One observes a linear dependence on the value $1/(\gamma-2)$ up to $\rho\approx 0.045$ which yields $\delta:=\gamma-2 \propto 1/\rho_p^0$ as predicted by the phenomenological analysis in Sec. \ref{theory_sec} (Eq. (\ref{delta_approx})). The exponents were obtained by power law regressions in the range $m\in[30,80]$ for $\rho<0.04$ and $m\in[30,100]$ for $\rho>0.04$. Error bars (obtained by varying fitting range) increase for $\gamma$ approaching the singular value 2. }
\end{center}
\end{figure}

The configuration for default density $\rho_p^0=0.04$ (Fig. \ref{config_disNetw_cl}) shows that well separated compact clusters exhibiting different sizes emerge (see also scaling in Fig. \ref{CD_dis_netw_lambdavar}). %while at $\rho_p^0=0.008$ no clusters are observed.
In a large parameter regime including the biological relevant default parameters (Table \ref{default_parameters}), the asymptotic decay of the CD is algebraic in contrast to the predominant exponential behavior on a regular network. For intermediate cluster sizes $m$, the CD follows a power law, $P(m)\sim m^{-\gamma}$, while at larger scales, there appears to be a crossover to a decreased exponent $\tilde \gamma<\gamma$. The exponent $\gamma$ depends explicitly on system parameters. It decreases with particle density $\rho_p^0$ and increases with the actin density $\rho_{act}^0$ which mainly determines the network density $\rho_s\approx\rho_{act}^0$ (see appendix). The dependence on $\rho_p^0$ indicates a behavior in form of $(\gamma-2) \propto 1/\rho_p^0$, which is consistent with analytical results in Sec. \ref{theory_sec} (see Fig. \ref{gamma_rho_wd=0.002} and Eq. (\ref{delta_approx})). The dependence on other parameters like $\omega_a$ and $\omega_d$ appears to be weak for default parameters. However, for lower $\rho_p^0$, or a larger value of $\omega_d/\omega_a$, the dependence on these parameters becomes more relevant, while varying other parameters does not lead to qualitative changes except in extreme regimes. In Fig. \ref{CD_Sq_vs_dis} cluster size distributions of a regular and inhomogeneous network are compared\footnote{Differences in effective rates due to the different spatial character of the system (discrete and continuous) are not significant since dependence on these parameters is weak.}. One observes that clustering is significantly enhanced in the inhomogeneous network. 

Due to the finite number of particles there is a cut-off at the upper end (e.g. in Fig. \ref{CD_dis_netw_lambdavar}). Fig. \ref{CD_disNetw_Lvar} shows that for increasing system size $L$ the cut off regime tends to larger values, indicating that this indeed is a finite size effect and asymptotically algebraic behavior prevails in the thermodynamic limit. 
%This finite size effect also results in a bulge that compensates the cut off and ensures normalization of the distribution. 
Though the exponent $\gamma$ of the algebraic decay varies for different particle densities, the algebraic form is a robust feature. This indicates that in the thermodynamic limit clusters on all size-scales exist. In contrast to regular networks, scale free clustering occurs even for moderate densities ($\rho_p^0\approx 0.01$) exhibiting a pattern of well separated clusters.
%%%%%%%%%%%%%%%%%%%%%%%%%%%%%%%%%%%%%%%%%%%%%%%%%%%%%%%%%%%%%%%%%%%%%%%%%%%%%%%%%%%%

\section{Phenomenological description of Cluster Formation in Inhomogeneous Networks}
\label{theory_sec}

In order to understand the distribution of cluster sizes $m$ in the inhomogeneous network theoretically, we analyze the capacity of intersections of the irregular network.  Thereby we consider dynamics of cluster initialization and stability of the cluster distribution in the stationary state. 

\subsection{Single queues}
\label{single_q_sec}

A necessary condition for cluster formation is that two particles moving along a filament encounter each other at an intersection. Then the two particles may block each other due to steric interactions and form a \emph{cluster seed}. Therafter other particles can attach to the filament moving towards the initial two-particle cluster and form a queue. 

Studying the queuing mechanism, we regard a single filament with an intersection occupied by a cluster seed. The filament can be considered as a one dimensional discrete system coupled to a reservoir of particles with
 %\emph{local} 
 density $\rho_p - \rho_b$, i.e. the density of unbound particle (where $\rho_p$ denotes the global particle density and $\rho_b$ the density of bound particles).
%\footnote{In this one dimensional consideration, $\rho^1_p$ is a line density} 

The effective attachment rate of a particle is the  attachment rate $\omega_a$ times the fraction of area that allows binding and the probability that there is space on the filament, i.e.
\begin{equation}
\tilde\omega_a \approx \omega_a 2d_b d_s \rho_s\left(1-\frac{\rho_p}{\rho_s} n_s\right)  \,\,\, \footnote{The binding area of a filament in continuous space is approximated by a rectangular shape since $d_s<d_b$. In the regular network $d_s=1,d_b=1/2$ and binding area = active tracks. } .
\end{equation}
Here $n_s$ is the number of binding sites that are not accessible if a particle occupies a filament and $\rho_s$ is the total density of filament subunits of length $d_s$ in the system, i.e. $\rho_s=\rho_{act}^0-\rho_{act}$. We assume $\rho_p \ll \rho_s$ which is the case for default parameters. In a regular discrete network $n_s=1$, while in the inhomogeneous one $n_s=5$ for $d_s=0.36\, lu, r_p=0.5\, lu$, since the distance of two particles must be at least $2r_p$\footnote{For default parameters, $\tilde\omega_a$ takes the value $\tilde\omega_a \approx 2/3 \omega_a$}. 
%Therefore the inflow of particles on a given site due to attachment is $j_a=\tilde\omega_a (\rho_p-\rho_b)/\rho_s(1-\rho_b)\approx 2 \omega_a(\rho_p-\rho_b)d_b d_s$. This approximation neglects terms of $O((\rho_p/\rho_s)^2)$, since $\rho_p \ll \rho_s$ for default parameters
%The effective attachment rate to a given site of the filament is given by  average number of particles, which can bind to the given site, times attachment rate, i.e.  $\tilde\omega_a\approx \omega_a (\rho_p-\rho_b) 2d_b d_s$
% where $\rho^1_b$ is line density of particles bound to a filament

Complete Detachment occurs with an effective rate $\tilde\omega_d$, comprising detachment and diffusing away, such that there is free space for particles behind to move on the filament. Therefore a detached particle may not reattach immediately and a subsequent diffusive step must be lateral to the filament. Diffusing can also be inhibited by a high density of free particles. We therefore write $\tilde\omega_d=\omega_d(1-\omega_a\Delta t)(1-\rho_p\pi r_p^2 \Delta t ) C_1(D)$ where the phenomenological factor $C_1<1$ reflects the inhibition of a diffusing step by other attached particles on the filament. This factor represents the angle sector that allows free diffusion and is assumed only to depend increasingly on the diffusion constant $D=l_D^2/2\Delta t$. A one dimensional system of this kind corresponds to the \emph{totally asymmetric simple exclusion process} with \emph{Langmuir kinetics}\cite{pff1}. In this system phase coexistence with a high and a low density domain, separated by a stationary domain wall (shock) is observed, if inflow of particles $J^1_{in}$ is larger than outflow $J^1_{out}$ of the initial cluster seed. High density domains correspond to queues, i.e. one dimensional clusters. On long filaments the density of attached particles quickly approaches the stationary density $\rho_b=\rho_p \tilde\omega_a /(\tilde\omega_a+\omega_d)$ (cf. \cite{lipowski_network}). Therefore the inflow on a single filament queue can be approximated by $J^1_{in}= p\,\rho_{b}/\rho_s(1-\rho_{b}/\rho_s) \approx p\, \tilde\omega_a/(\omega_d+\tilde\omega_a)\,\rho_p/\rho_s$, neglecting $(\rho_b/\rho_s)^2$. Outflow by detaching particles is $J^1_{out}=\tilde\omega_d l$ where $l$ is the number of particles in the queue. The condition that a stationary queue of length $l$ establishes is $J^1_{in} = J^1_{out}$ if a two-particle cluster has established. Hence we have $\rho_p p /(\rho_s(\tilde\omega_a+\omega_d)) = \tilde\omega_d l$
\begin{equation}
\label{rho_crit}
\rho_p(l)\approx \frac{\tilde\omega_d l \rho_s}{p}\left(1+\frac{\omega_d}{\tilde\omega_a}\right)
\end{equation}
If the queue does not cross other filaments, a finite queue of length $l_0$ establishes, while the shock, i.e. the end of the queue performs fluctuations around the mean value $l_0$ \cite{ludger_pff}. 

% The probability distribution of queue lengths hence is of the form $P(l)\sim \exp((l-l_0)^2/\sigma^2)$. In any part of the system where $\rho_p < \rho_p(a)$ clusters therefore exhibit a L-shaped form and distribution of cluster sizes $m$ corresponding to the distribution of single queues, therefore $P(m)\sim \exp(-(m-m_0)^2/\sigma^2)$. 

\subsection{Cluster branching}

If a queue spans over intersections connecting the filament with other ones, it acts as an obstacle for particles moving along crossing filaments. This obstacle serves as a nucleation seed for other queues on respective filaments in a same manner like at the initial two-particle cluster seed. It leads to a branching of the queue and can initialize a cascade of queues that constitute a large connected cluster. At first glance, we neglect freely diffusing particles in the neighborhood of the queues which can also be part of clusters by the definition in sec. \ref{cluster_char_sec}, since their effect on cluster in- and outflow can be treated by the local particle density $\rho_p$ and effective detachment rate $\tilde\omega_d$. Then the full cluster is constituted by the connected set of these individual queues, $m=\sum_i l_i$ where the index $i$ runs over all filaments covered by the cluster.   

The intersections not only initialize new cluster branches, but also serve as defects for particle hopping, since at these points the hopping rate $p$ is effectively lowered. This affects the structure of the queues. The TASEP with Langmuir kinetics and defect sites, which corresponds to this problem is treated in \cite{pff_dis}. If inflow is larger than the transport capacity of a defect, as in the pure system a macroscopic high density domain emerges limited by a stationary shock. Though at defects, small diluted regions after defect sites occur, we can assume the queues to be connected on a coarse graining scale $\lambda > r_p$ and by particles diffusing in the neighborhood of the filament\footnote{The considerations in \cite{pff_dis} are for large systems where attachment and detachment rates scale like $\omega_{a,d}\sim 1/L$. However, qualitative results do not change while relative fluctuations of shockpositions and boundary layers increase for small systems.}. If $\rho_p<\rho^*(a)$ so that queues do not span other intersections, clusters consist of two queues each on one of the filaments at the cluster seed's intersection.
Since the length of the queues in TASEP-LK models is always finite, there is a finite mean value $\bar l$. Therefore the total cluster size can be estimated by
\begin{equation}
m\approx n_F  \bar l\,\, ,
\end{equation}
where $n_F$ is the number of filaments it covers. 

The considerations of this and the last subsection apply in an analogue way to a regular network if $d_b$ is replaced by $1/2 \,lu$ and $d_s=1 \,lu$. In the following subsections, however, the explicit statistics of the inhomogeneous network are considered.

%%%%%%%%%%%%%%%%%%%%%%%%%%%%%%%%%%%%%%%%%%%%%%%%%%%%%%%%%%%%%%%%%%%%%%%%%%%%%%%%%
\begin{figure}
\begin{center}
\resizebox{\columnwidth}{!}{\includegraphics{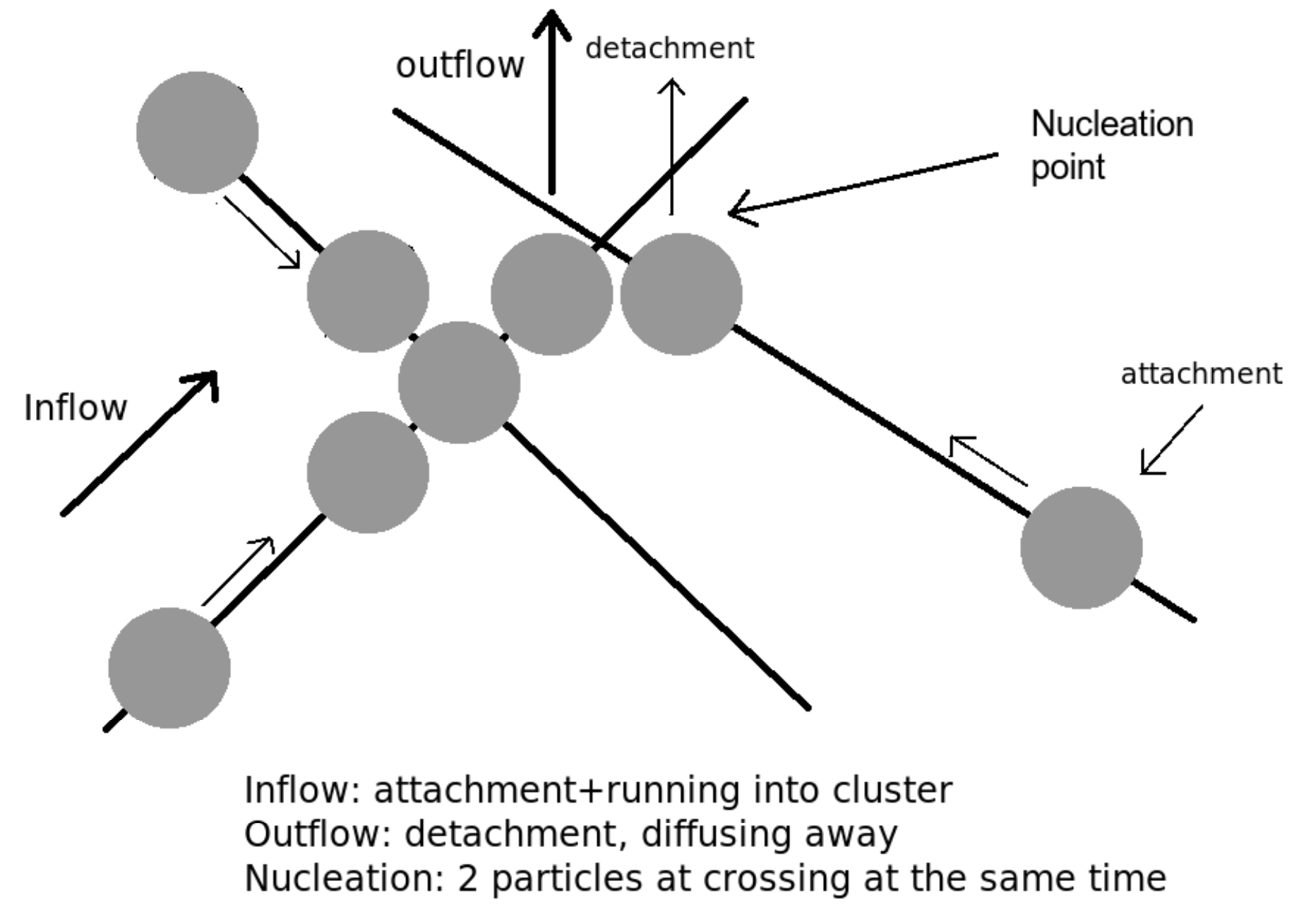}}
\caption{\label{illust_clustering_dis_netw} Illustration of the mechanism that leads to the formation of clusters in a inhomogeneous network. If the density of intersections is high, existing clusters serve as additional obstacles for particles on other filaments, enhancing cluster formation.}
\end{center}
\end{figure}

\subsection{Graph approximation}

The above considerations show that the statistics of filament crossings appear to be crucial for cluster dynamics if the network is disordered. In the following we want to introduce a coarse graining (length-) scale $\xi$ and study the distribution of filaments on this scale. %determine the probability distribution of filaments crossing an area spanning a length $\xi$ in the inhomogeneous network. 
This provides the basis for a coarse graining procedure where the filament network is approximated by a \emph{graph} where dynamics of particles are approximated by effective transition rates between nodes of a topological network. 

In the following considerations on inhomogeneous networks we assume that filament lengths $L_F$ are large compared to the scale $\xi$ of the target area we consider. 

The probability that a given filament with arbitrary orientation, position and length $L_F \gg \xi$ intersects an area of diameter $\xi$ is 
\begin{equation}
\label{prob_single_fil}
p\approx\frac{\xi}{L}\,\frac{L_F}{L}
\end{equation}
The average length of filaments is $\langle L_F\rangle=\rho_s L^2 d_s/N_F$ and $N_F$ is the number of filaments. Averaging over filament lengths, one obtains
\begin{equation}
p=\frac{\xi\rho_s d_s}{ N_F} \quad . 
\end{equation}
Obviously the probability that $n_F$ filaments cross an area of diameter $\xi$ is equivalent to the probability distribution of a Poisson process with individual hit probability $p$.
\begin{equation}
\label{filaments_distr}
P(n_F)=\frac{\sigma^{n_F}}{n_F!}e^{-\sigma}
\end{equation}
with expectation value and standard deviation $\sigma=p\,N_F=\xi\rho_s d_s$. Therefore the average number of intersecting filaments grows linearly with the diameter of the considered area and one can define a linear filament density $\rho_F=\langle n_F\rangle /\xi = \rho_s d_s$. The average subunit density is related to the actin density by $\rho_s=\rho_{act}^0-\rho_{act} \approx \rho_{act}^0$ (for $\omega_s/ \omega_g \ll \rho_{act}^0$, cf. sec. \ref{disNetw_sec}). In uncorrelated filament networks, the average distance between nodes, i.e. the mesh size $a=2/(\pi \rho_s d_s)$ \cite{frey_fibernetw}, i.e. the dependence on the actin density is $a\sim 1/\rho^0_{act}d_s$. Since the structure of the clusters is one dimensional, its linear scale is $\xi\propto m r_p$. Therefore the number of filaments a cluster covers is $n_F\sim m \, r_p/a$. On the other hand we have $n_F\approx m/\bar l$. Therefore the average queue length scales like the mesh size, i.e. $\bar l \sim a / r_p$.

In order to describe particle dynamics on a coarse grained level, we approximate the filament network by a topological network (graph) consisting of nodes connected by links, representing the filaments. On the topological network the particle dynamics is described by hopping from node to node with given rates, assuming that particle transport is dominated by active transport. In this approximation we assume that most particles are bound to filaments, i.e. $\rho_b=\rho_p \,1/(1+\omega_d/\tilde\omega_a) \approx \rho_p$, which is justified for $\omega_d \ll 2\omega_a \rho_s d_s d_b$ \footnote{For default parameters $\omega_d/2\omega_a \rho_s d_s d_s \approx 0.12$}. This marks the limit of the \emph{graph approximation}. Diffusive phases are assumed to be short but can lead to a change of the filament, i.e. changing travel direction.
 
Similar to the considerations above, the network structure is coarse grained by virtually subdividing filaments into segments of length $\xi$ representing the nodes of the network. Segments from different filaments that overlap at intersections are treated as one node. A filament hosting segments of two nodes $i$ and $j$ directed from $i$ to $j$ corresponds to a link from $i$ to $j$. It mediates a net particle drift from $i$ to $j$. The full network can then be represented by the adjacency matrix $\mathbf A$ whose components $A_{ij}$ denotes the number of links between $i$ and $j$. Note that in this view two nodes can be connected by more than one link. We denote the number of outgoing links from a node $i$ by $K_i^{out}=\sum_j A_{ij}$ and the number of ingoing links by $K_i^{in}=\sum_j A_{ji}$ (out-degree and in-degree respectively). 
Each filament crossing a node without ending inside provides exactly one link in and one out of the node. If filament lengths are large compared to the length scale $\xi$ as assumed above, we can neglect filament ends inside a node. Therefore the number of incoming links is approximately equal to the number of outgoing links. Their number is given by the number of filaments, i.e. $K_i^{in}\approx K_i^{out}=n_F$. 

In \cite{noh_rw_network} it has been shown that for noninteracting particles performing a random walk on a topological network with undirected links and at most one link between two nodes, the density of particles in the stationary state is proportional to the number of links 
\begin{equation}
\label{dens_network}
\rho_i = K_i/\mathcal N
\end{equation}
 where the normalization factor $\mathcal N=\sum_i K_i$ is the total number of links. However, it is easy to show that this is also valid for directed networks as long as $K_i^{in}=K_i^{out}$ for all $i$. If any particle moves within one time interval $\tau$ from one node to an adjacent node via a link, the master equation yields 
\begin{equation}
\rho_i(t+\tau)=\sum_j \frac{A_{ji}}{K_j^{out}}\rho_j(t)
\end{equation}\footnote{The outflow term cancels since it is exactly $\rho_i(t)$.}
Inserting (\ref{dens_network}) and applying $K_i=K_i^{in}=K_i^{out}$, one obtains
\begin{equation}
\rho_i(t+\tau)=\sum_j \frac{A_{ji}}{K_j^{out}} \frac{K_j^{out}}{\mathcal N}=\frac{\overbrace{\mbox{$\sum_j$} A_{ji}}^{=K_i}}{\mathcal N}=\rho_i(t) 
\end{equation}
therefore (\ref{dens_network}) is a stationary state also for this network structure.

We can transfer the results from topological networks to this one and can state that the density of free particles $\rho_p^f$ inside a node, i.e. particles that are not associated to a cluster,  is proportional to the number of links which is given by the number of filaments $n_F$ crossing it. The spatial distribution of (local) free particle density therefore is proportional to the distribution of filaments $\rho_p^f = \bar\rho_p^f n_F/ \langle n_F \rangle$, where $n_F$ is the number of filaments in a node and $\bar\rho_p^f$ is the \emph{average} free particle density. This distribution however is scale dependent and the selection of the appropriate scale must be justified by other means.

In the graph approximation the system is modelled by a hopping of particles from one node to another. This introduces a time scale $\tau$ which is the time, a particle needs to travel from one node to another. If we use the average distance between intersections (mesh size of the network) $a$ as the length scale, the effective hopping time can be related to the node distance by $a= p d_s \tau_b$, where $\tau_b$ denotes the time a particle is bound to a filament, i.e. $\tau_b=\tau / (1+\omega_d/\tilde\omega_d)$ (see sec. \ref{single_q_sec}). Hence $\tau=a/p d_s(1+\omega_d/\tilde\omega_d)$. For this coarse graining scale nodes in the corresponding graph are connected on average by two filaments with adjacent nodes, i.e. $\langle n_F \rangle=2$ and the distribution of links (and therefore local densities) is governed by a Poisson distribution with mean value $\sigma=4$. If queues branch, queue cascades constitute a \emph{large} cluster that covers a number of filaments $n_F$ proportional to its size $m$. The structure of the graph must hence be adjusted since large clusters are able to span over more than one node as defined above. In order to treat clusters as single objects we consider all nodes a cluster covers as a single one. Then the network consists of cluster-nodes%made up by large clusters 
and free nodes where there are no stable clusters. We denote the number of cluster nodes by $N_{cl}$ and the cluster density $\rho_{cl}=N_{cl}/(L/a)^2$.

The free particle density $\rho_i$ in (\ref{dens_network}) corresponds to the probability that after long times a particle inserted anywhere will be at node $i$. Since the size $m_i$ of a cluster $i$ is proportional to the number of filaments $n_F$ it covers, it is proportional to its connectivity $K_i$ and therefore $\rho_i$. The probability of new inserted particles to end up in cluster $i$ is hence proportional to its size, i.e. the growth rate of a cluster is proportional to its size $m$.

\subsection{Cluster size distributions}

%Particles that are not part of clusters encounter only rarely for moderate densities and there dynamics can be treated as if they were non-interacting. The distribution is similiar to the one of non-interacting ones though their average density is decreased due to the amount of particles inside clusters. 
%{\bf Hier und auch spŠter sollte man det. Balance durch stationarity ersetzen und die Formulierung der n"achsten beiden S"atze noch einmal "uberarbeiten! Equilibrium sollte auch durch Stationarity o."a. ersetzt werden}

Due to particle conservation, the total inflow of particles in clusters $J_{in}$ must balance outflow $J_{out}$ in the stationary state. In the following we denote the total portion of particles associated to clusters by $\tilde N$ and free particles by $N_{f}=N-\tilde N$. The outflow can be expressed by $J_{out}=\omega_d^{eff} \tilde N$, where the effective rate $\omega_d^{eff}$ comprises the rate of particle detachment from a filament and attachment to another one in order to be moved to a free node. We can write $\omega_d^{eff}=\tilde\omega_d C$. The factor $C<1$ denotes the inference by interactions with particles on \emph{other} queues of the cluster and other filaments directed into the cluster that allow reattachment to the cluster. The factor $C$ is a mean value and depends explicitly on the structure of the clusters, but not explicitly on system parameters\footnote{At least in the network approximation where particles are assumed to be attached to filaments most of the time. For higher values of $\omega_d/\omega_a$ an explicit dependence on $\omega_a$ is assumed.}. It becomes small if very large clusters are present that can confine particles within their structure
(see paragraph on large clusters below). The flow of particles into clusters is given by $J_{in}\approx N_{f} \,\rho_{cl}\,\langle n_F \rangle/\tau$. The stationarity condition is $J_{out}=J_{in}$ and with $\tilde N = N_{cl} \langle m \rangle$, $\langle n_F \rangle=\langle m \rangle/\bar l$ we obtain
\begin{eqnarray}
\label{rho_free}
&&\tilde\omega_d C N_{cl} \langle m \rangle \approx \, N_{f} \frac{N_{cl}}{(L/a)^2} \, \frac{\langle m \rangle}{\bar l} / \tau \\ \nonumber
&\Leftrightarrow& \bar \rho_p^f=N_{f}/L^2 = \frac{\tilde\omega_d C\,\, \bar l}{a p d_s}\left(1+\frac{\omega_d}{\tilde\omega_a}\right)
\end{eqnarray} 
The corresponding number of particles associated to a cluster is $\tilde N = N-N_{f}$. For large $\omega_d^{eff}$ this quantity could reach zero so that there are no clusters left. This suggests a condensation transition between a free phase and a phase exhibiting clusters. However, we have to be careful since on the one hand the graph approximation does not work for large $\omega_d$ and on the other hand we neglected aster-like configurations of nodes where filament ends are arranged to point only into a node. These configurations can become relevant in this situation and allow clusters even for lower particle densities. We therefore assume that for $L\to\infty$ large clusters can be present even for small densities. However since we have seen in the last sections that most particles are in clusters for default parameters, we can assume that $\omega_d^{eff} \ll a^2\rho_p^0/(\tau\bar l)$ which suggests $C_1 C \ll 1$.
%The number of free particles in the stationary state is the rate of detaching particles, that do not immediately reattach multiplied with the time $\tau$ they stay free on average, i.e. $N_{free}=\omega_d^{eff} \tilde N \tau$, where $\tilde N=N-N_{free}$ ist the number of particles that are part of any cluster and $\omega_d^{eff}$ is the effective detachment rate from clusters. Note that $\omega_d^{eff}$ is \emph{not} identical to $\tilde\omega_d$ since it implies that particles do not rebind on \emph{another} filament of the same cluster. When particles are not attached to a cluster they perform a combination of normal diffusion and active transport. According to \cite{lipowski_network} this can be described on large time scales by an enhancd effective diffusion constant. Since a change of direction usually occurs by detachment and reattachment to another filament according to \cite{lipowski_network} we conclude that the diffusion constant $D_{eff}\sim \frac{(p d_s)^2}{\omega_d}\frac{\omega_a}{\omega_a+\omega_d/(\rho_F d_b)}$\footnote{This corresponds to regime (II) in \cite{lipowski_network}.}. The mean square distance a particle travels by diffusion is $\langle r_D^2 \rangle = D_{eff}\tau$, while the average distance between clusters is given by the cluster density : $d^2_{cl}=1/\rho_{cl}$ inserting this, we obtain $N_{free}=(1-\tilde N)=\tilde \omega_d^{eff}/\rho_{cl}D_{eff}\tilde N$.

In order to determine the probability distribution of cluster sizes $P(m)$, we apply an expansion of the system size. Assume the system to be in the stationary state. Increasing the system size by a small area $\Delta A$, while always remaining in the stationary state, $\Delta N=\rho_p^0 \Delta A$ new particles are inserted. The portion of cluster-associated particles hence is
\begin{equation}
\Delta\tilde N = \Delta N-\Delta N_{f} \approx \Delta A(\rho_p^0-\tilde\omega_d C \tau \bar l/a^2)
\end{equation}
which is the number of particles that are effectively added to the clusters. However, not only particles are added to the clusters but also $\Delta N_{cl}=\rho_{cl}\Delta A/a^2$  new clusters emerge within the new area $\Delta A$.
Thus for each new cluster that emerges $\Delta \tilde N/\Delta N_{cl}= \langle m \rangle = a^2 (\rho_p^0-\tilde\omega_d C \tau \bar l)/\rho_{cl}$ new particles are distributed among the clusters, while the probability that a particle is associated to a given cluster is proportional to its size $m$ as argued above. This process corresponds to a generalized Yule process, where between two cluster initialization events, $\langle m \rangle$ objects are distributed among the clusters (see e.g. the review \cite{power_laws}). The stationary state of the Yule process exhibits a distribution which approaches for large $m$ asymptotically a power law distribution $P(m)\sim m^{-\gamma}$ with an exponent $\gamma=2+\frac{m_0}{\langle m \rangle}$ where $m_0$ is the initialization value of the clusters. Hence the exponent yields
\begin{eqnarray}
\label{gamma}
\gamma &=& 2+\delta \\ \nonumber 
\delta &\approx& \frac{m_0 \rho_{cl} }{\rho_p^0-\tilde \omega_d C \tau \bar l/a^2}
\end{eqnarray}
If distances between clusters are large, the cluster density $\rho_{cl}$, i.e the number of clusters per area unit $\xi^2\approx a^2$, corresponds to the probability $P_{cl}$ that cluster seed nucleates at an intersection. In order to maintain a stable queue with at least one particle on a filament (i.e. a initial two-particle cluster), according to (\ref{rho_crit}) the line density of particles on the filament must be $\rho_p^1 = \bar\rho_p^f n_F/\langle n_F \rangle > \tilde \omega_d \rho_s/p  (1+\omega_d/\tilde\omega_a)$. Inserting (\ref{rho_free}) and $\rho_s=2/(\pi \,a d_s)$, this yields a condition on the local filament density
\begin{equation}
\frac{n_F}{\langle n_F \rangle} = \frac{2}{\pi \bar l C} \sim \frac{1}{a C} \sim \rho_s \,r_p \,d_s
\end{equation}
where the fact that $\bar l$ scales like $a/r_p$ was used. Since we are only interested in the dependence on system parameters, we neglect any prefactors that do not depend explicitly on them, like $C$. Since the filaments are Poisson-distributed with mean value $2$, this probability can be given by the cumulative Poisson distribution with mean $\sigma=2$:
\begin{equation}
P_{cl}=P\left(n_F > \frac{2}{C \bar l}\right) \sim \frac{\Gamma(2/C\bar l+1,2)}{\Gamma(2/C \bar l)} 
\end{equation}
and the dependence of $\gamma=2+\delta$ is given by
\begin{equation}
\delta \sim \frac{P_{cl}(\rho_{act}) \rho_{act}^2}{\rho_p^0 \left(1 - \frac{\pi \tilde\omega_d \bar l C\rho_s}{2 p \rho_p^0}\right)}
\end{equation}
by inserting $a= 2/(\pi\rho_{s}d_s)\sim 1/\rho_{act}$, considering $r_p$ and $d_s$ to be fixed by biological reasons. As argued above, we assume $C_1 C \ll 1$, hence the term $\pi \tilde\omega_d \bar l C\rho_s / 2 p \rho_p^0$ can be neglected for small $\omega_d$, and one can further simplify: 

\begin{equation}
\label{delta_approx}
\gamma-2 = \delta \sim \frac{P_{cl}(\rho_{act})\rho_{act}^2}{\rho_p^0}  \,\, .
\end{equation}

\begin{figure}
\begin{center}
\resizebox{\columnwidth}{!}{\includegraphics{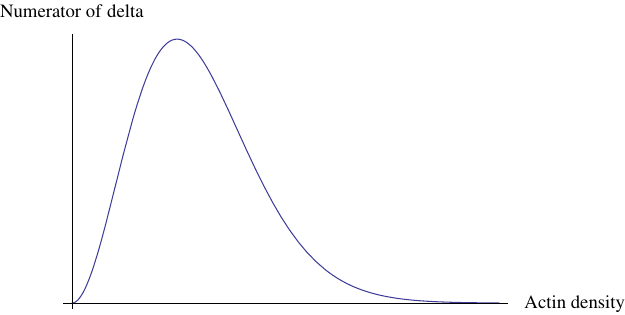}}

\caption{\label{P_cl} Sketch of the dependence of $\delta$ on the network density. }
\end{center}
\end{figure}

The dependence of $\delta=\gamma-2$ on $\rho_p^0$ appears to be a quite good approximation as can be seen in fig. \ref{gamma_rho_wd=0.002}. In fig. \ref{CD_disNetw}, one observes only a weak monotonic dependence on $\omega_d$ as long as $\omega_d$ is small, though this dependence becomes stronger for large $\omega_d$ as expected.  The numerator depends only on $\rho_{act}$ and has the form as displayed in fig. \ref{P_cl} (the scale is not defined, since the prefactor is not given). However, only for small values of $\rho_{act}$ the cluster density can be approximated by the probability $P_{cl}$. For large network densities, if clusters on average cover many intersections, the effective cluster number is smaller since clusters nucleating on different intersections can merge. Therefore we assume that the the relevant values are restricted to the lower branch which attains a monotonic growth. This appears to be valid as is shown in Fig. \ref{CD_disNetw}.

%%%%%%%%%%%%%%%%%%%%%%%%%%%%%%%%%%%%%%%%%%%%%%%%%%%%%%%%%%%%%%%%%%%%%%%%%%%%%%%%%%%%
%%%%%%%%%%%%%%%%%%%%%%%%%%%%%%%%%%%%%%%%%%%%%%%%%%%%%%%%%%%%%%%%%%%%%%%%%%%%%%%%%%%%

Summarizing, we can say that for small $\omega_d$ relative to $\tilde\omega_a$ and mean particle density $\rho_p^0$, $\delta$ is proportional to $1/\rho_p^0$ (Fig. \ref{CD_disNetw}) and depends increasingly on $\rho_{act}$ (though not linear in general). In these limits there is no dependence on other system parameters (see fig. \ref{CD_disNetw}). For larger $\omega_d$, the influence of $\omega_d$, $\omega_a$, $l_D$ and $p$ becomes relevant.

%($\rho_s^{\rm perc}\approx 5.8$ filaments/$L_F^2$ \cite{sornette_fiberperc}). 

So far, we neglected free particles that also contribute to clusters. This approximation yields good results for low particles densities and if clusters are not too large. Then cluster structure is mainly one dimensional, made up by queues and only some free particles in their neighborhood whose influence on detachment can be comprised in $\tilde\omega_d$. Large clusters however rather have a two- than a one dimensional structure since detaching particles can completely fill the cavities engulfed by queues. Then the full cluster size is $\tilde m\sim m^2$. Cluster attachment and detachment are still determined by the one dimensional fraction $m$ constituted by the queues, thus arguments from above remain valid. Hence the distribution of 2D-clusters yields
\begin{equation}
\label{gamma_crossover}
P(\tilde m)=\left. P(m)\right|_{m=\tilde m^{1/2}} \frac{\partial m}{\partial \tilde m}\sim \tilde m^{-\gamma'}
\end{equation}    
with $\gamma'=\gamma/2+1/2$. The distribution is also described by a power law but with a decreased exponent $\gamma'$. This result is consistent with simulation results in fig. \ref{CD_disNetw}(a). Note that high filament densities suppress this effect since no big cavities between filaments are present (cf. \ref{CD_disNetw}(d)). High $\omega_d$ / low $\omega_a$ enhance the effect, since due to a large number of unbound particles their contribution to clusters is enhanced.

%If the mean density $\rho_p^0$ approaches $\rho^*$, the average distance of clusters becomes smaller so that they can overlap and fuse. These cases are not reproduced by the approach introduced here. In contrast to (\ref{gamma}) where $\gamma$ is restricted to values larger than two, merging clusters reduces the number of clusters while increasing cluster sizes. This corresponds to a negative $\Omega$ which enables values of $\gamma<2$ and might explain the crossover for large $\rho_p^0$ where exponents $\gamma<2$ are observed at large values of the cluster size distribution.  

In principle the above considerations are also valid for a regular network, while in those systems the particle density is homogeneously distributed and $a=const$. If the density for cluster initialization (l=1) is exceeded, clusters can emerge anywhere in the system. As long as the density is lower than the critical density $\rho^*(a)$ to form queues of length $a$, only small L-shaped clusters, consisting of two queues, emerge (see configuration in fig. \ref{config_reg_netw}(a)). These are characterized by a centered distribution $P_{m_0}$ exhibiting fluctuations around a mean value $m_0$%an Gaussian distribution $P(m)\sim e^{-(m-m_0)^2/\sigma^2}$
, while no cluster branching occurs. Since the scale of queue lengths is in the same order of magnitude as random clustering (cf. fig. \ref{random_clusters_discr}), the exponential background of free particles must be added, hence $P(m)\sim P_{m_0}(m)+1/m_r\exp(-m/m_r)$ where $m_r$ is the scale of random clusters\footnote{Note that due to the attractive interaction of filaments a lateral aggregation of particles is induce that locally increases the density compared to the average density $\rho_p^0$. This leads to an increased size scale of random clusters}. This behavior corresponds to fig. \ref{CD_reg_netw} where a bulge over an exponential is exhibited. However, if $\rho^*(a)$ is exceeded the critical density is exceeded cluster cascades can develop. However, due to the homogeneous density clusters can branch at any intersection and due to the high cluster density, they merge forming a mesh shaped structure. This leads to a percolative behavior yielding a scale free distribution, while clusters are not well separated (cf. fig. \ref{config_reg_netw}(b)), thus the cluster size distribution depends sensitively on the coarse graining scale. Hence cluster distributions do not follow the same scheme as disordered networks.

While scale free clustering in regular networks only emerges for $\rho_p^0>\rho^*$, in inhomogeneous networks this can occur also for small densities $\rho \ll \rho^*$ since the distribution of the particle density and filament distances determining the critical density is wide. Only few regions where the critical density is exceeded are needed for scale free clustering. This corresponds to a Griffith phase where only locally critical values are exceeded exhibiting an ordered structure, in contrast to the case when the full system is clustered\footnote{Note that this does not imply percolation. There are dilute regions behind intersections emerging that can tear clusters apart, if there are only few free particles.}. Therefore clusters are well separated and do not depend significantly on the coarse graining scale.

\section{Discussion}

We examined transport of hard core particles (discs) on regular and inhomogeneous networks embedded in a two dimensional diffusive environment.
% in comparison to a system without network that includes attractive interactions and exclusion. 
The models consist of regions where particles perform diffusive motion and \emph{one dimensional stripes (filaments)} to which particles can attach and perform directed motion. In most parameter regimes clusters are observed. As a reference for clustering without transport networks we examined cluster features of a model with attractive interaction of particles.

A detailed analysis of cluster size distributions (CDs) shows that there are qualitative differences between the regular and inhomogeneous network structure. On regular networks one typically observes a typical size-scale to the clusters at low densities. Algebraic CDs are only observed at large densities when single clusters merge to form large mesh-shaped cluster complexes. In this case, however, clusters are not well separated.

In contrast disordered networks, where filament lengths and orientations are randomly distributed, exhibit algebraic cluster size distributions in a wide range of the parameter space, indicating that clusters on all size-scales exist. It is important to notice that the algebraic CDs are observed one order of magnitude below the respective densities in reference systems.
In the disordered network, clusters are well separated exhibiting a compact structure. The transport driven clustering on disordered networks therefore appears to be an alternative mechanism to generate scale free cluster size distributions compared to e.g. the one studied in \cite{meakin_family} which is driven by irreversible cluster-cluster-attachment. In contrast to that model, however, here the distribution is even scale free in the stationary state. 

% Switching off the hard core potential results in a strong suppression of large clusters, while the distribution of particles in the disordered network still remains inhomogeneous. Hence, inhomogeneity of the network structure as well as exclusion interaction strongly enhance the formation of macroscopic clusters. 

Clusters are assumed to nucleate if two particles encountering at an intersection of two filaments block each other such that they cannot move, while other particles running towards the intersection form queues. Below a critical local particle density, queues remain small, exhibiting fluctuations around the mean length. However, if the particle density exceeds a critical density $\rho^*$, queues covering multiple filaments induce branching of queues into large clusters, while each branch contributes to cluster inflow. Since the number of filaments a cluster covers is proportional to its size, the growth rate of a cluster is proportional to its size. Dynamics of this kind can be described by a Yule process (see e.g. \cite{power_laws}) which yields a power law distribution $P(m) \sim m^{-\gamma}$ with an exponent depending on the microscopic parameters (see also preferential attachment in scale free networks \cite{BA_networks}). A thorough investigation shows that for small detachment rate and moderate network densities, the exponent merely depends on the particle density $\rho_p^0$ and the network density represented by $\rho_{act}$, while dependence on other parameters is weak.    

In regular networks, the mesh size $a$ and particle density $\rho_p^0$ are homogeneous. If the particle density is smaller than the critical density $\rho^*$ no clusterbranching occurs and only small clusters, each consisting of two queues emerge. For $\rho_p^0 > \rho^*(a)$ there are large clusters distributed on all scales. In latter case clusters can emerge at any point of the system such that they are crowded and not well separated. In contrast in disordered networks the inhomogeneous distribution of filaments leads to an inhomogeneous particle- and mesh size distribution. Even for $\rho_p^0 \ll \rho^*$, locally, the critical value can be exceeded, to induce cluster branching. These clusters are well separated for moderate densities and exhibit the scale free distribution explicated above. The regime $0 < \rho_p^0 < \rho^*$ corresponds to a Griffith phase where only small parts of the disordered system are in the cluster phase.  

The analysis of cluster size distributions in the system of diffusive particles with attractive interaction shows clustering on a finite size-scale and a maximum in the cluster size distribution. The size scale itself increases with time. In contrast to the diffusion driven irreversible aggregation modelled in \cite{meakin_family} which exhibits scale free distributions at transient states, this model exhibits reversible clustering, which appears more realistic for interactions of membrane proteins \cite{gil_memb_prot}.  

In summary, we have found that the microscopic particle dynamics as well as the network structure have significant influence on the qualitative form of the cluster distribution. From our point of view this observation is of great importance for the analysis of biological systems since it is often not possible to identify the underlying microscopic mechanisms leading to experimentally observed aggregation. In these cases the analysis of the cluster distribution on larger scales may answer the question whether observed patterns are the result of active transport or of aggregation due to attractive interactions.

\begin{acknowledgement}
We thank M. Schmitt, A. Schadschneider, O. Pulkkinen, S. Dmitrieff and O. Markova for fruitful discussions and the German Science Foundation under grant number DFG GK 1276/1 for financial support.
\end{acknowledgement}

\begin{appendix}

\section*{Appendix A: Filament dynamics and default parameters}
\label{appendix}

\begin{table*}
\begin{center}
\begin{tabular}{ccc}
Process & Description & Probability \\ \hline \\
\emph{Nucleation} & 
\begin{minipage}[t]{8cm} Initialization of filaments with arbitrary direction at an arbitrary point in the system. The (-)-end receives a cap inhibiting shrinking.  \end{minipage} & $\omega_n\,\rho_{act} \,\rho_{ARP}$ \\
\emph{Branching} & 
\begin{minipage}[t]{8cm} New filaments are initialized at an existing one (not necessarily the (+)-end; angle between parent filament and branch=$70^o$\cite{arp_nuc+branch}.  \end{minipage} & $\omega_b\,\rho^2_{act} \,\rho_{ARP}$ \\
\emph{Growth} & 
\begin{minipage}[t]{8cm} New subunits are generated at the (+)-ends of filaments \end{minipage}. & $\omega_g\rho_{act}$ \\
\emph{Shrinking} & 
\begin{minipage}[t]{8cm} Subunits are removed at the (-)-end of filaments if the end is not capped. \end{minipage} & $\omega_s$ \\
\emph{Uncapping} & 
\begin{minipage}[t]{8cm} Caps are removed. \end{minipage} & $\omega_u$ \\
\end{tabular}
\caption{\label{dis_netw_dyn_tab} Dynamics of the filament network.}
\end{center} 
\end{table*}
%%%%%%%%%%%%%%%%%%%%%%%%%%%%%%%%%%%%%%%%%%%%%%%%%%%%%%%%%%%%%%%%%%%%%%%%%%%%%%%%%%%%

%%%%%%%%%%%%%%%%%%%%%%%%%%%%%%%%%%%%%%%%%%%%%%%%%%%%%%%%%%%%%%%%%%%%%%%%%%%%%%%%%%%%
\begin{table*}
\begin{center}
\begin{threeparttable}
\begin{tabular}{cccc}
Parameter name & Reference & Reference Value & model parameters \\ \hline \\
\emph{Filament dynamics}: &  &  &   \\
nucleation rate $\omega_n$ & \cite{actin_dynamics_exp1} & $8.7\, \times 10^{-5} \mu M^{-2}s^{-1}$ & $1.0 \times 10^{-5} \,lu^{-6} \,tu^{-1}$  \\ 
growth rate $\omega_g$ & \cite{actin_dynamics_exp1} & $8.7\mu M^{-1}s^{-1}$ & $0.5 \,lu^3 \,tu^{-1}$   \\ 
shrink rate $\omega_s$ & \cite{actin_shrinkrate} & $4.2 s^{-1}$ & $0.075 \,tu^{-1}$  \\
branch rate $\omega_b$ & \cite{actin_dynamics_exp1} & $5.4\,\times10^{-4}\,\mu M^{-3} \,s^{-1}$ & $0.0001 \,lu^9 \,tu^{-1}$   \\ 
%attachment $\omega_a$ & & ? &  \\ 
%detachment $\omega_d$ & & $\approx 2 \mu M$ & $0.05-0.1/tu$  \\ 
%diffusive step $l_D$  & & $D\approx 0.1-1\mu m^2$ & 1 $lu$  \\ 
uncap rate $\omega_u$ & \cite{actin_dynamics_exp1} & $0.0018\,s^{-1}$ & $0.0001 \,tu^{-1}$  \\
actin density $\rho_{act}^0$ & \cite{actinmesh_pics} & meshsize: $0.1-1\mu m$ & $2 \,lu^{-3}$\,\,\tnote{1}   \\
ARP2/3 density $\rho_{ARP}^0$ & \cite{actindyn1} & $0.1\mu M$ & $0.1 \,lu^{-3}$  \\
\emph{Particle dynamics}: &  &  &   \\
particle radius $r_{p}$ & \cite{vesicle_radius} & 42.5nm (average) & 0.5 $lu$  \\
binding distance $d_b$ & \cite{lipowski_network} & 1 site (50nm) & 0.5 $lu$  \\
subunit distance $d_s$ & \cite{alberts} & 36nm & 0.36 $lu$  \\
attachment $\omega_a$ & \cite{lipowski_network} & 1/4 of diffusive steps & 0.25 $tu^{-1}$  \\ 
detachment $\omega_d$ & \cite{lipowski_network} & $0.8s^{-1}$ & $0.02 \,tu^{-1}$  \\ 
diffusive step length $l_D$  & \cite{lipowski_network} & 1 per time step & 0.5 $lu$  \\ 
step rate $p$ & \cite{lipowski_network} & $20s^{-1}\Rightarrow v=1\mu m/s$ & $0.75 \,tu^{-1}$  \\
particle density $\rho_p^0$ & \cite{govindan1995_vesicles} & 10-60 vesicles in bud ($\sim$0.75$\mu$m radius) & $0.04\,lu^{-3}$
\end{tabular}
\caption{\label{default_parameters} Default parameters of the model which are biologically motivated by transport of vesicles by myosin on actin filaments. The referenced values are either based on experimental data or existing models for intracellular transport \cite{lipowski_network} and filament dynamics \cite{actindyn1}. Model parameters are chosen to be in the order of magnitude of referenced values, fitted to time and space scale of the simulations. Length scale: $1\,lu$ = 100nm = $2r_{p}$   $\Rightarrow 1\mu M=0.6\,lu^{-3}$.
Time scale: $1 \,tu=\Delta t=0.025s$. By default we consider square systems of system length $L=200\,lu$ and a layer thickness of $1\,lu$. In theoretical considerations the layer thickness however is neglected and the system is treated two dimensional.}
\begin{tablenotes}
\item[1]We adjusted $\rho^0_{act}$ such that the mesh size was in the order of magnitude as in the referenced work.
\end{tablenotes}
\end{threeparttable}
\end{center} 
\end{table*}
%%%%%%%%%%%%%%%%%%%%%%%%%%%%%%%%%%%%%%%%%%%%%%%%%%%%%%%%%%%%%%%%%%%%%%%%%%%%%%%%%%%%

%%%%%%%%%%%%%%%%%%%%%%%%%%%%%%%%%%%%%%%%%%%%%%%%%%%%%%%%%%%%%%%%
\begin{figure}
\resizebox{\columnwidth}{!}{\includegraphics[width=8cm]{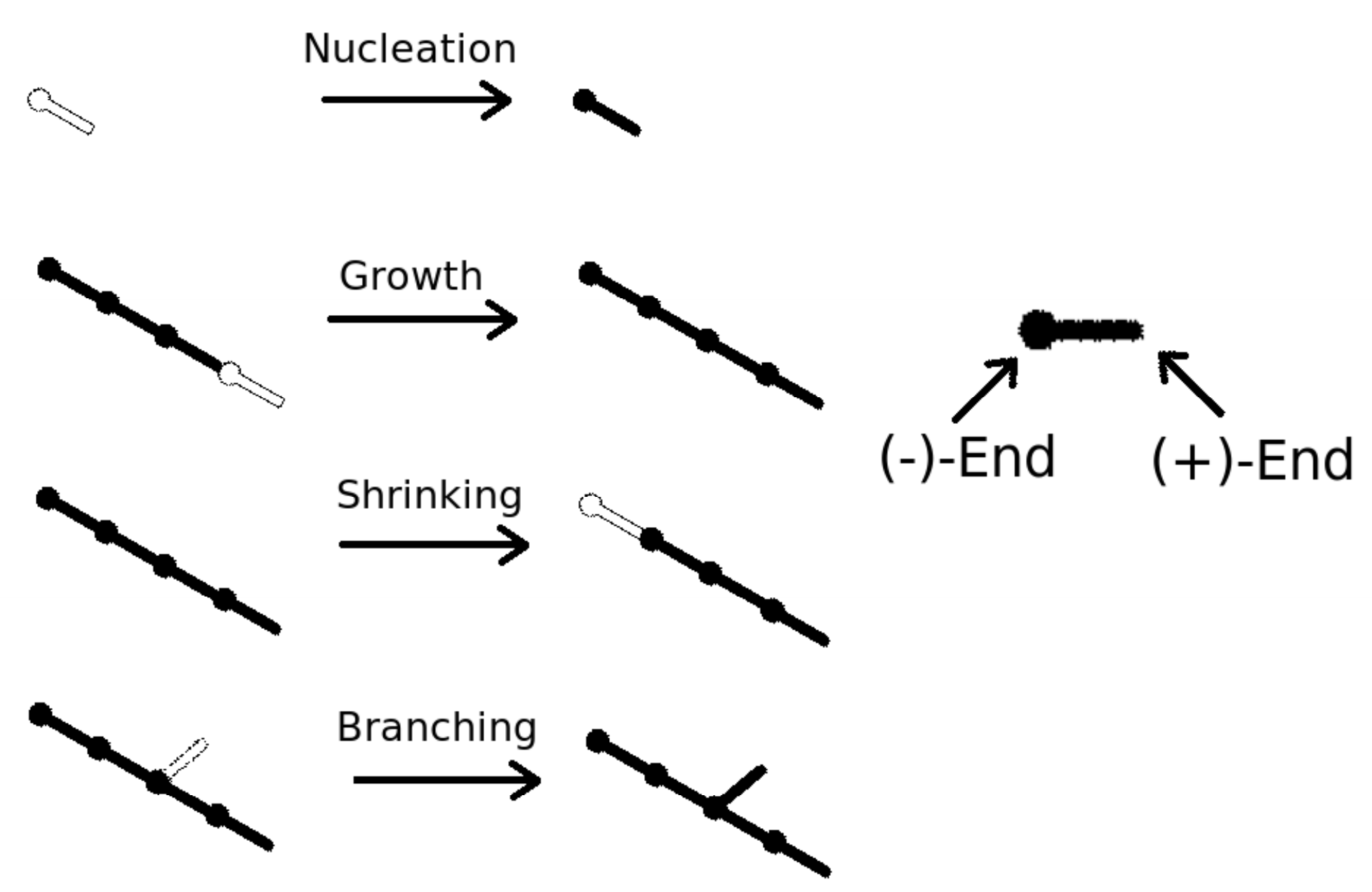}}
\caption{Illustration of the filament dynamics. Filaments are implemented as sequences of subunits (small dots, corresponding to actin monomers) separated by a distance $d_s$ (short bars). Filament subunits are polarized, with a (+)-end where subunits are created to elongate, and a (-)-end where subunits dissociate causing shrinking.}
\label{dis_netw_filam_illust}
\end{figure}
%%%%%%%%%%%%%%%%%%%%%%%%%%%%%%%%%%%%%%%%%%%%%%%%%%%%%%%%%%%%%%%%%%%%%%%%%%%%%

%%%%%%%%%%%%%%%%%%%%%%%%%%%%%%%%%%%%%%%%%%%%%%%%%%%%%%%%%%%%%%%%%%%%%%%%%%%%%%%%%%%%
\begin{figure}
\begin{center}
\resizebox{\columnwidth}{!}{\includegraphics{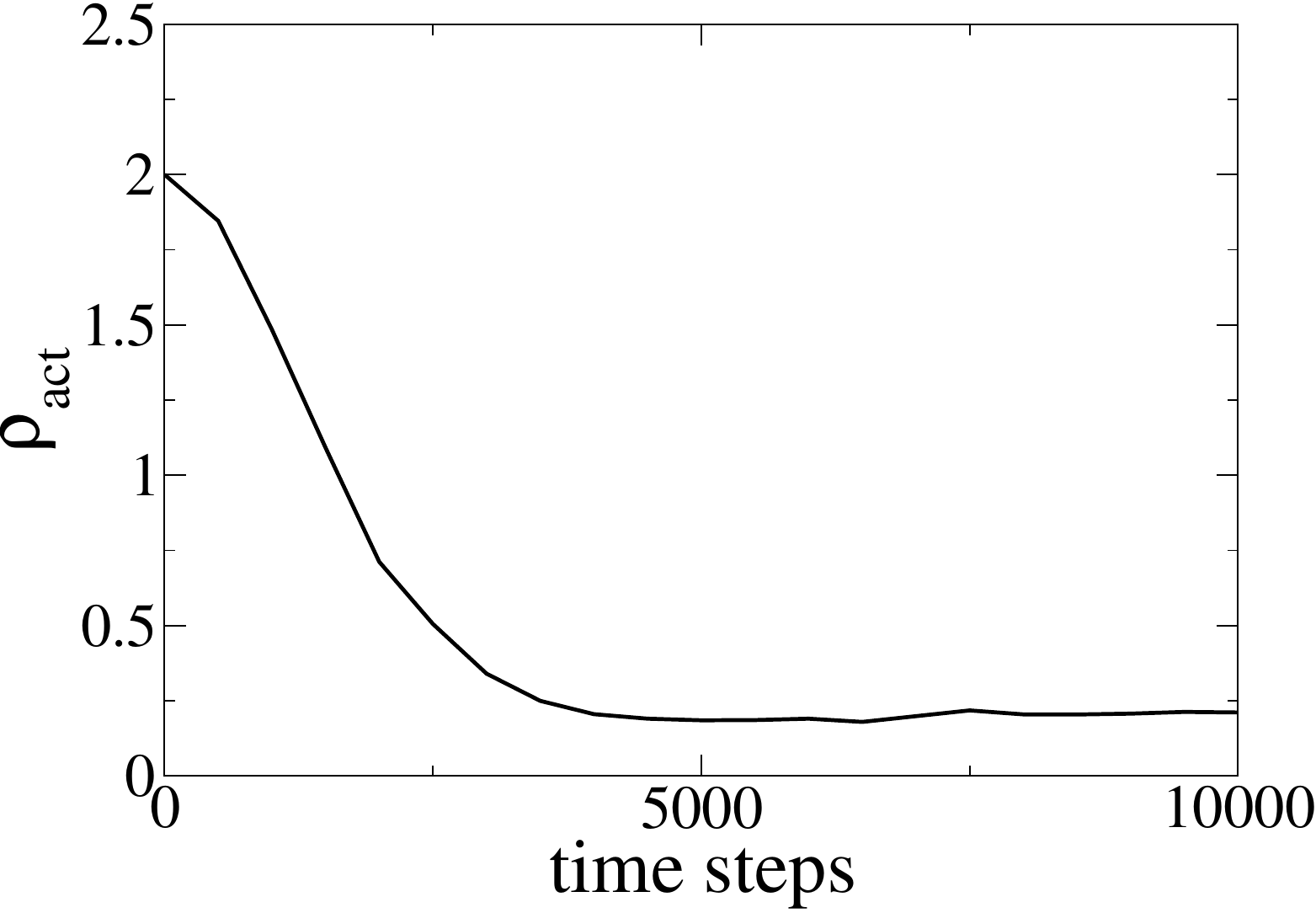}}
\caption{\label{nodenumber(t)} Density of free actin $\rho_{act}$ in dependence on time. After $5000$ time steps a stationary state is reached.}
\end{center}
\end{figure}
%%%%%%%%%%%%%%%%%%%%%%%%%%%%%%%%%%%%%%%%%%%%%%%%%%%%%%%%%%%%%%%%%%%%%%%%%%%%%%%%%%%%

While we assume that the qualitative properties of the system only require randomness in filament directions- and lengths, the explicit algorithm of network generation must be explicated.  In order to be comparable to real biological situations, we generated the networks structure by stochastic processes that mimic the growth dynamics of real actin networks \cite{alberts}. Therefore we implemented dynamics as described in Table \ref{dis_netw_dyn_tab}, which is illustrated in Fig. \ref{dis_netw_filam_illust}.

The quantity $\rho_{ARP}$ introduced in Table \ref{dis_netw_dyn_tab} represents the density of free \emph{$ARP2/3$-complexes} that serve as nucleation and branching seeds for filaments, while the actin density $\rho_{act}$ corresponds to the density of free actin subunits constituting the filaments. Their initial values are $\rho^0_{ARP}$ and $\rho^0_{act}$ which corresponds to the case that all monomers are dissociated. The densities decrease with the growing filament network as shown in fig. \ref{nodenumber(t)}. After 5000 steps the actin density attains a stationary value. We therefore stop network dynamics at this point. Since in the stationary state association and dissociation of subunits must balance, $\omega_g\rho_{act}\approx \omega_s \, \Rightarrow \,\rho_{act}\approx \omega_s/\omega_g$ \footnote{The contribution of filament nucleation can be neglected for average filament lengths $\gg 1$ } (cf. \cite{alberts} for actin networks). The density of filament-subunits hence is $\rho_s=\rho_{act}^0-\rho_{act}\approx\rho_{act}^0$ for $\omega_s \ll \omega_g$, therefore the network density is mainly determined by $\rho_{act}^0$. In order to keep dynamics simple but retaining the crucial features of disordered networks, branching and the dynamics of ARP2/3 were neglected in Sec. \ref{actin_netw_sec} to obtain a network of uncorrelated filament orientations. However, we resume these dynamics appendix B.

Although we do not consider a particular biological system, we choose parameters to fit the typical order of magnitude in real vesicular transport. If not stated differently, we will use default parameters displayed in table \ref{default_parameters} for our simulations. The referenced works used experimental and modeling techniques to obtain the data given in the third column. For particle dynamics, we choose the parameters to be consistent with the discrete model introduced in the last section relying on the model in \cite{lipowski_network}.

\section*{Appendix B: Networks with branching filaments}
\label{with branching}

In actin networks, branching of filaments takes place quite frequently, resulting in a dendritic network structure. In Sec. \ref{actin_netw_sec} branching of filaments was neglected in order to avoid correlations of filament orientations. If one is interested in the dynamics of vesicle transport on submembranal actin networks, one has to consider this process as well. We checked CDs in a system with finite branching rate $\omega_b$ (here: branching probability $\omega_b \Delta t=0.0035$) including the dependence of growth dynamics on the ARP2/3-density $\rho_{ARP}$ (cf. section \ref{disNetw_sec}). In this system, filament orientations are highly correlated.

\begin{figure}
\begin{center}
\resizebox{\columnwidth}{!}{\includegraphics{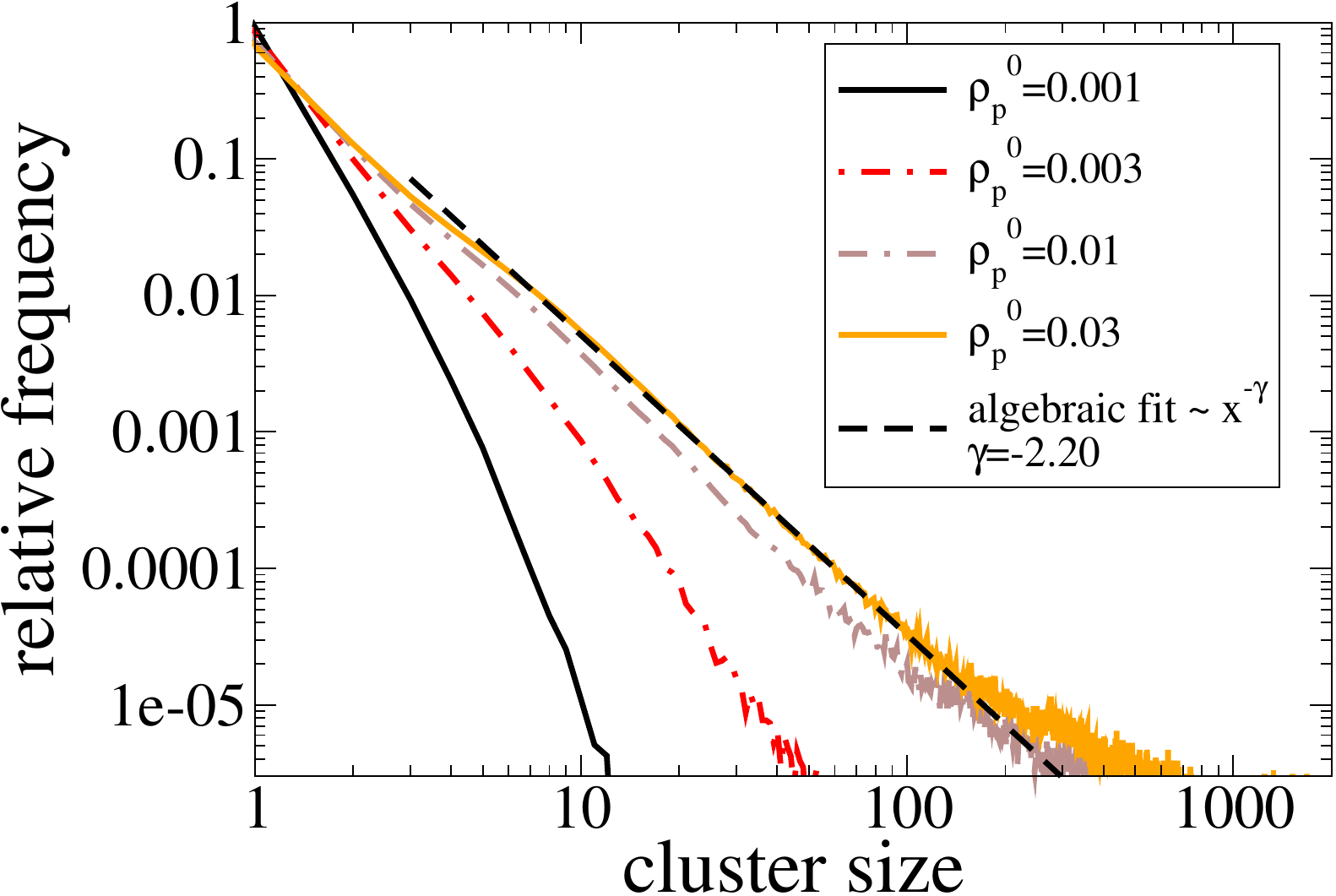}}

\caption{\label{CD_actinnetw_rhovar} Cluster size distribution for branch rate $\omega_b>0$. Like in the system with non-correlated filaments, the CD exhibits algebraic decay for large clusters. }
\end{center}
\end{figure}

%One system where quasi two dimensional disordered networks are realized is the submembranal actin network of cells. By giving the branch rate $\omega_b$ a finite value include the dependence on the variation of $\rho_{ARP}$, the system can be interpreted as a model for growth dynamics of actin networks. These networks are governed by dendritic tree-like structures that exhibit strong correlations of filament orientations. 
In fig. \ref{CD_actinnetw_rhovar} we display cluster size distributions for different particle densities $\rho_p^0$. As in the more basic case of uncorrelated filaments, $\omega_b=0$, one observes an algebraic decay as well. This indicates that the scale free behavior of the clusters is a robust feature of inhomogeneous transport networks and active particles exhibiting mutual steric interaction.

%%%%%%%%%%%%%%%%%%%%%%%%%%%%%%%%%%%%%%%%%%%%%%%%%%%%%%%%%%%%%%%%%%%%%%%%%%%%%%%%%%%%
%%%%%%%%%%%%%%%%%%%%%%%%%%%%%%%%%%%%%%%%%%%%%%%%%%%%%%%%%%%%%%%%%%%%%%%%%%%%%%%%%%%%

\end{appendix}

\bibliography{all}

\end{document}